\documentclass{egpubl}
\usepackage{egsr18}

\usepackage{booktabs} 

\usepackage{import}

\usepackage{t1enc,dfadobe}

\usepackage{cite}

\usepackage{tabularx} 
\usepackage{tabulary}
\usepackage{amsmath}
\usepackage{multirow}

\usepackage{caption}
\usepackage{subcaption}
\hfuzz=\maxdimen
\usepackage{graphicx}   
\usepackage{balance}

\title[Polycube Deformation]%
{Robust Edge-Preserved Surface Mesh Polycube Deformation}

\author[H.Zhao   X.Gu  N.Lei X.Li P.Zeng \&K.Xu ] 
{\parbox{\textwidth}{\centering Hui Zhao$^{1}$     Na Lei$^{4}$ Xuan Li $^{2}$   Peng Zeng $^{1}$  and Ke Xu$^{3}$  Xianfeng Gu$^{2}$   }
	\\
	{\parbox{\textwidth}{\centering     $^1$Tsinghua University, China\\  
			$^2$State University of New York at Stony Brook, USA\\
			$^3$Beijing University of Technology, China\\
		    $^4$Dalian University of Technology, China
		} 
	}
}

\begin{document}

     \teaser{
	 \includegraphics[width=0.9\linewidth]{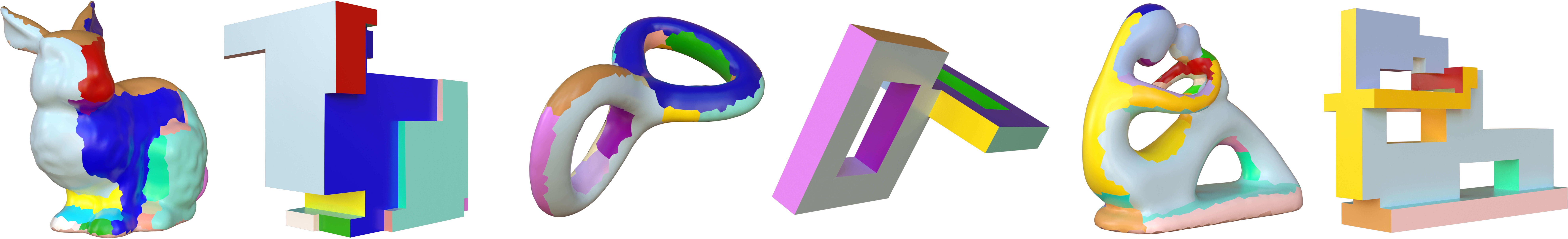}
	  \centering
	  \caption{Three surface models and their corresponding polycube shapes}
	  \label{fig:teaser} 
	 }
	\maketitle


	\begin{abstract}
The problem of polycube construction or deformation is an essential problem in computer graphics.
In this paper, we present a robust, simple, efficient and automatic algorithm to deform the meshes of arbitrary shapes into their polycube ones. 
We derive a clear relationship between a mesh  and its corresponding polycube shape. Our algorithm is edge-preserved, and works on surface meshes with or without boundaries.  
Our algorithm  outperforms previous ones in speed, robustness, efficiency.  Our method is simple to implement. To demonstrate the robustness and effectiveness
of our method, we apply it to hundreds of models of varying complexity
and topology. We demonstrate that our method compares favorably to other state-of-the-art
polycube deformation methods.


\end{abstract} 
	\section{Introduction} 

Polycube denotes a special geometric shape whose face normals are aligned with the one of the six axis of a specific orthonormal coordinate frame.   A polycube captures overall and global shape features of a mesh, and remove its local details. 

Polycube  is coined and firstly proposed in \cite{tarini2004polycube} to extend cube mapping to general shapes. This kind of special shapes generalize   Geometry Images \cite{gu2002geometry} to allow geometry and texture stored efficiently.  Due to its highly regular structure and special global parametric domain,   polycube is applied in many kinds of graphics applications, such as: surface texturing   \cite{tarini2004polycube}, volumetric texturing \cite{chang2010texture}, parameterization  \cite{yao2008adaptive,garcia2013interactive}, reconstruction \cite{wang2008user}, hexahedral remeshing \cite{gregson2011all,xia2010direct,han2010hexahedral,Livesu:2013:PolyCut}, shape morphing\cite{fan2005mesh}, spline constructions \cite{wang2008polycube,wang2008user}, volumetric mapping\cite{li2007harmonic,he2008harmonic} and T-mesh construction \cite{liu2015feature}.

Constructing polycube shapes from meshes  is a challenging problem. In earlier works, the polycube meshes are constructed manually \cite{tarini2004polycube,yao2008adaptive}, which needs a lot of tedious, labor-intensive user interaction with great care. After the polycube is available, these methods need  another extra algorithm  to achieve the cross-surface map between the polycube and the mesh \cite{wang2008polycube}. The cross-surface map is also a challenge problem itself \cite{wang2008user}. If we change the polycube, we need rebuild the map.

In this paper, we propose a  novel, automatic polycube deformation algorithm applied on surface mesh. Our method  separates the polycube construction process into   three  components explicitly: segmentation, polycube topology  and polycube geometry. 
Our major contribution is on the polycube geometry step, we propose a face normal rotation based deformation method.
The technique we implement can process all kinds of meshes: such as different genus, orientation or non-orientation,  with  or without boundaries.
  Compared to previous methods, our algorithms is   more efficient, robust, fast, accurate     and  can  deform an arbitrary mesh  of complicated geometry
and arbitrary topology into its polycube. There is no pre-processing or post-processing cleanup processing,  and topological degeneracies in our experimental results.
As we achieve the polycubes by deforming original meshes,   we get  a direct cross-surface parameterization between the meshes and their corresponding polycube shapes automatically.

	\section{Related Works}

Recently some automatic algorithms are proposed in \cite{he2009divide,lin2008automatic,gregson2011all,huang2014ell}.  The authors  uses a segmentation method to patch the input mesh, then use box-primitives to approximate it coarsely in \cite{lin2008automatic} . But this method fails in complicated models.  \cite{he2009divide} applies a distance-based, divide-and-conquer algorithms to build the polycube. The method in \cite{he2009divide} generates over-refined polycubes and is sensitive  to off-axis features. The algorithms in \cite{he2009divide,lin2008automatic} are based on surface meshes and can not build the cross-surface map automatically.  While the algorithms in \cite{gregson2011all,huang2014ell} are volume mesh based, these two algorithms look for a specific polycube which minimizes the distortion of the volumetric map.

Polycube is a an axis aligned shape which mimics the original shape,  but with a geometric characterization that its every face normal is  aligned with one of the axis of an orthonormal coordinate frame respectively. Therefore the value of the $\ell_1$-norm of every unit face normal of the polycube shape is equal to $1$  \cite{huang2014ell}.

 Based on this observation,  \cite{huang2014ell} defines  a $\ell_1$-norm energy which is weighted by triangle areas, then proposes a variational method to deform an input triangle
  mesh into its polycube shape by the minimization of  this energy.  The  $\ell_1$ norm term  of the mesh's face normals and the weight term of triangle areas  both are non-linear in the mesh position. Then they change the unconstrained system into a constrained minimization to make it   well-behaved, yet numerically
  tractable and efficient. Finally complicated numerical method is resorted to solve the minimization problem in  \cite{huang2014ell}.    
  To decrease the distortion, a volumetric mesh is created from surface mesh. Then an as-rigid-as-possible volumetric distortion energy \cite{alexa2000rigid} is used to regularize the system. There are a lot of minimum for the deformation energy, therefore  one regularization approach is also used to single out the desired polycube \cite{huang2014ell}.

 Because the resulting polycube is different with varying orientation, they also introduce an energy term to find the optimal global orientation for the polycube. This energy is integrated into their whole system.  However, the optimal polycube in all orientation is an ill-defined concept. We do not think there should be an optimal polycube.

The polycubes achieved from above minimization often have some spurious topological degeneracies. A post-processing cleanup step is used to fix the problem in \cite{huang2014ell}. Their method also requires the input mesh is closed, however our method can process open boundary meshes.

 The method in \cite{gregson2011all} proposes to align the surface normals of an input mesh with one of six axes($ \pm X, \pm Y, \pm Z$) gradually, this step is called rotation-driven deformation. However the result can not be a perfect polycube and second position-driven deformation is required to align each polycube face with the corresponding axis exactly and
 enforce planarity.   This method is similar to ours, but our algorithms uses different method to compute rotations, therefore ours can converge to enforce the faces to be planer and do not need other post-processing steps .

The algorithm in \cite{wan2011topology} requires an existing polycube, then they can optimize it to match the desired quality. 
This method in \cite{Livesu:2013:PolyCut} attacks the polycube segmentation problem. They use graph-cut based approach to achieve a polycube base-complex to satisfy  certain quality requirement. The algorithms can balance parameterization distortion against the singularity count of a polycube. We use the same segmentation step as theirs \cite{Livesu:2013:PolyCut,Fu2016PC} , we propose a different polycube geometry deformation method from theirs.

In \cite{Fu2016PC}, they propose another polycube method which is similar to ours. Their algorithm is also a normal-driven method.      Given a polycube with complicated topology, a simplification method is proposed in \cite{cherchi2016polycube}.

 Poisson system based deformation  \cite{yu2004mesh} is well-known technique.  After the rotations of all triangle faces are known, the triangles can be rotated into the new orientation, then the Poisson system is used to blend the triangle soup together and reconstruct a consistent mesh  into its new shape. The rotations can be computed according to the different application. In  \cite{xu2006poisson} the rotation is achieved by the interpolation from two corresponding meshes. In \cite{zayer2005harmonic}, they interpolate the rotations of  triangles.

	\section{Our Algorithms}

\textbf{Polycube} is a shape formed by joining several rectangle face against face and the surface normals
of a polycube are axis-aligned.
Polycube is also called as orthogonal polyhedra \cite{Livesu:2013:PolyCut,eppstein2010steinitz}. We observe that there are three  components in deforming a mesh into its polycube shape: segmentation, polycube topology and polycube geometry.

\begin{figure} [h!] 
	\centering   
	\begin{subfigure}[b]{0.115\textwidth}
		\includegraphics[width=\textwidth]{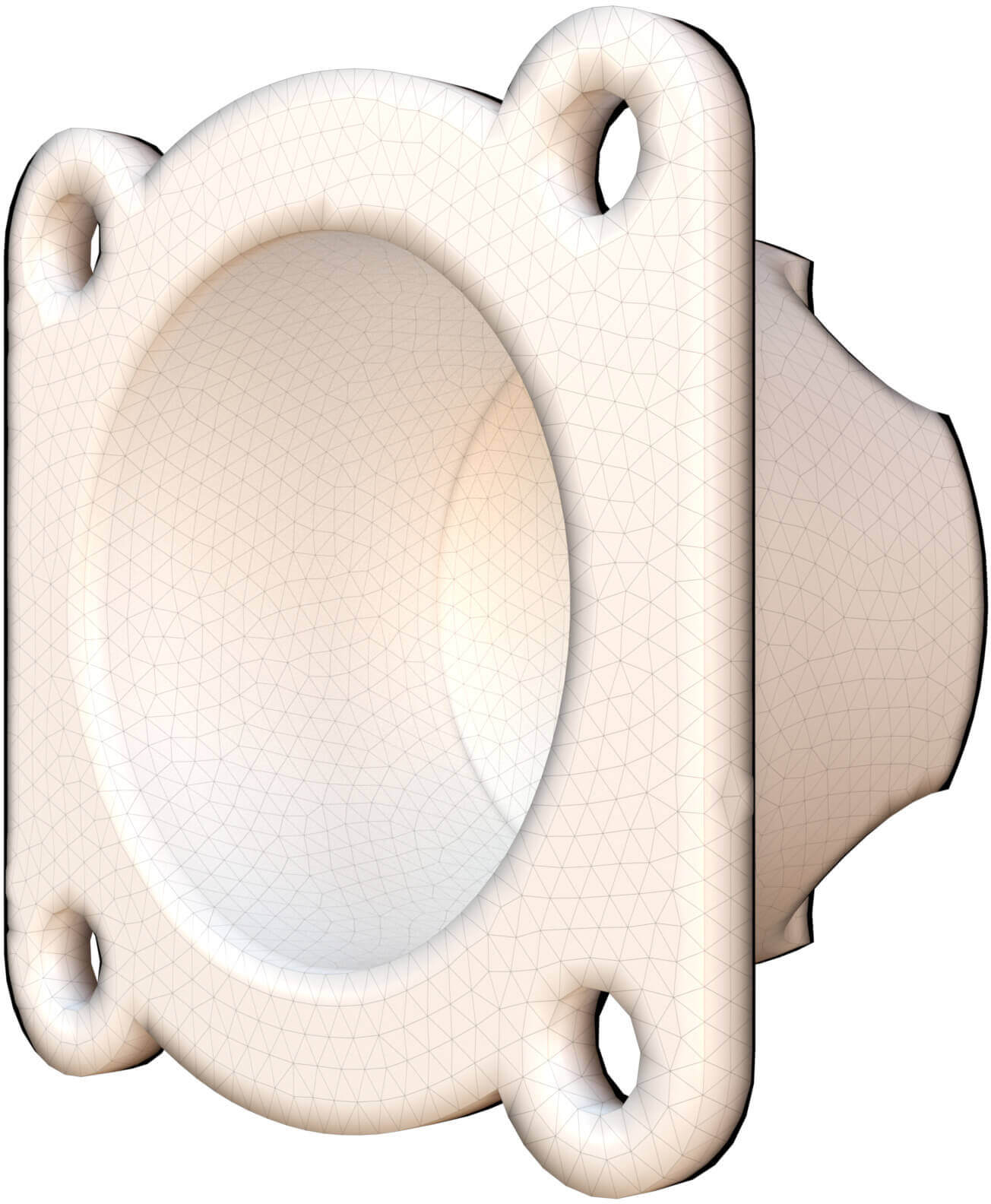}
		\caption*{ }   
	\end{subfigure}
	\begin{subfigure}[b]{0.115\textwidth}
		\includegraphics[width=\textwidth]{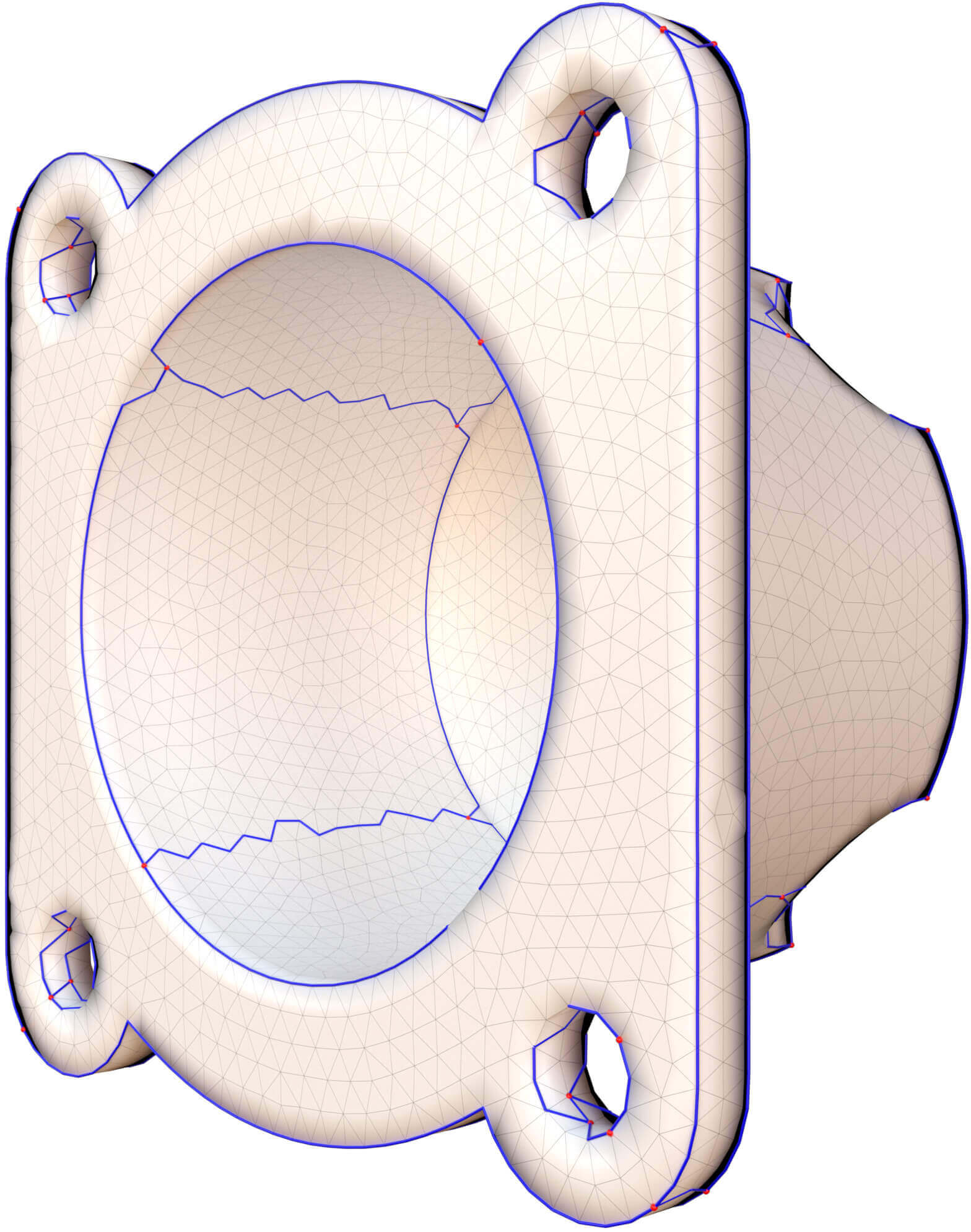}
		\caption*{ }    
	\end{subfigure}
	\begin{subfigure}[b]{0.115\textwidth}
		\includegraphics[width=\textwidth]{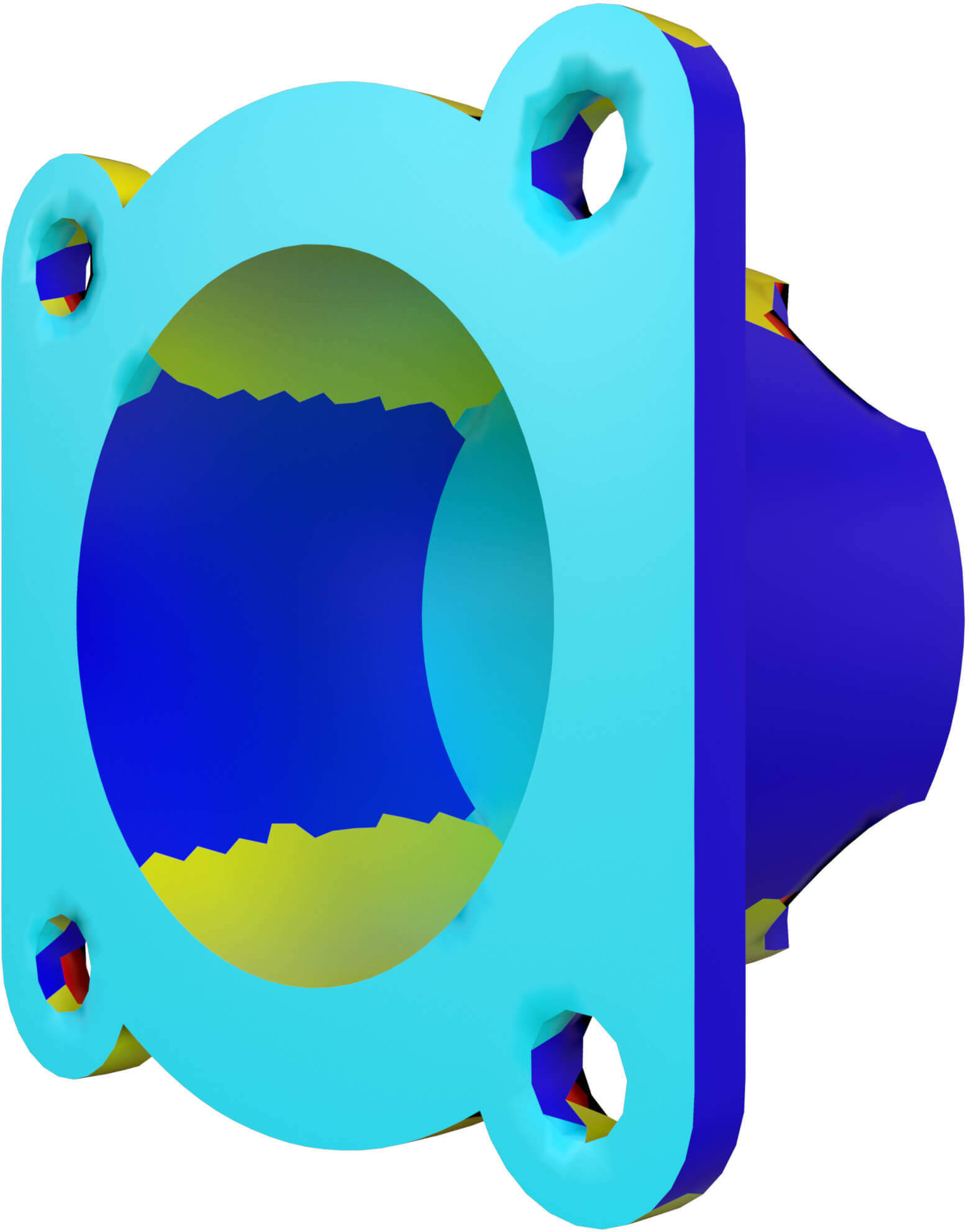}
		\caption*{ }  
	\end{subfigure}   
	\begin{subfigure}[b]{0.115\textwidth}
		\includegraphics[width=\textwidth]{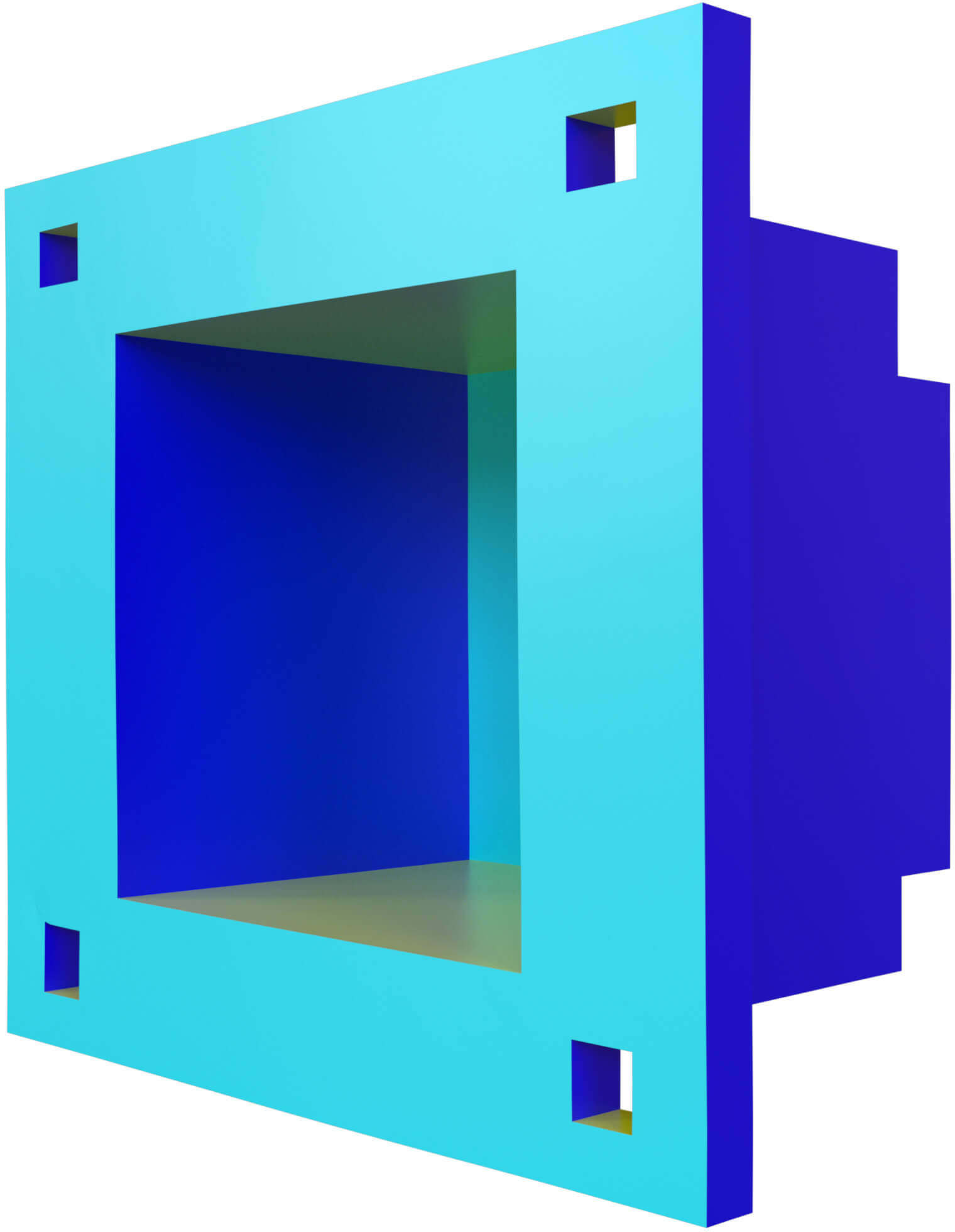}
		\caption*{ }     
	\end{subfigure} 

    	\begin{subfigure}[b]{0.115\textwidth}
    	\includegraphics[width=\textwidth]{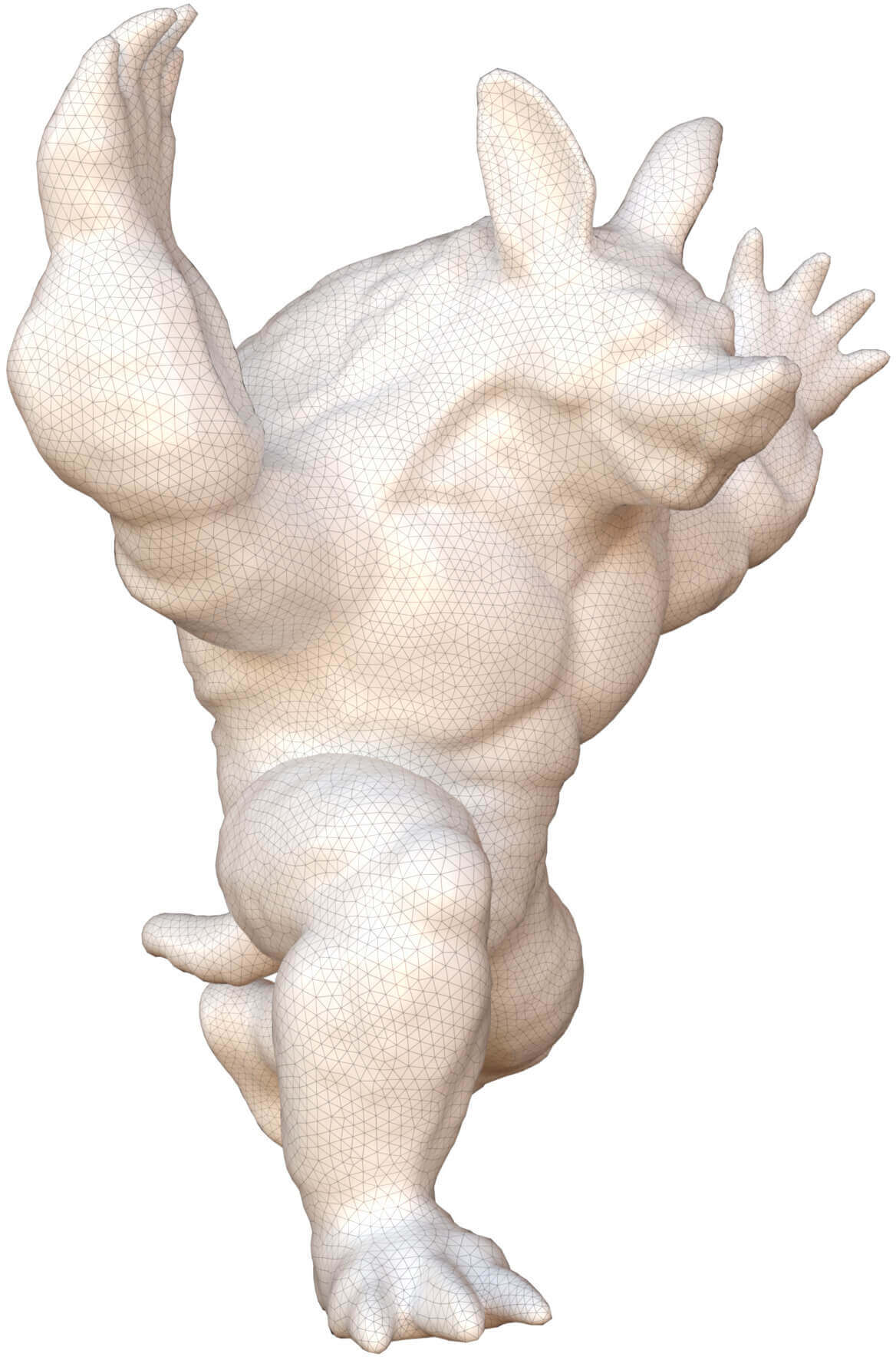}
    	\caption{original}   
    \end{subfigure}
    \begin{subfigure}[b]{0.115\textwidth}
    	\includegraphics[width=\textwidth]{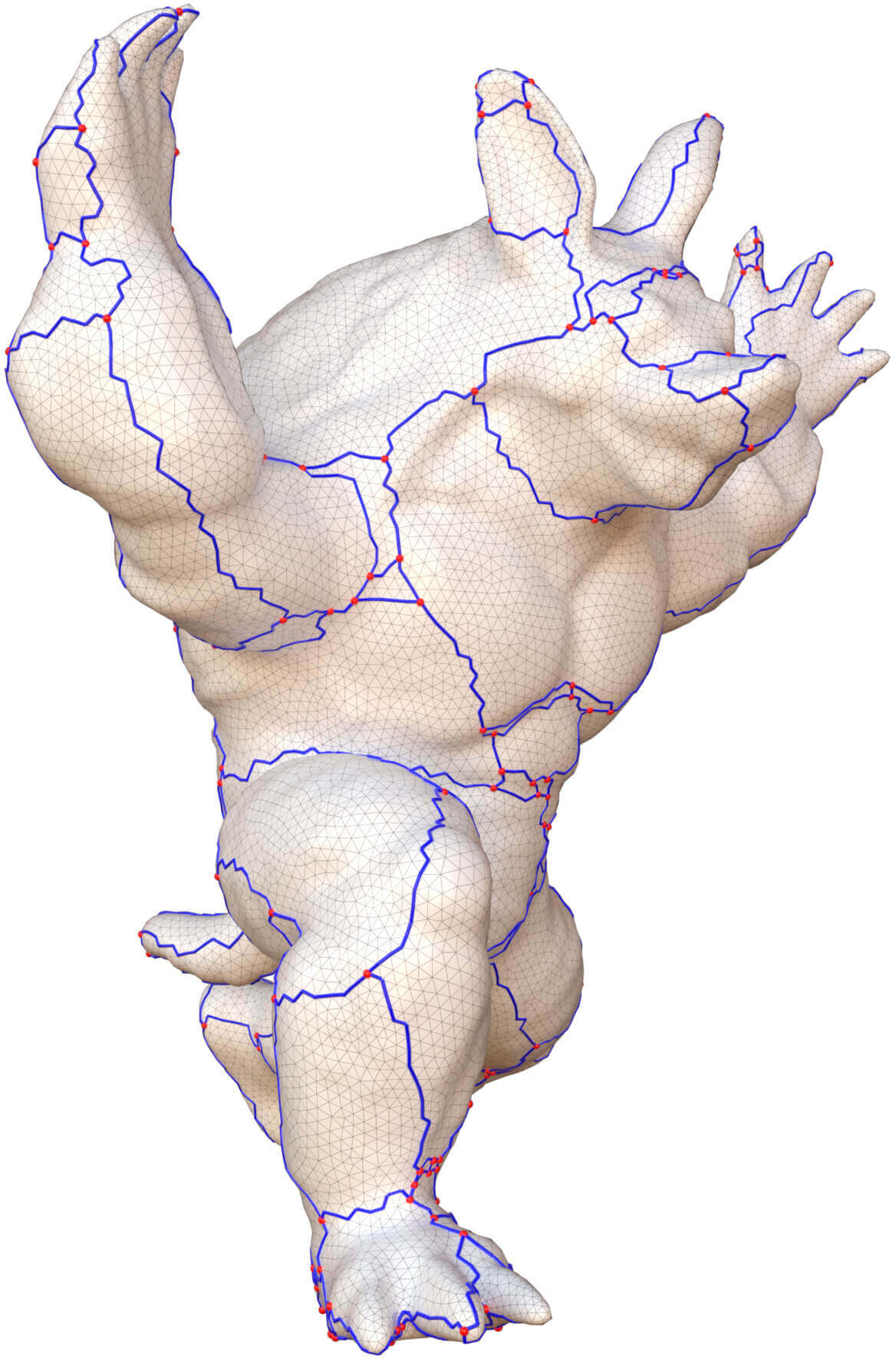}
    	\caption{segmentation}    \label{fig:seg} 
    \end{subfigure}
    \begin{subfigure}[b]{0.115\textwidth}
    	\includegraphics[width=\textwidth]{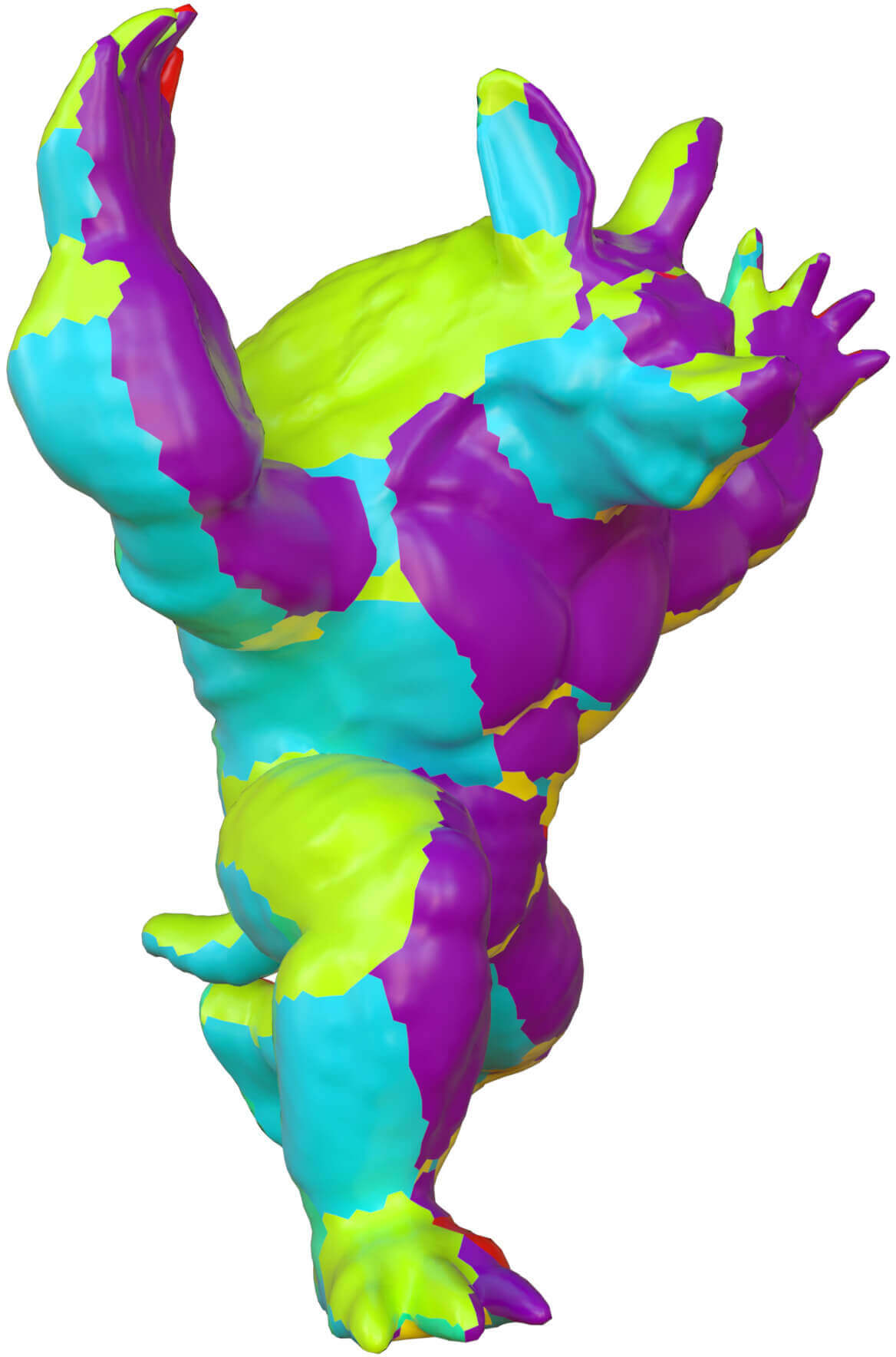}
    	\caption{topology}  \label{fig:polycubeT} 
    \end{subfigure}   
    \begin{subfigure}[b]{0.115\textwidth}
    	\includegraphics[width=\textwidth]{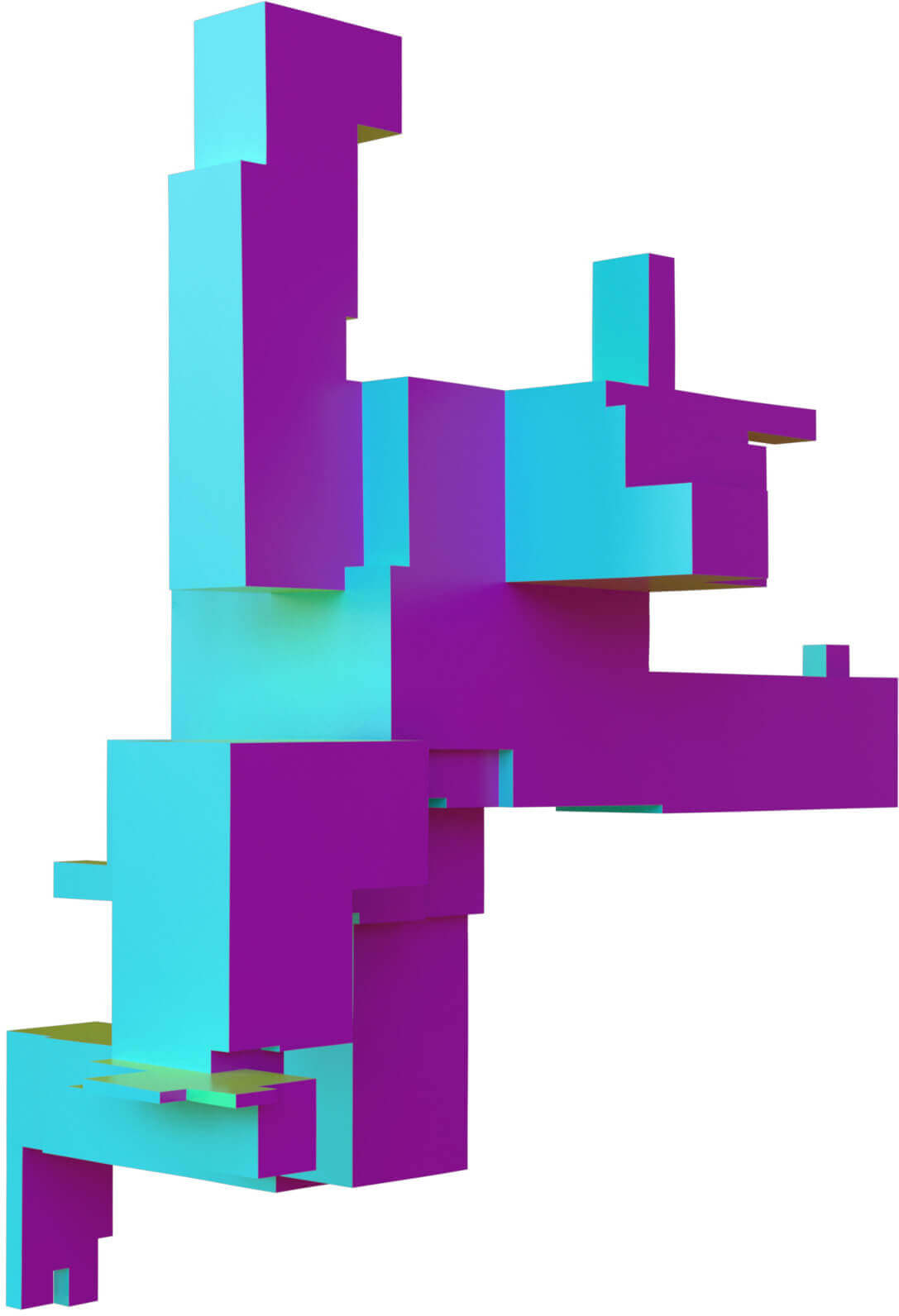}
    	\caption{geometry}     \label{fig:polycubeG}
    \end{subfigure}   
	\caption{The segmentation, polycube topology and geometry.}\label{fig:polycubeTG}
\end{figure} 

These three steps can be independent from each other. The first step divides a mesh into several different charts. The second step determines the polycube  topology of the mesh, and the third step fixes the polycube geometry. In Figure \ref{fig:polycubeTG}, the two models are segmented into some parts in Figure \ref{fig:seg}; then based on the parts, we assign every model a polycube topology in Figure \ref{fig:polycubeT}; finally our algorithm can obtain an exact polycube geometry in Figure \ref{fig:polycubeG}.  The previous algorithms mix these two or three steps into one. Our algorithm separates them explicitly.  In this paper, we focus on the polycube geometry step. Given a model with a valid polycube topology, such as using the method proposed in \cite{Fu2016PC}, our algorithm produces a final shape with  perfect polycube geometry.

 In the first step,   the whole mesh is separated into several charts and the chart boundary is monotone \cite{Livesu:2013:PolyCut}.  
 It is well known the necessary  conditions of the segmentation for a  valid polycube  is still an open problem \cite{eppstein2010steinitz}. However there are three sufficient conditions \cite{eppstein2010steinitz} : 

a) one single patch of a polycube has at least four other neighboring charts; 

b) two neighboring polycube patches must not have opposite labels; 

c) the valence of every polycube vertex is three.

In this paper, we use   the same polycube topology validated data in \cite{Fu2016PC} for the comparison.

In the segmentation step, we must guarantee that there are only three parts which meet on a point to satisfy the above third topological requirement. The PolyCut method in \cite{Livesu:2013:PolyCut} can be applied in this step. In this paper, we use also the same polycube topology validated data in \cite{Fu2016PC} for the comparison.

\subsection{Polycube Topology}
 The second step is determining the polycube topology. After  a specific  segmentation of a mesh, this step  labels or associates  each triangle with one of six axis ($+X$,$-X$,$+Y$,$-Y$,$+Z$,$-Z$), such as in Figure \ref{fig:polycubeT}, the six different colors  represent ($+X$,$-X$,$+Y$,$-Y$,$+Z$,$-Z$) respectively.

A valid polycube topology assigns a target normal to every triangle face, and divides the whole mesh into different patches. In each patch, the triangles have the same target normal. Our algorithm  rotates all triangles to their corresponding target normal directions.   There is no explicit constraints between the patches. Every patch  is independently rotated. However there is an implicit global topological  constraints between them due to the polycube topology.

\begin{figure} [th!] 
	\centering   
	\begin{subfigure}[b]{0.125\textwidth}
		\includegraphics[width=\textwidth]{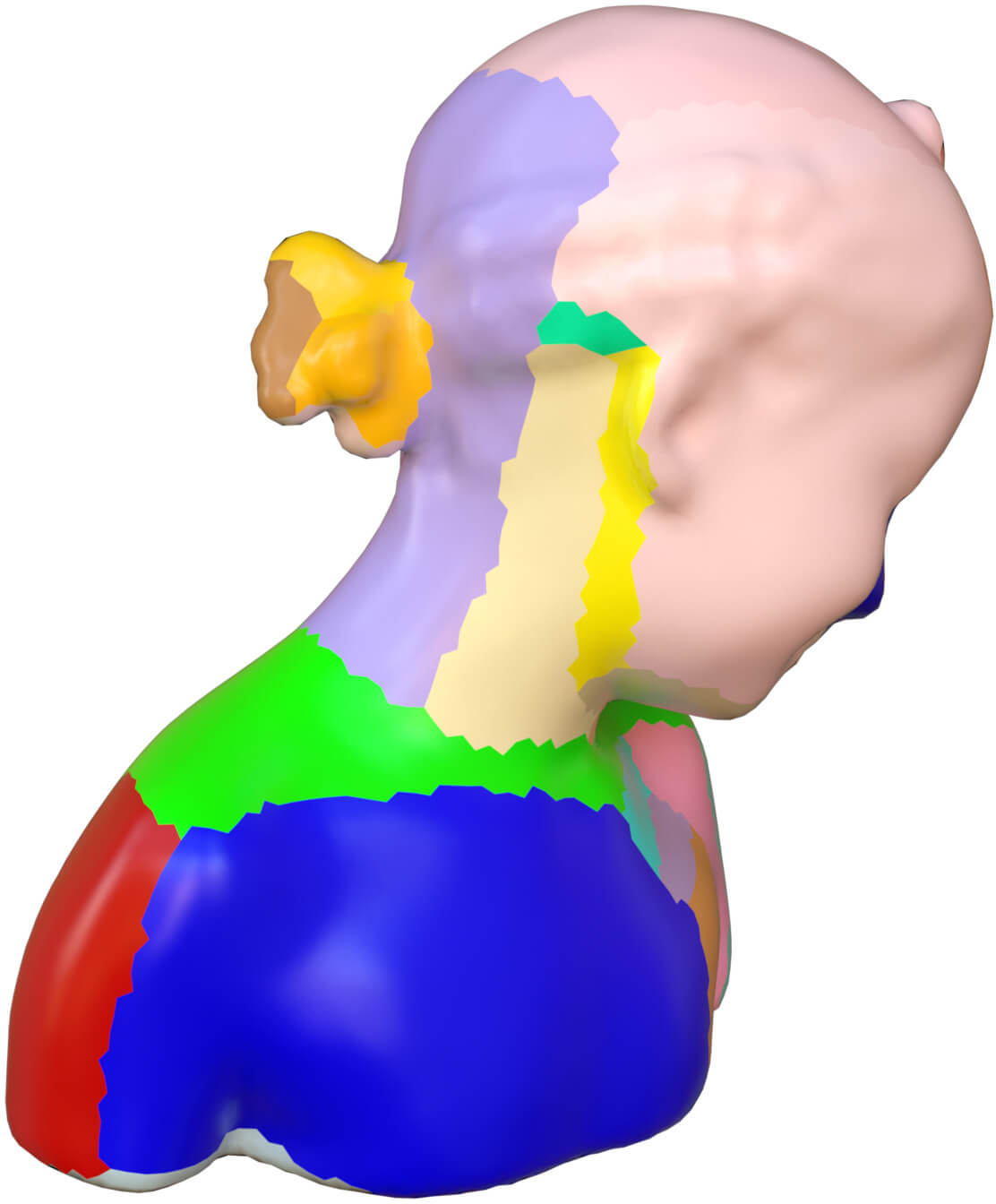}
		\caption{topology 1}   \label{fig:bimba01}
	\end{subfigure}
	\begin{subfigure}[b]{0.105\textwidth}
		\includegraphics[width=\textwidth]{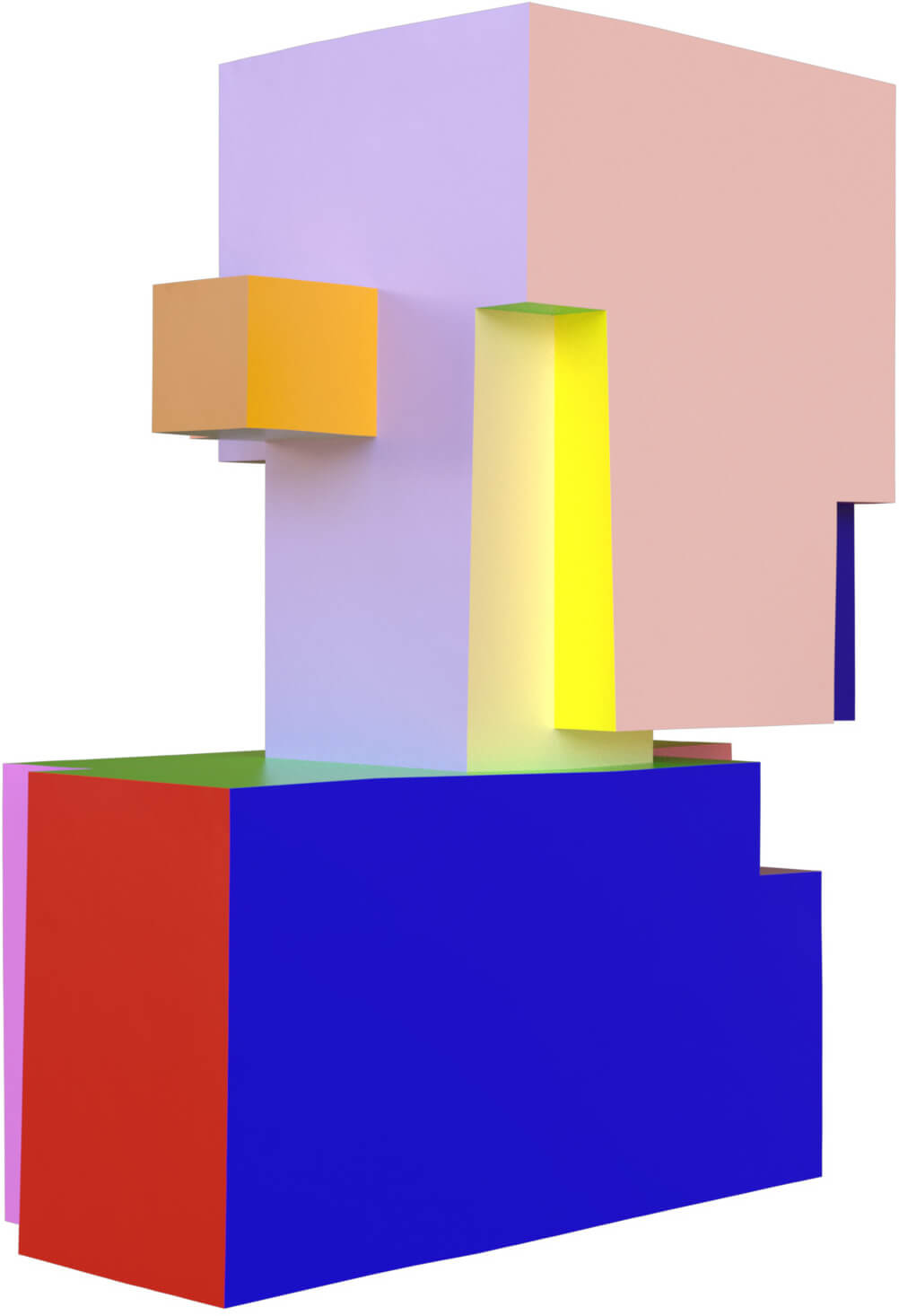}
		\caption{polycube 2}   \label{fig:bimbapolycube01}
	\end{subfigure}   
	\begin{subfigure}[b]{0.125\textwidth}
		\includegraphics[width=\textwidth]{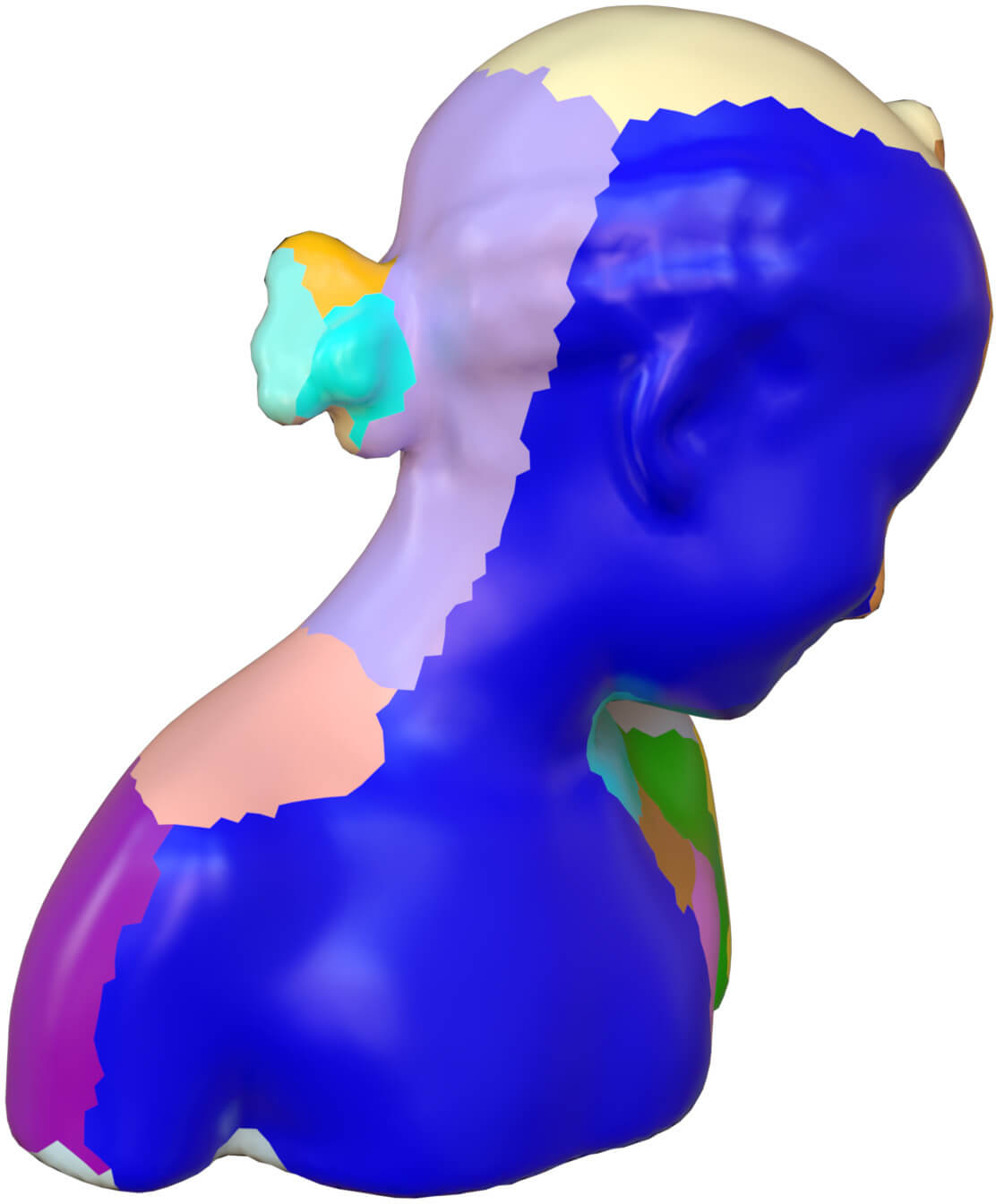}
		\caption{topology 1}     \label{fig:bimba02}
	\end{subfigure}
	\begin{subfigure}[b]{0.105\textwidth}
		\includegraphics[width=\textwidth]{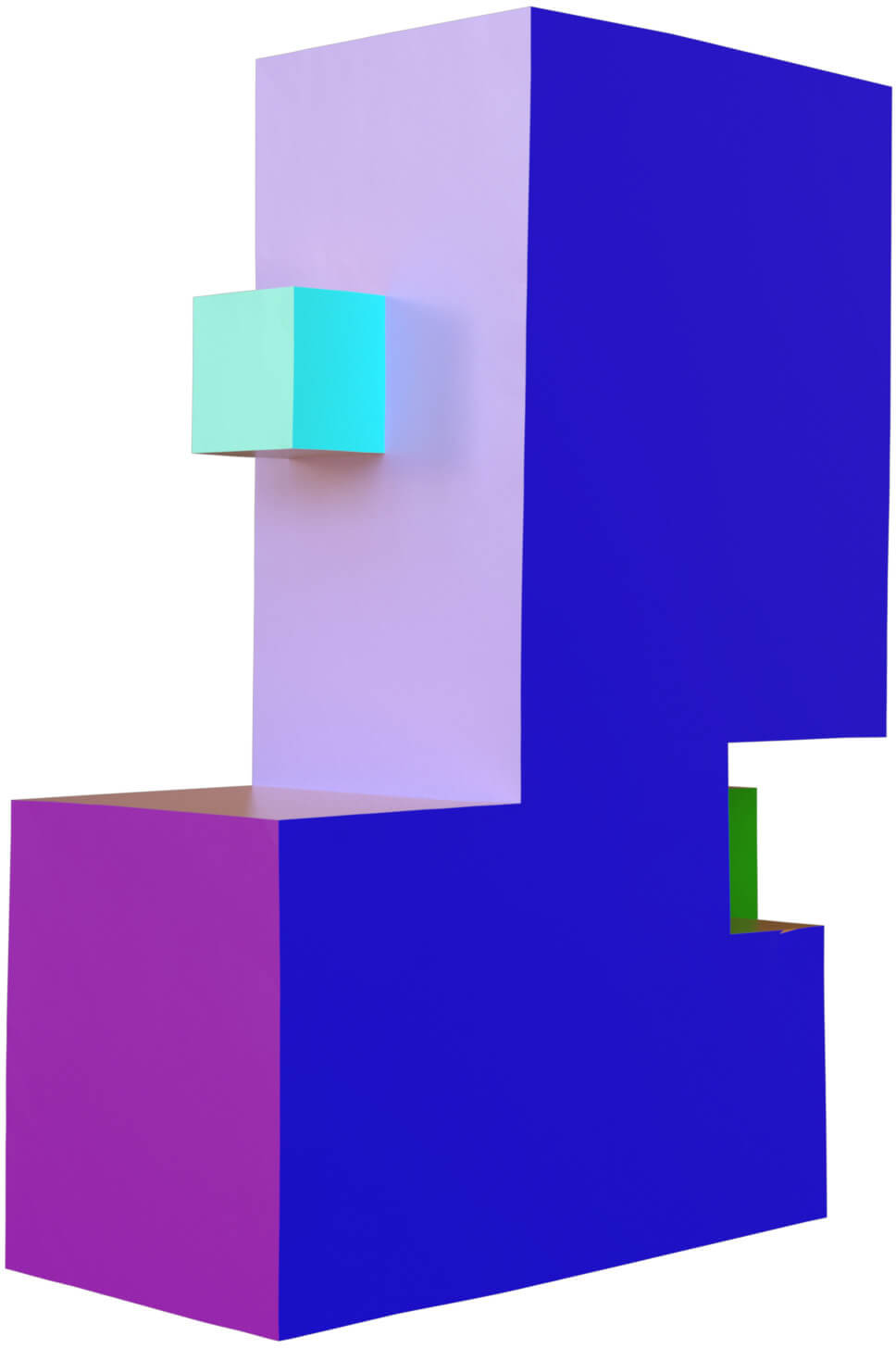}
		\caption{polycube 2}    \label{fig:bimbapolycube02}
	\end{subfigure}      
	\caption{The model  and two different polycube topologies.}\label{fig:polyDiffTopo}
\end{figure}

\begin{figure} [th!] 
	\centering	
	\begin{subfigure}[b]{0.115\textwidth}
		\includegraphics[width=\textwidth]{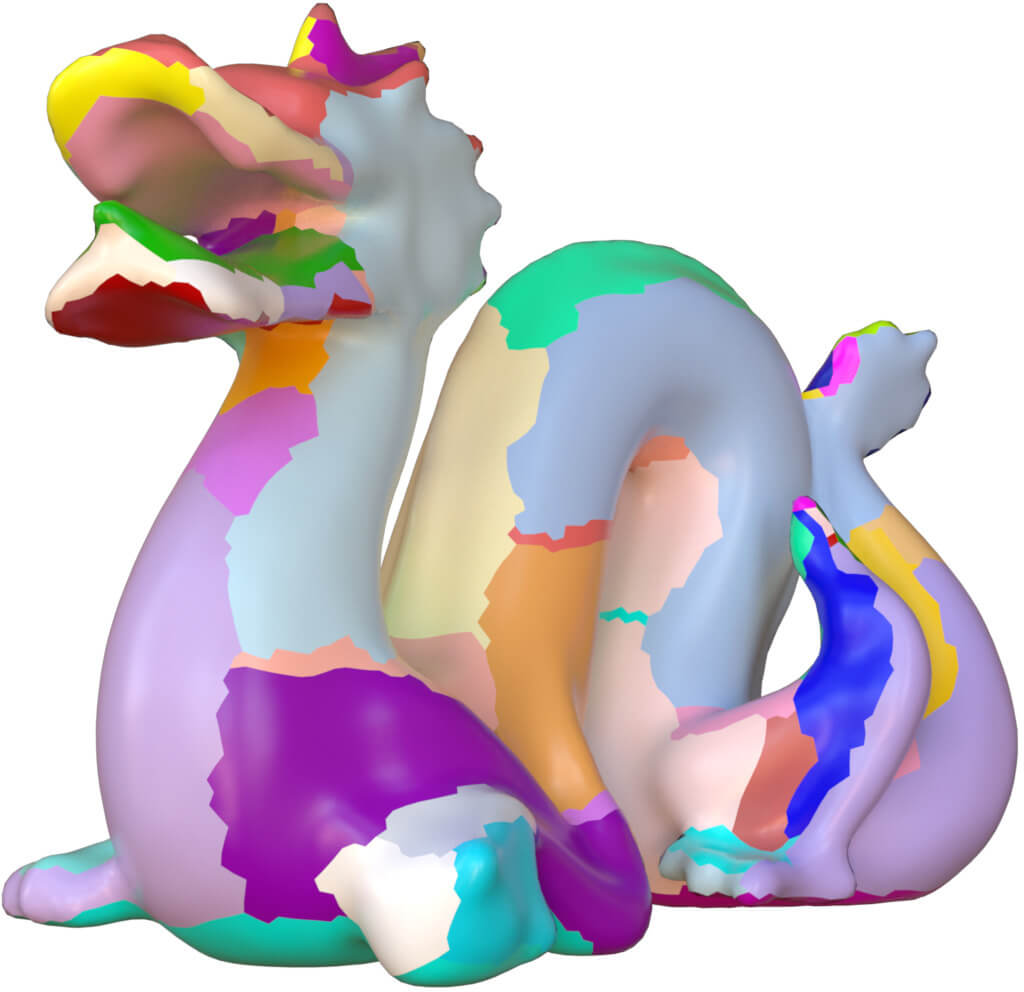}
		\caption{topology 1}   
	\end{subfigure}
	\begin{subfigure}[b]{0.115\textwidth}
		\includegraphics[width=\textwidth]{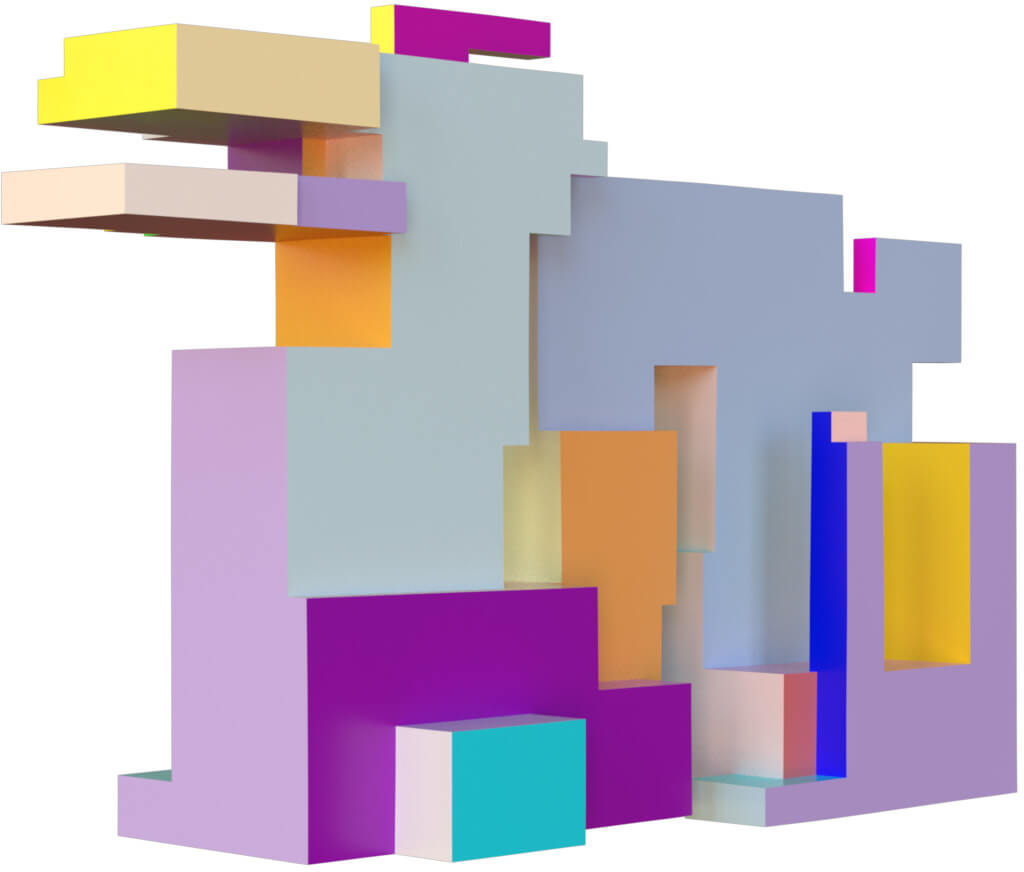}
		\caption{polycube 1}   
	\end{subfigure}    
	\begin{subfigure}[b]{0.115\textwidth}
		\includegraphics[width=\textwidth]{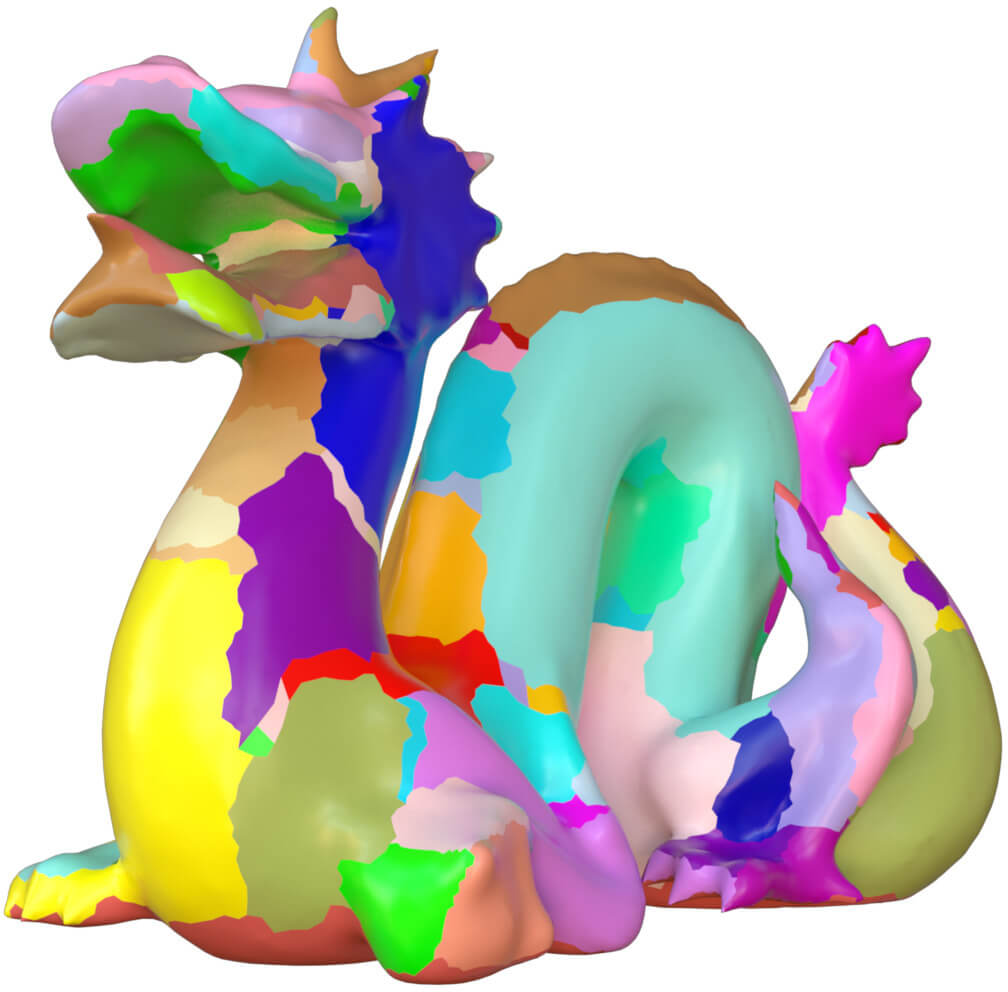}
		\caption{topology 2}  
	\end{subfigure}
	\begin{subfigure}[b]{0.115\textwidth}
		\includegraphics[width=\textwidth]{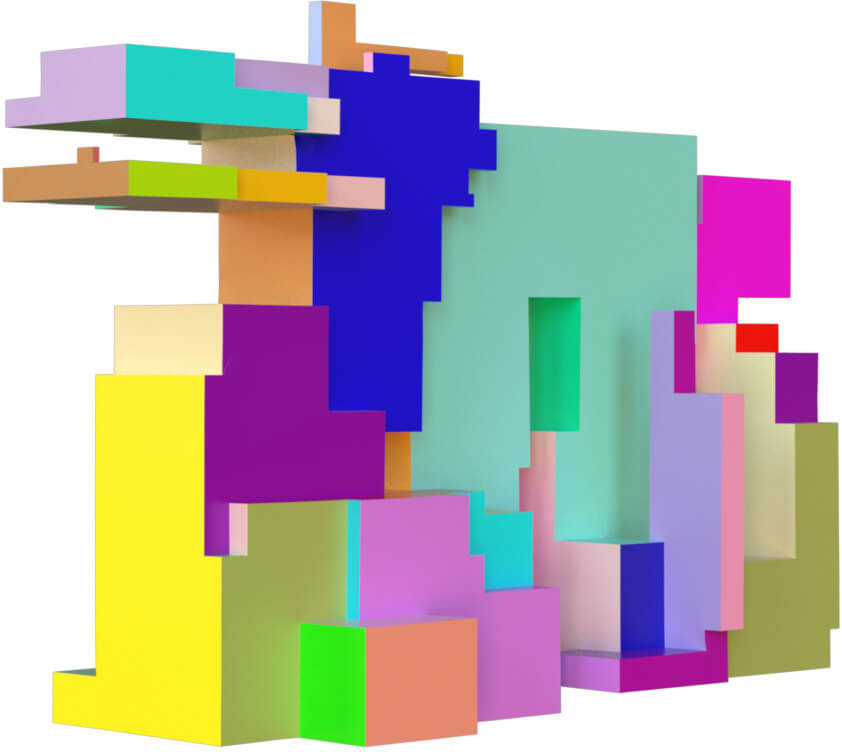}
		\caption{polycube 2}  
	\end{subfigure}      
	\caption{Another model  and two different polycube topologies.}\label{fig:polyDiffTopo2}
\end{figure}

\textbf{Different polycube topologies}. If the same model is assigned several different polycube topologies, then we will have different polycube shapes. In Figure \ref{fig:polyDiffTopo} and \ref{fig:polyDiffTopo2}, we demonstrate this conclusion. Figure  \ref{fig:bimba01} and \ref{fig:bimba02} show the same "bimba" model, but with different polycube topologies, and Figure \ref{fig:bimbapolycube01} and \ref{fig:bimbapolycube02} are their corresponding polycube shapes.


\begin{figure*} [th!] 
	\centering   
	\begin{subfigure}[b]{0.135\textwidth}
		\includegraphics[width=\textwidth]{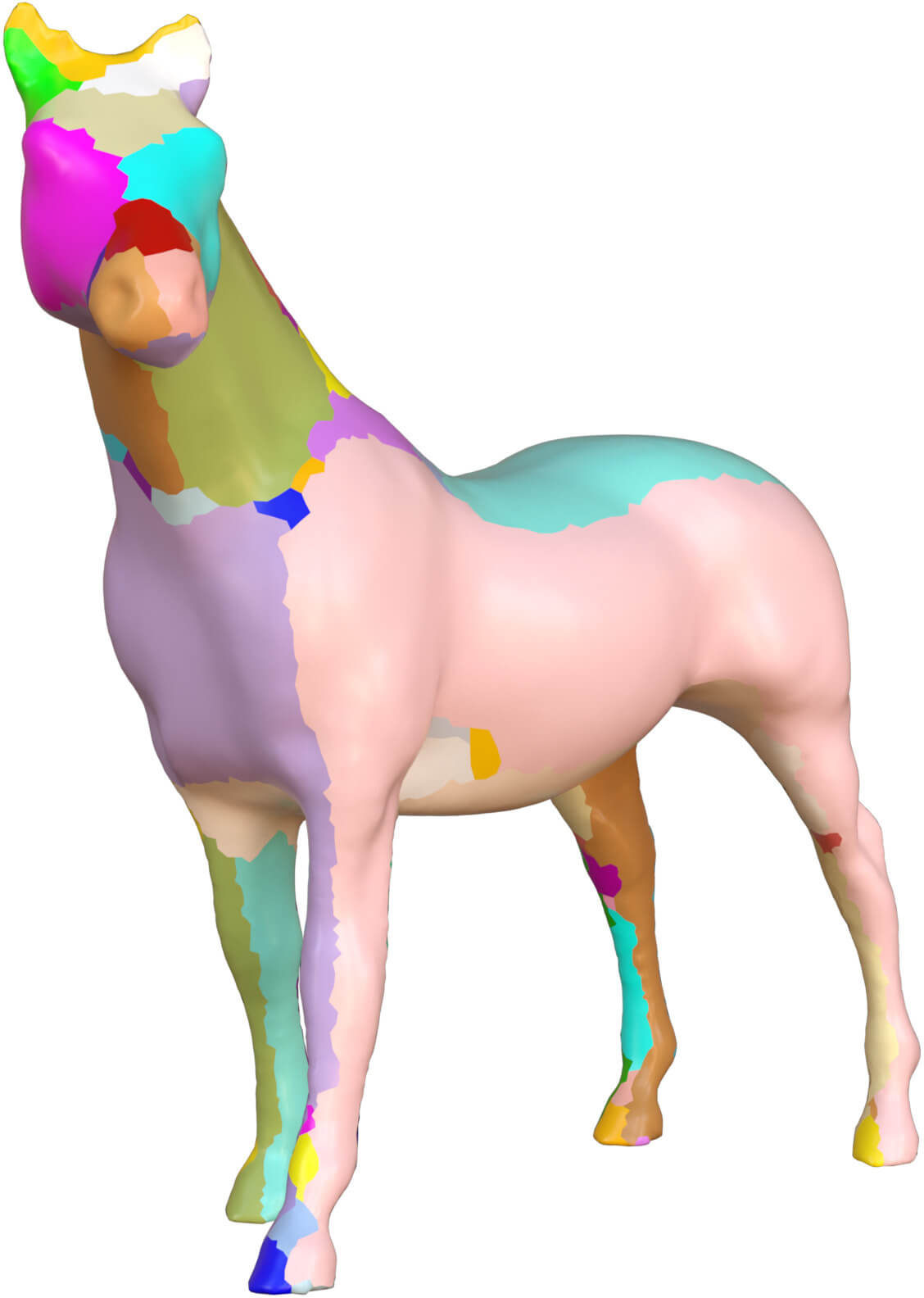}
		\caption*{ }   
	\end{subfigure}
	\begin{subfigure}[b]{0.135\textwidth}
		\includegraphics[width=\textwidth]{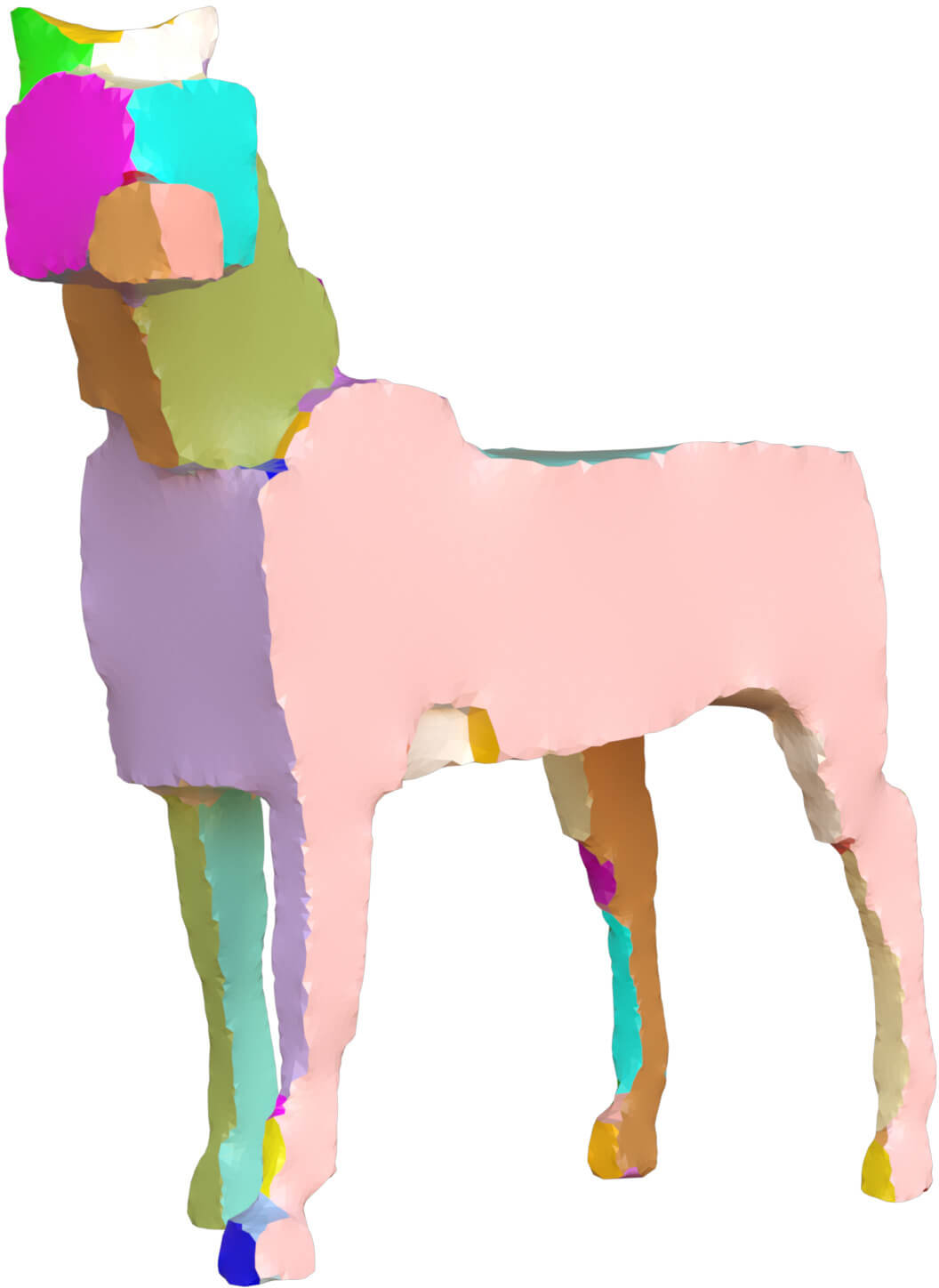}
		\caption*{ } 
	\end{subfigure}           
	\begin{subfigure}[b]{0.135\textwidth}
		\includegraphics[width=\textwidth]{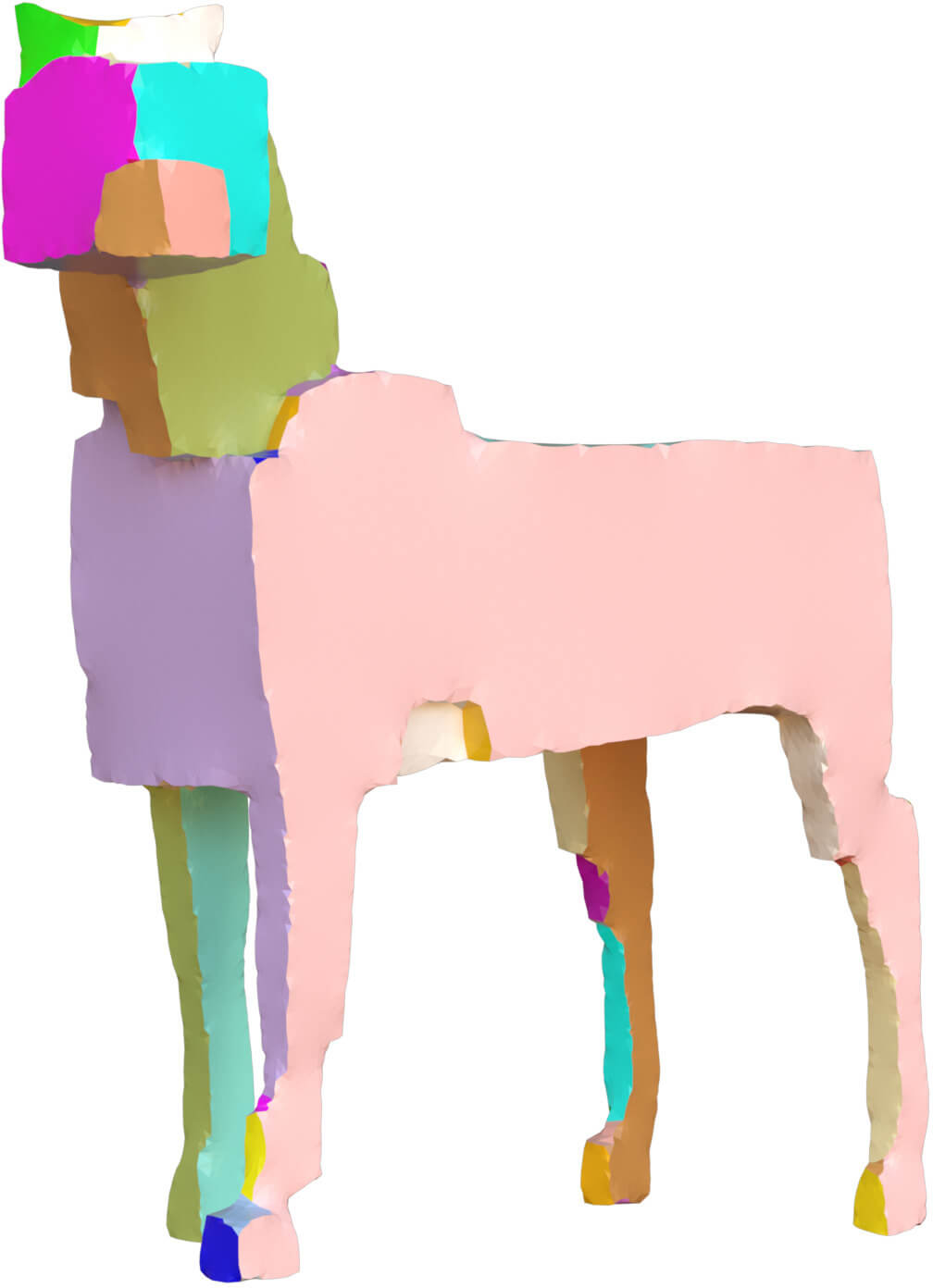}
		\caption*{ }     
	\end{subfigure}         
	\begin{subfigure}[b]{0.135\textwidth}
		\includegraphics[width=\textwidth]{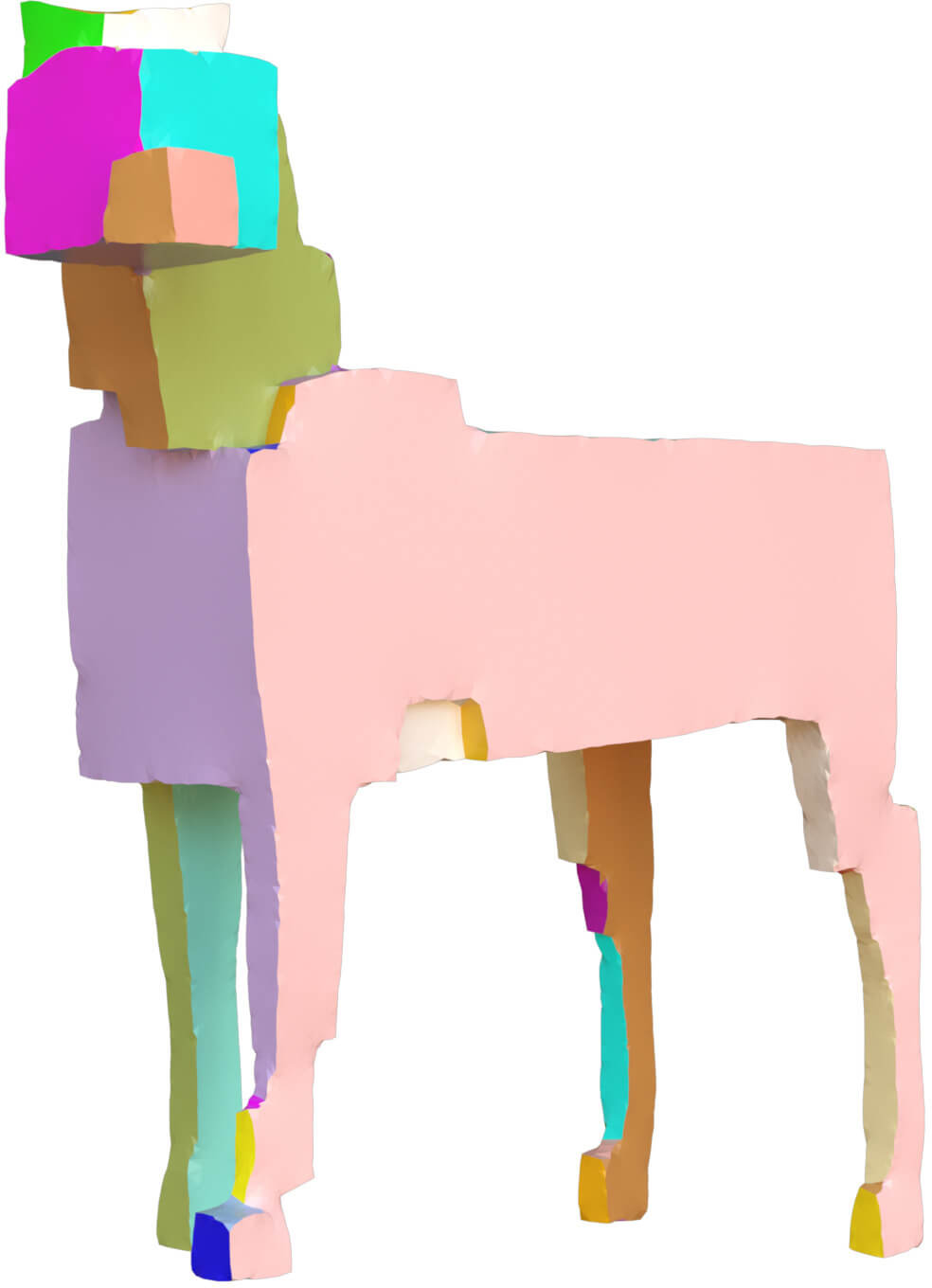}
		\caption*{ }        
	\end{subfigure}
	\begin{subfigure}[b]{0.135\textwidth}
		\includegraphics[width=\textwidth]{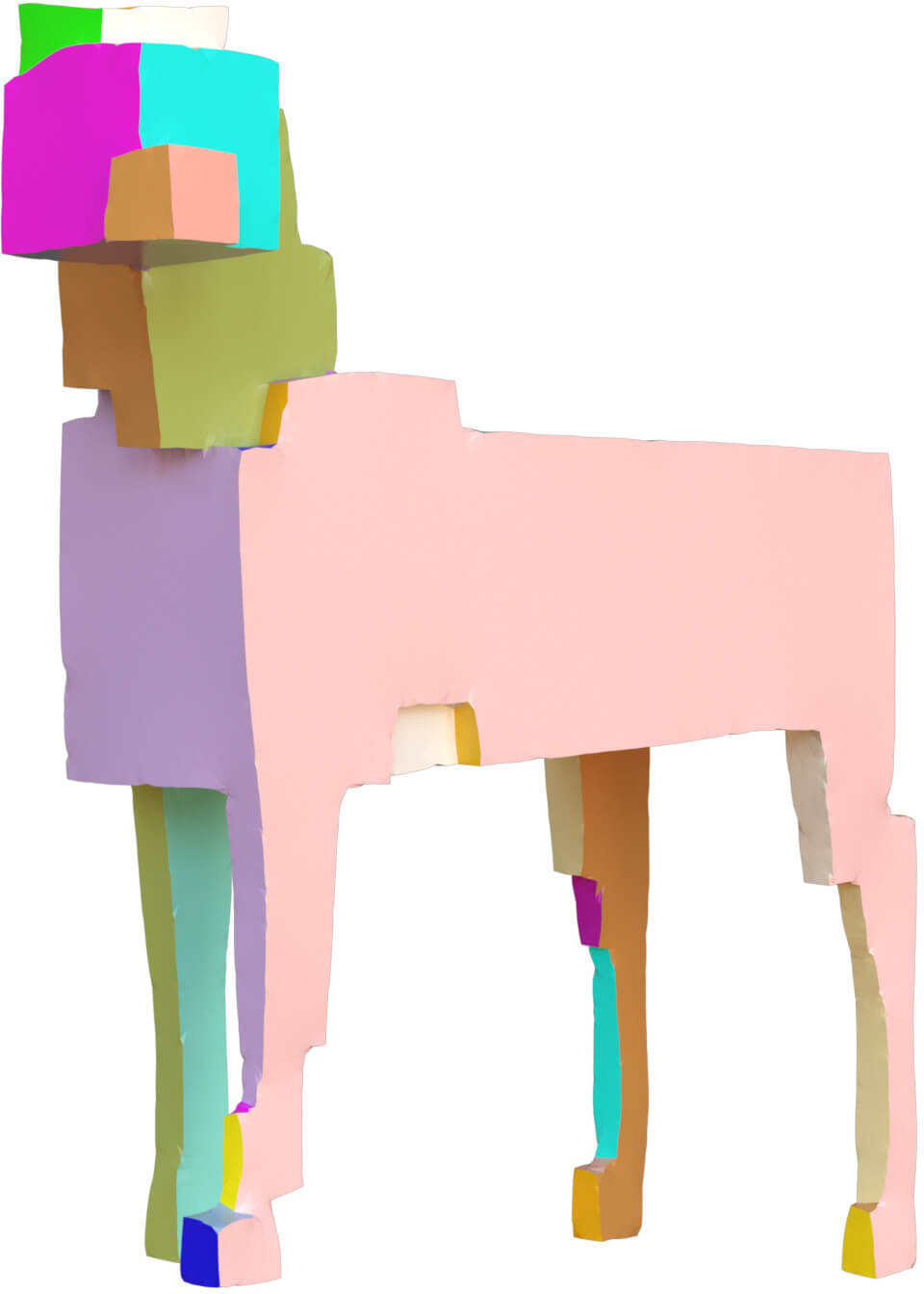}
		\caption*{ }          
	\end{subfigure}
	\begin{subfigure}[b]{0.135\textwidth}
		\includegraphics[width=\textwidth]{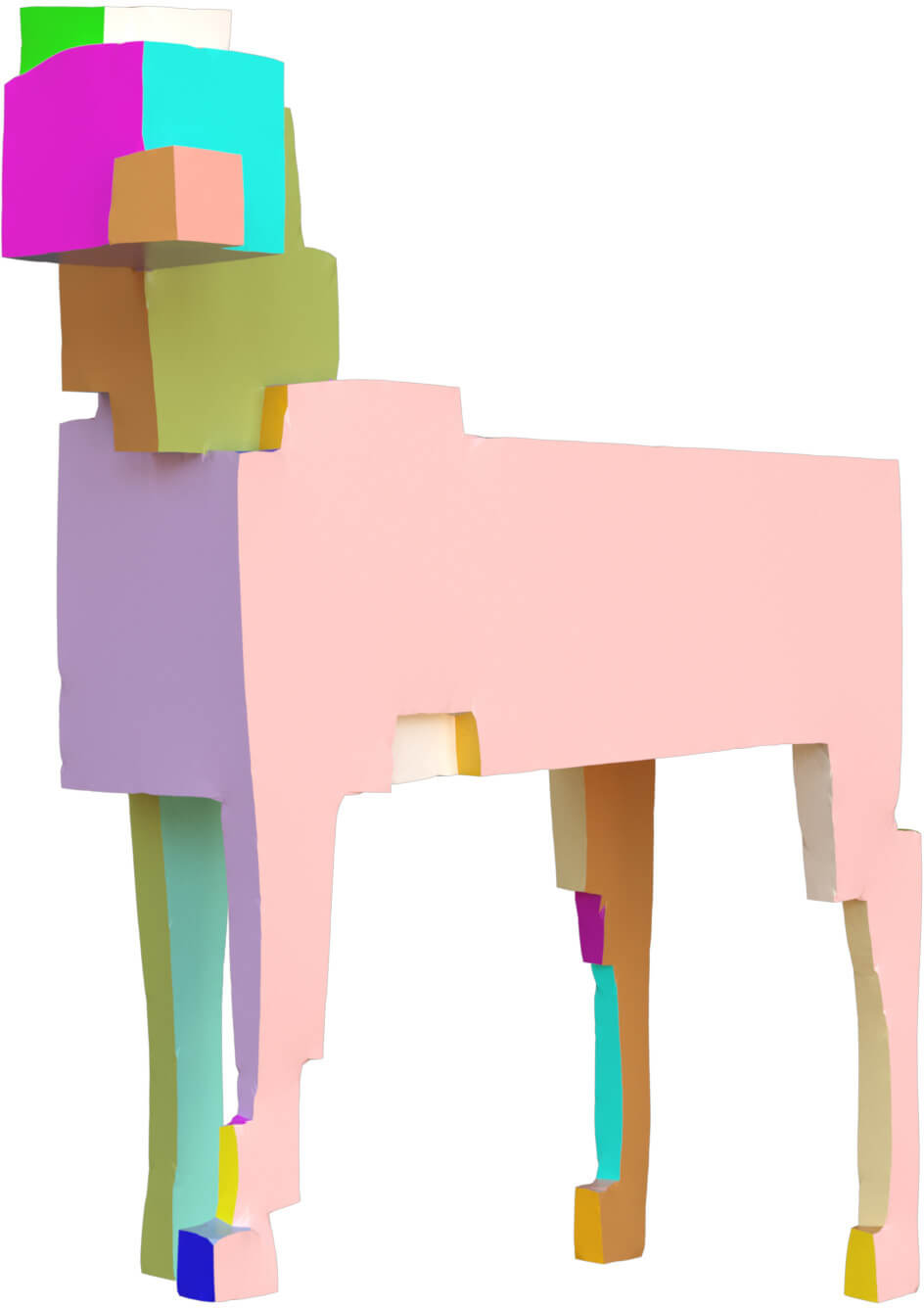}
		\caption*{ }          
	\end{subfigure}     	
	\begin{subfigure}[b]{0.135\textwidth}
		\includegraphics[width=\textwidth]{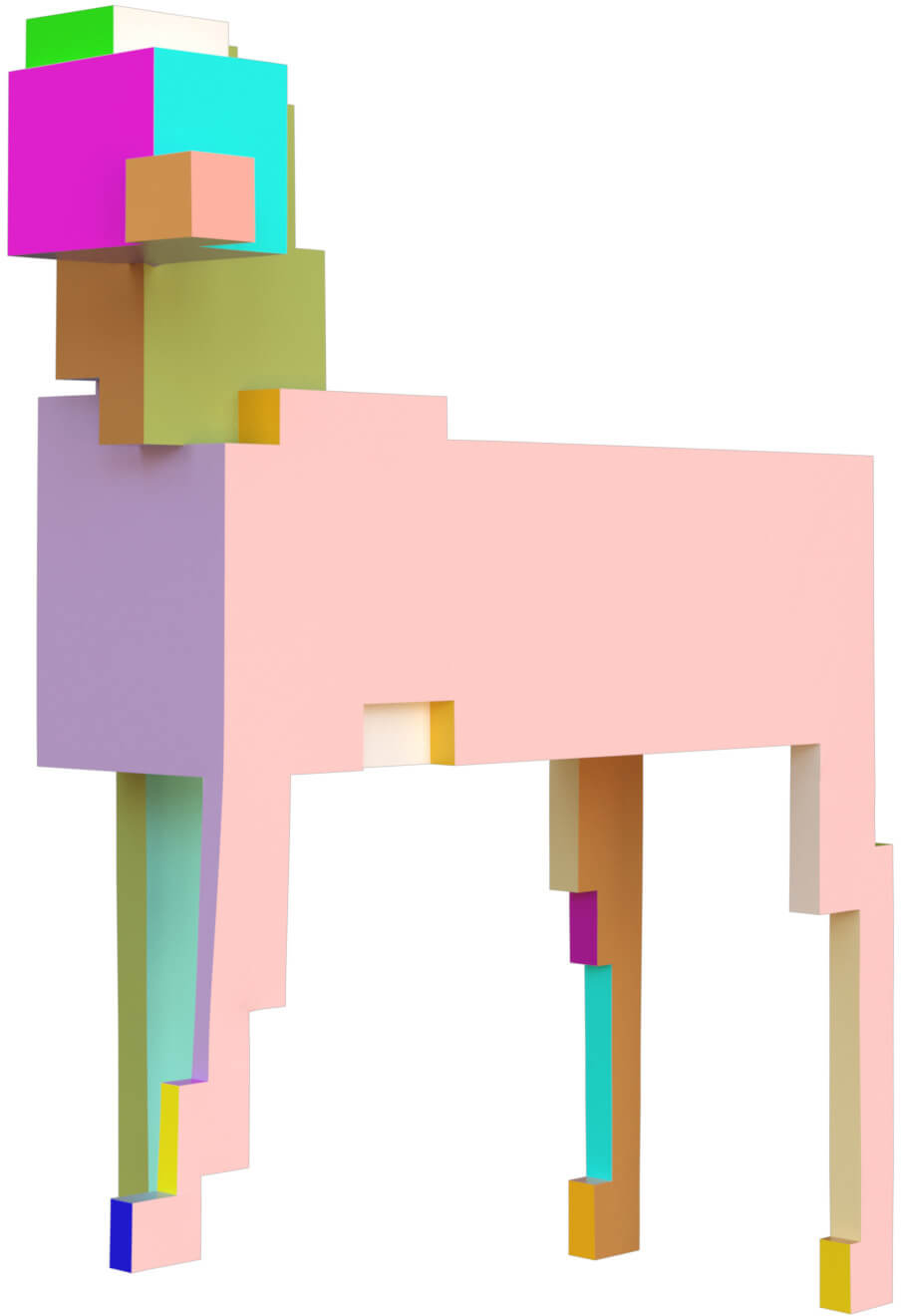}
		\caption*{ }    
	\end{subfigure}
	\begin{subfigure}[b]{0.135\textwidth}
		\includegraphics[width=\textwidth]{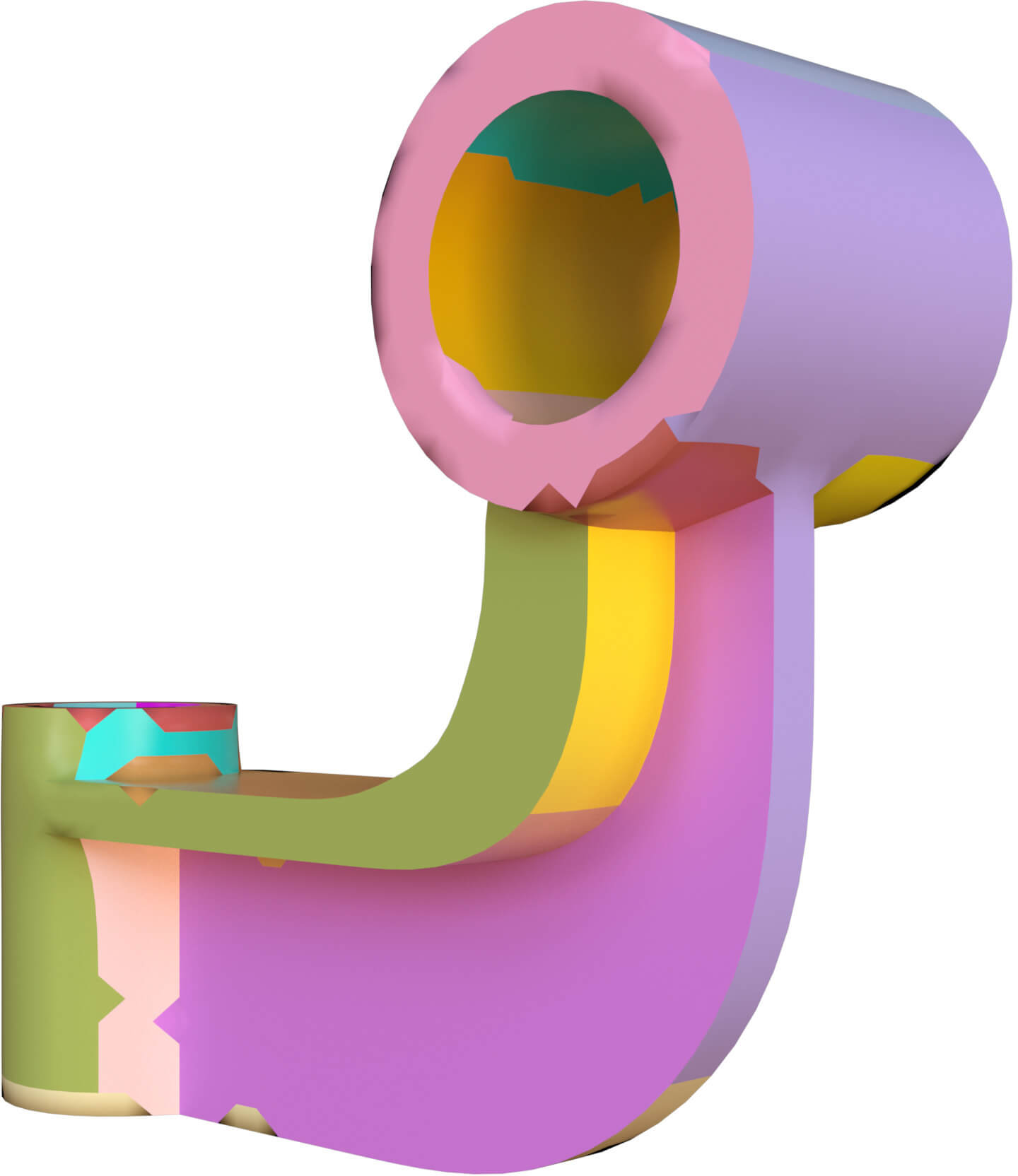}
		\caption{original}   
	\end{subfigure}
	\begin{subfigure}[b]{0.135\textwidth}
		\includegraphics[width=\textwidth]{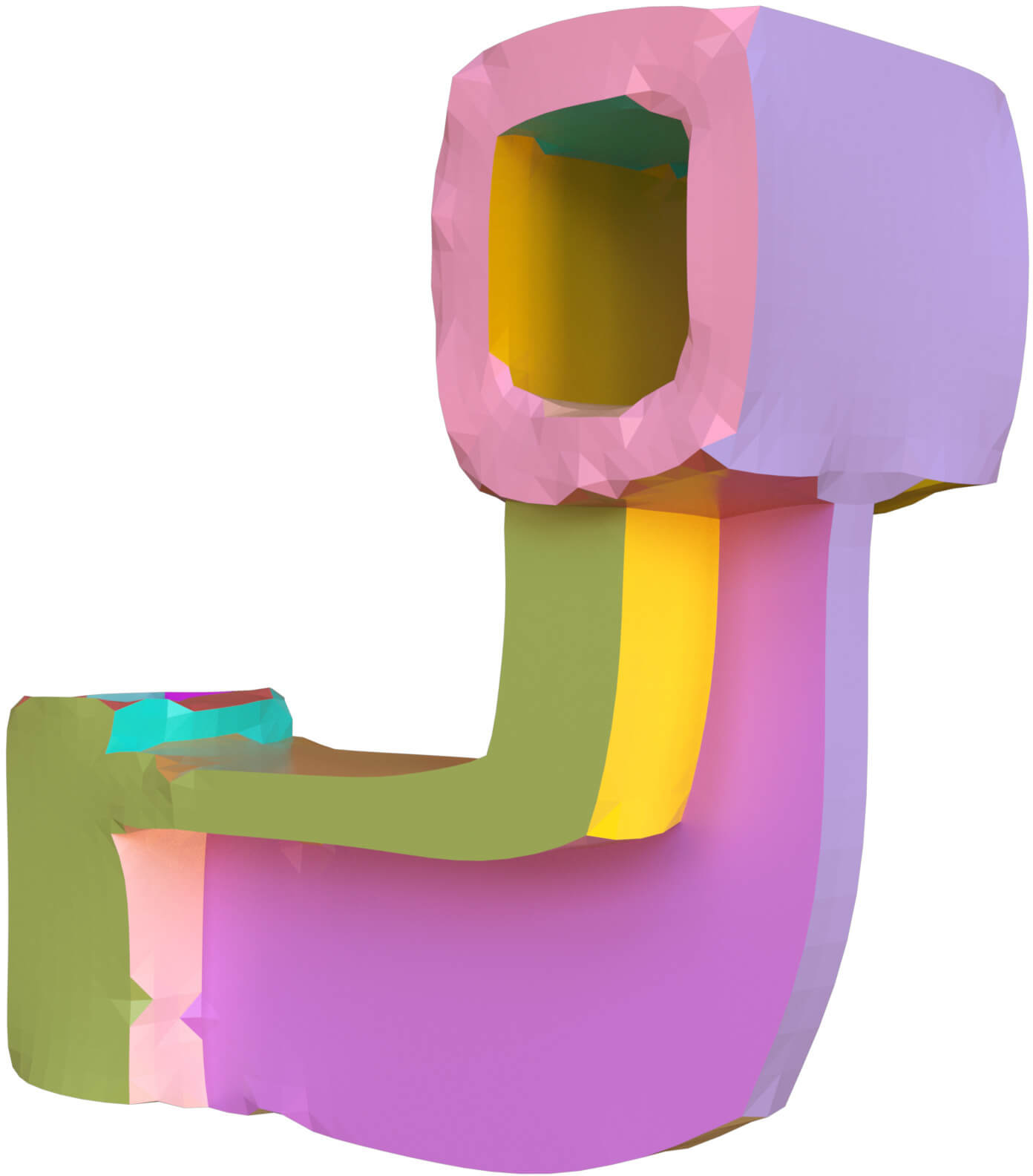}
		\caption{1} \label{fig:iterfirst}  
	\end{subfigure}           
	\begin{subfigure}[b]{0.135\textwidth}
		\includegraphics[width=\textwidth]{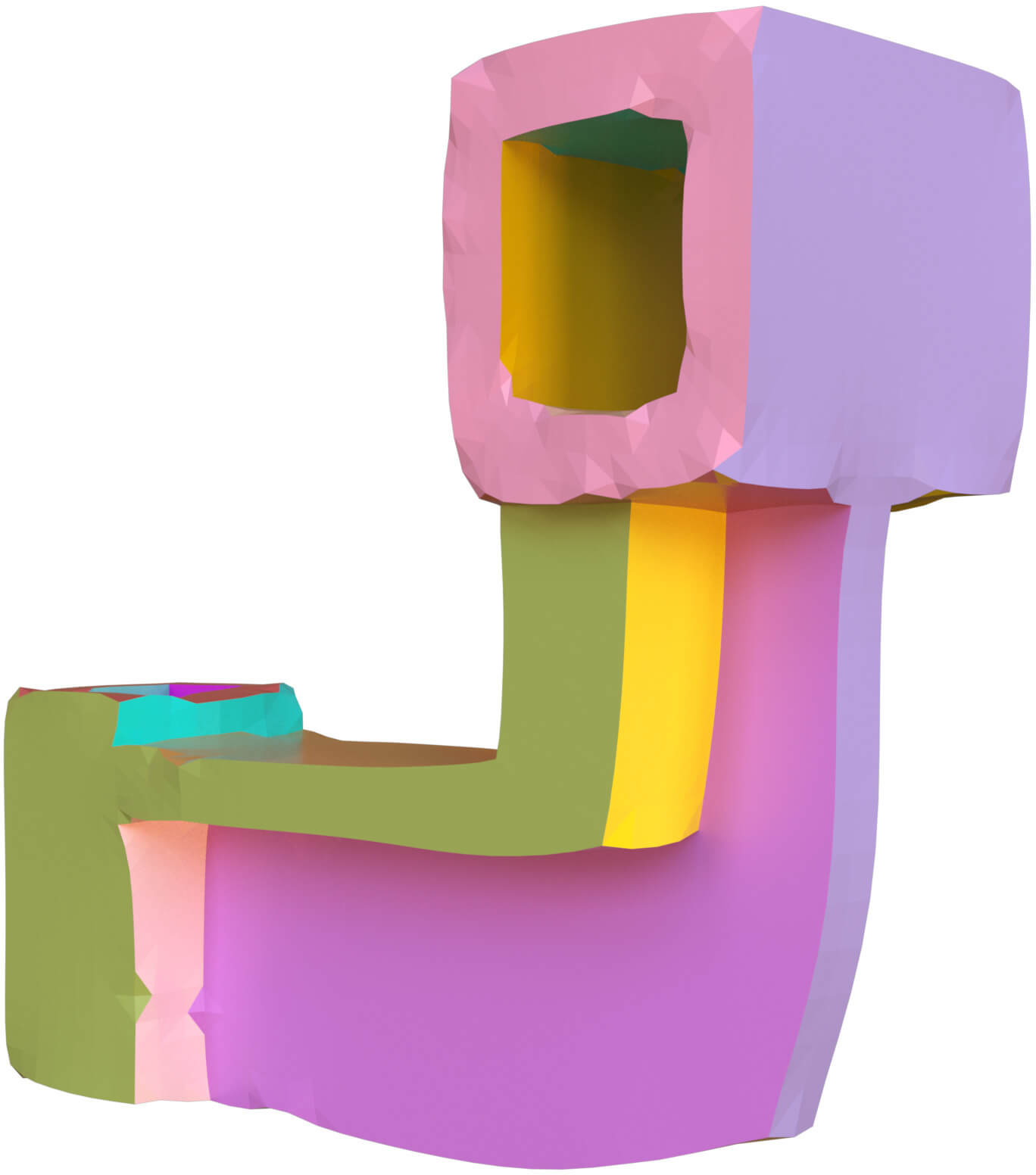}
		\caption{2}     
	\end{subfigure}         
	\begin{subfigure}[b]{0.135\textwidth}
		\includegraphics[width=\textwidth]{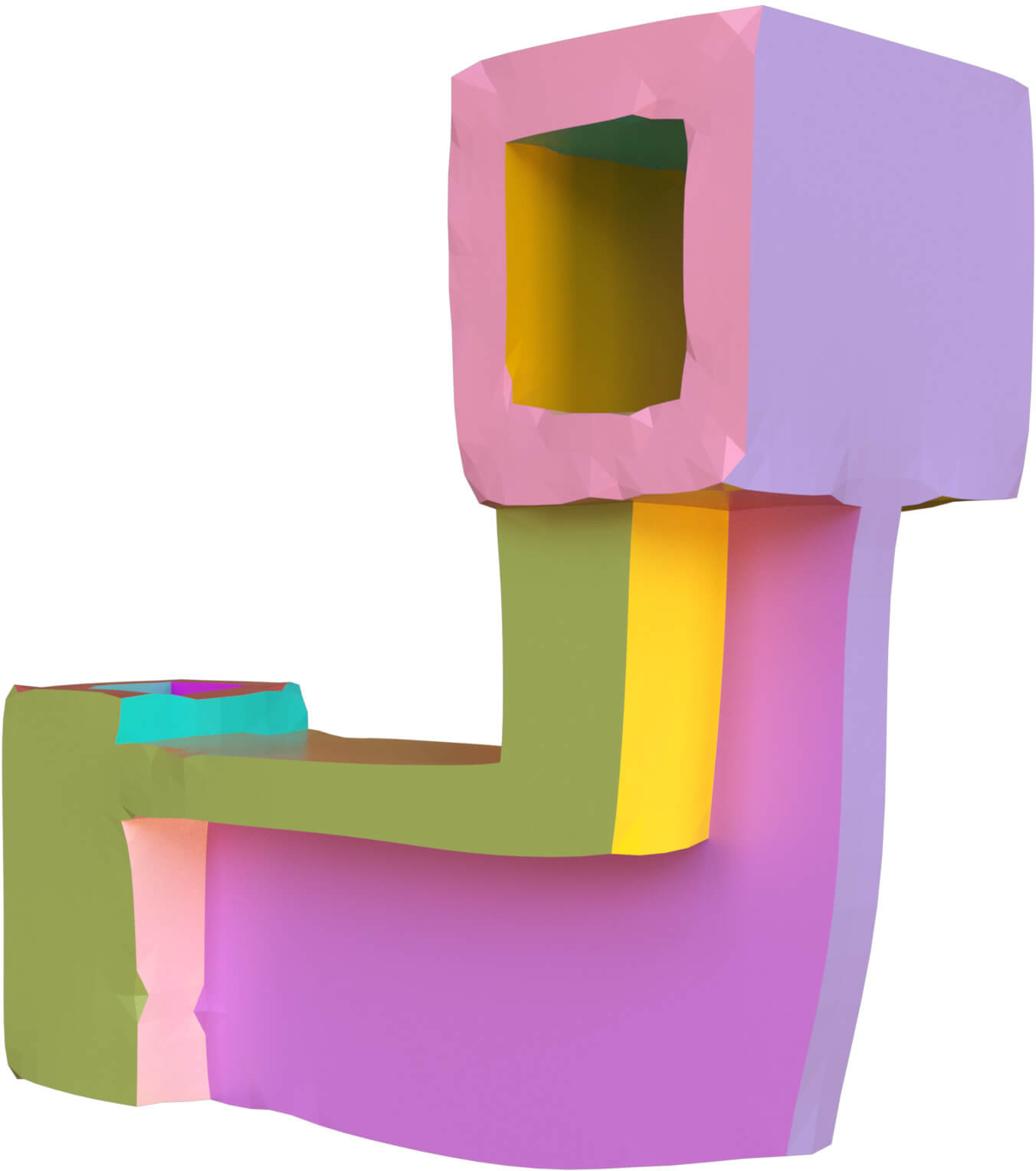}
		\caption{3}       
	\end{subfigure}
	\begin{subfigure}[b]{0.135\textwidth}
		\includegraphics[width=\textwidth]{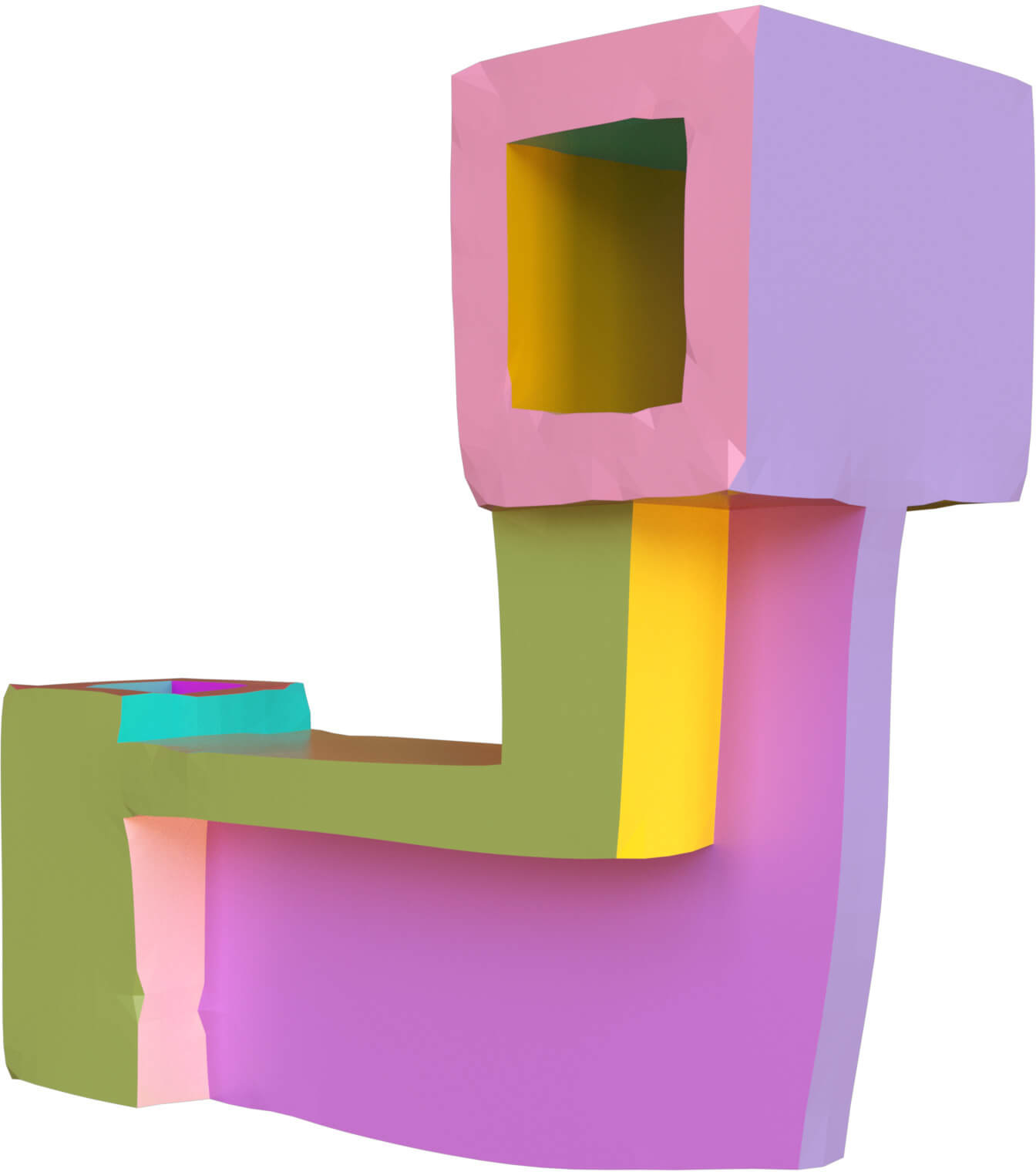}
		\caption{4}         
	\end{subfigure}
	\begin{subfigure}[b]{0.135\textwidth}
		\includegraphics[width=\textwidth]{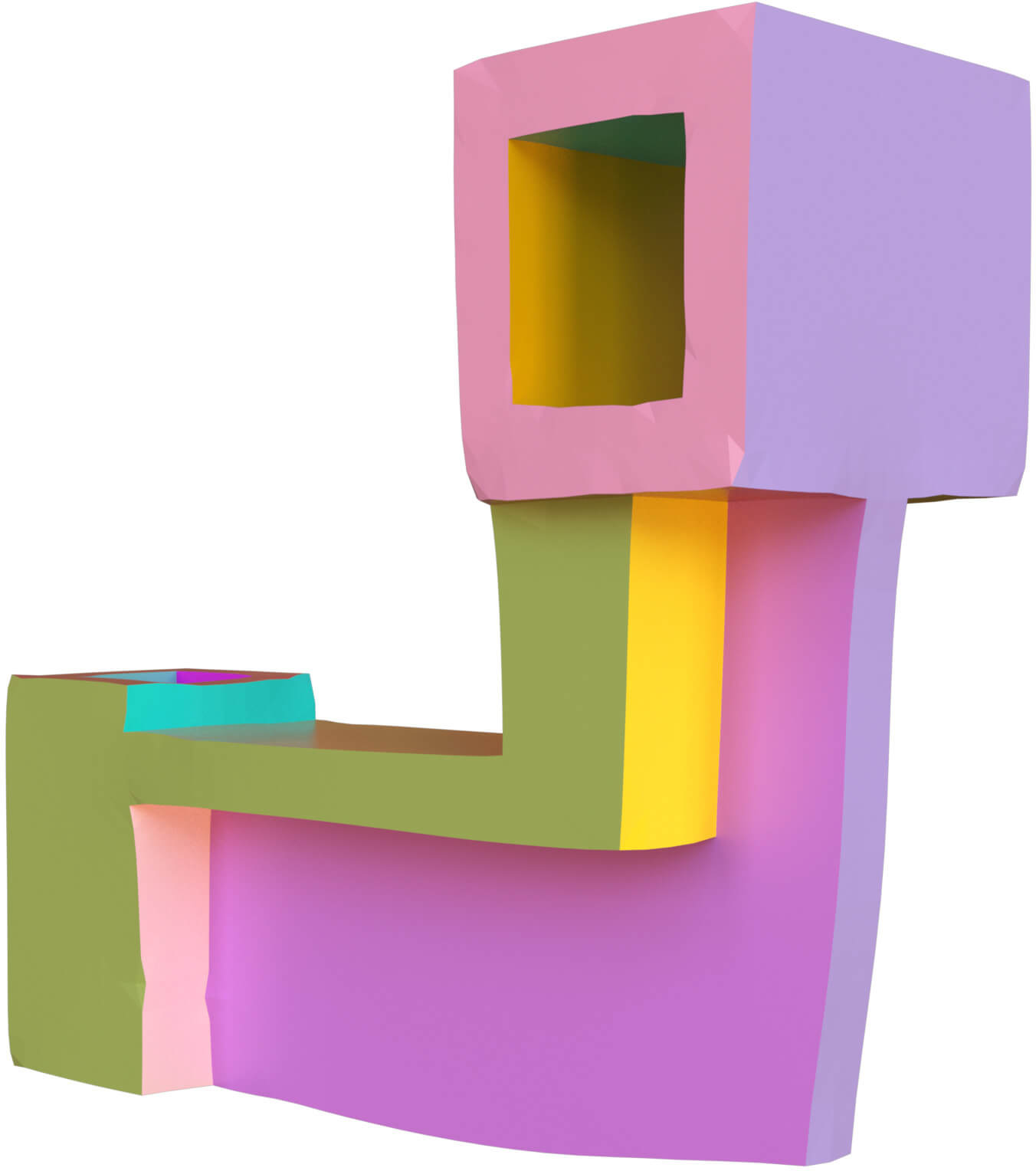}
		\caption{5}        
	\end{subfigure}     	
	\begin{subfigure}[b]{0.135\textwidth}
		\includegraphics[width=\textwidth]{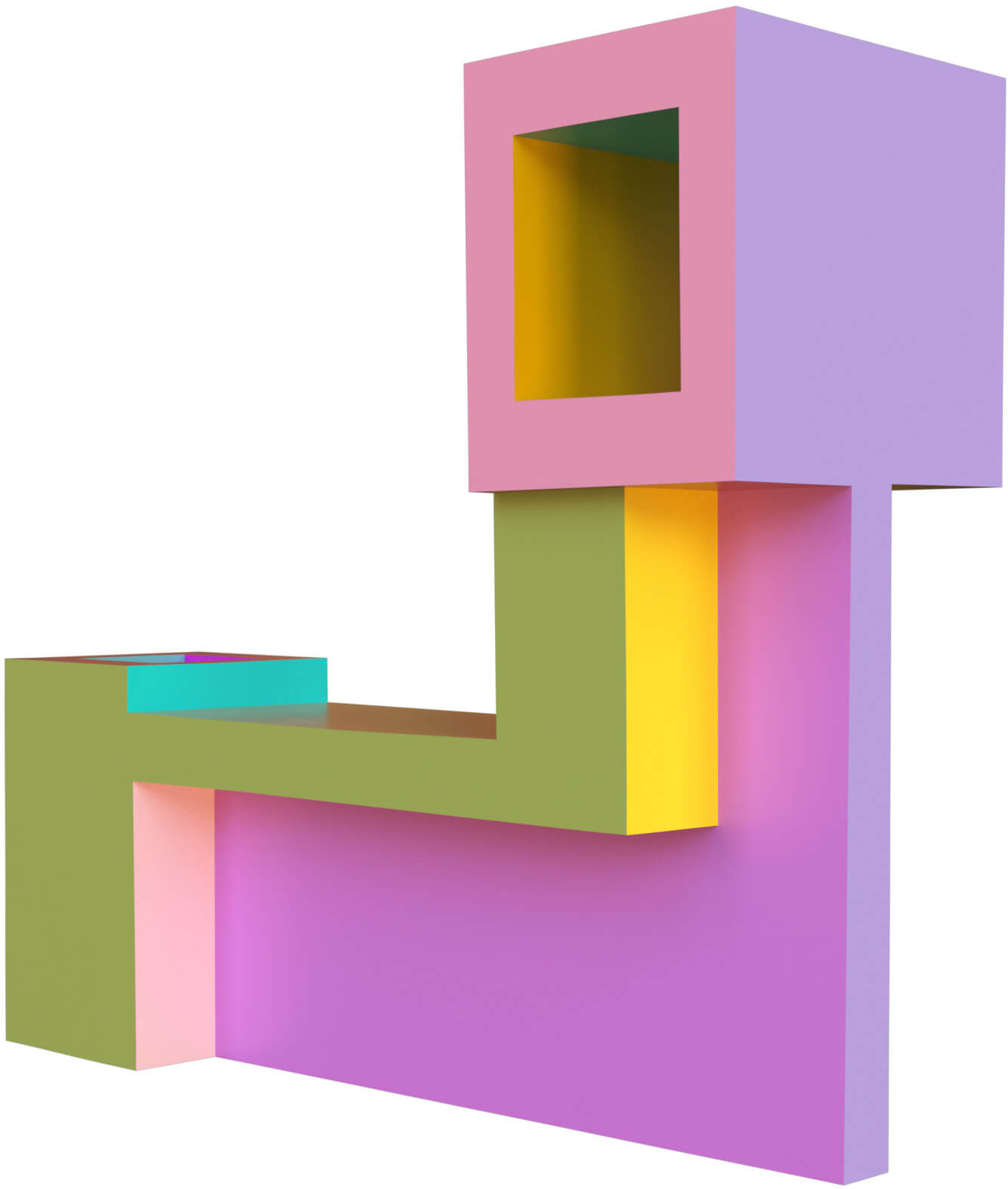}
		\caption{150}  
	\end{subfigure}
	\caption{The iteration of the polycube deformation of two models.}\label{fig:polycubeIteration}
\end{figure*}

\begin{figure*} [th!] 
	\centering   
	\begin{subfigure}[b]{0.08\textwidth}
		\includegraphics[width=\textwidth]{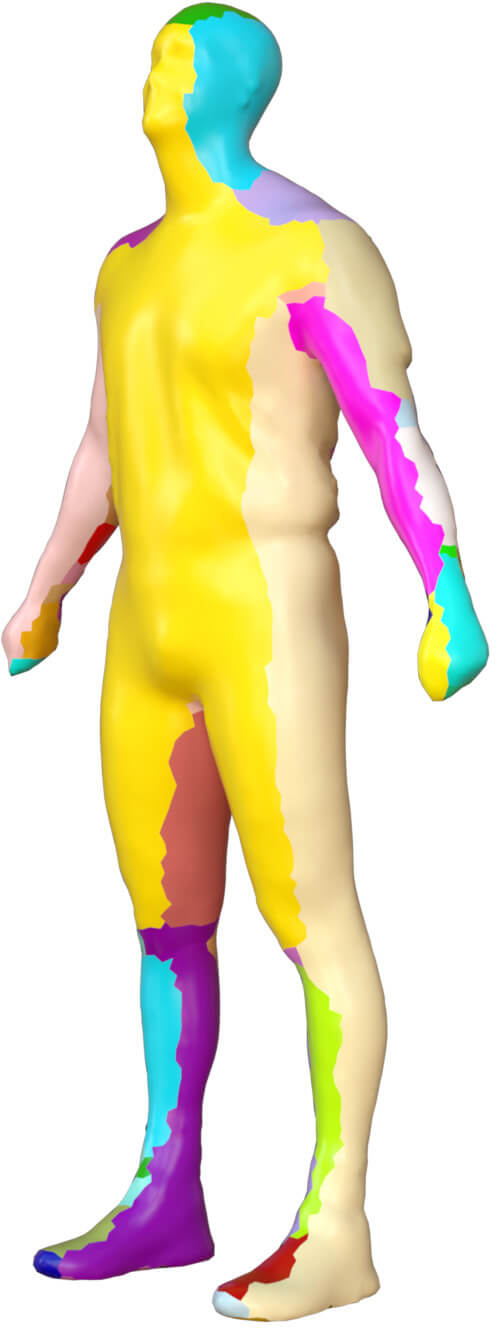}
		\caption*{ }   
	\end{subfigure}
	\begin{subfigure}[b]{0.18\textwidth}
		\includegraphics[width=\textwidth]{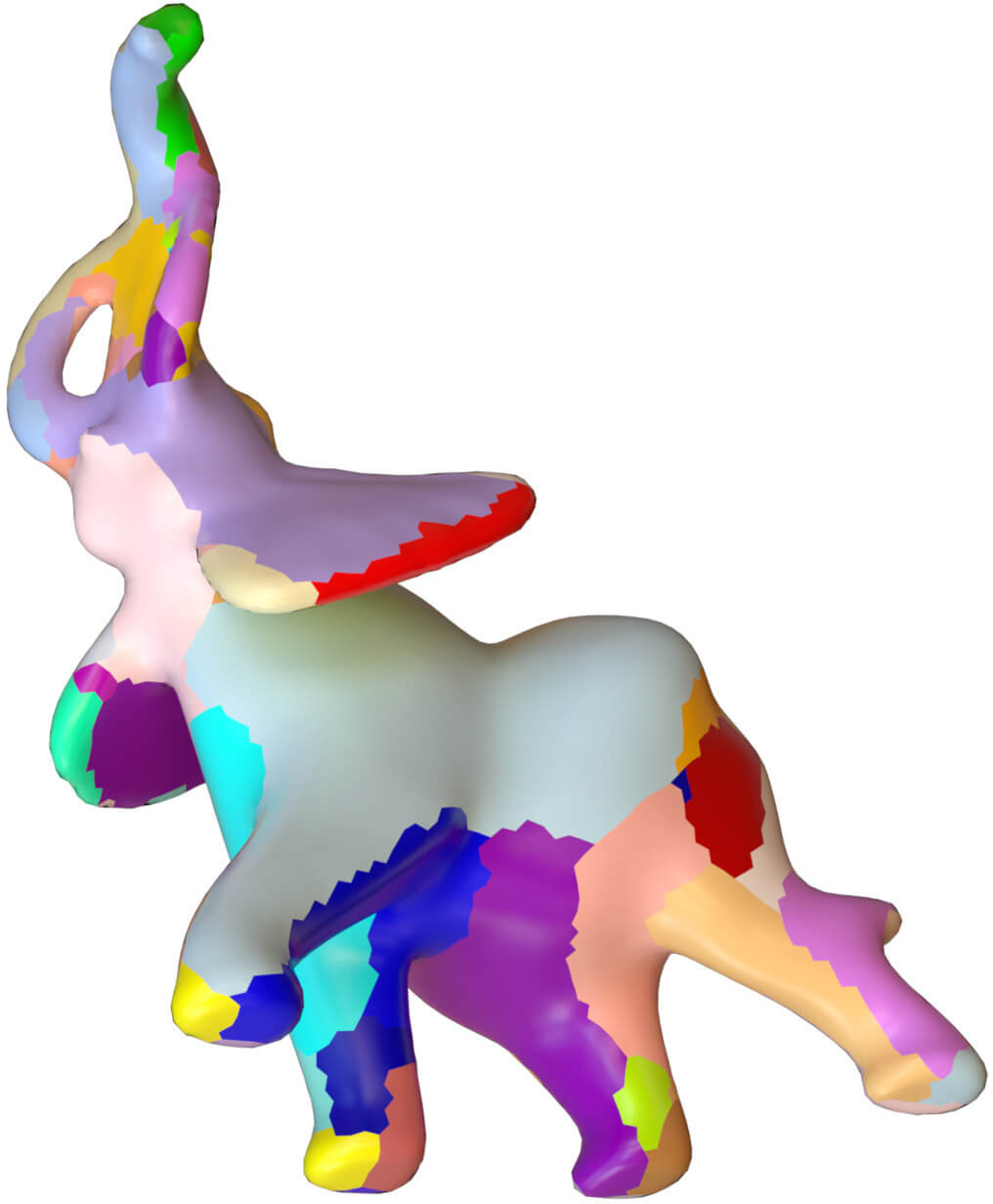}
		\caption*{ }   
	\end{subfigure}           
	\begin{subfigure}[b]{0.18\textwidth}
		\includegraphics[width=\textwidth]{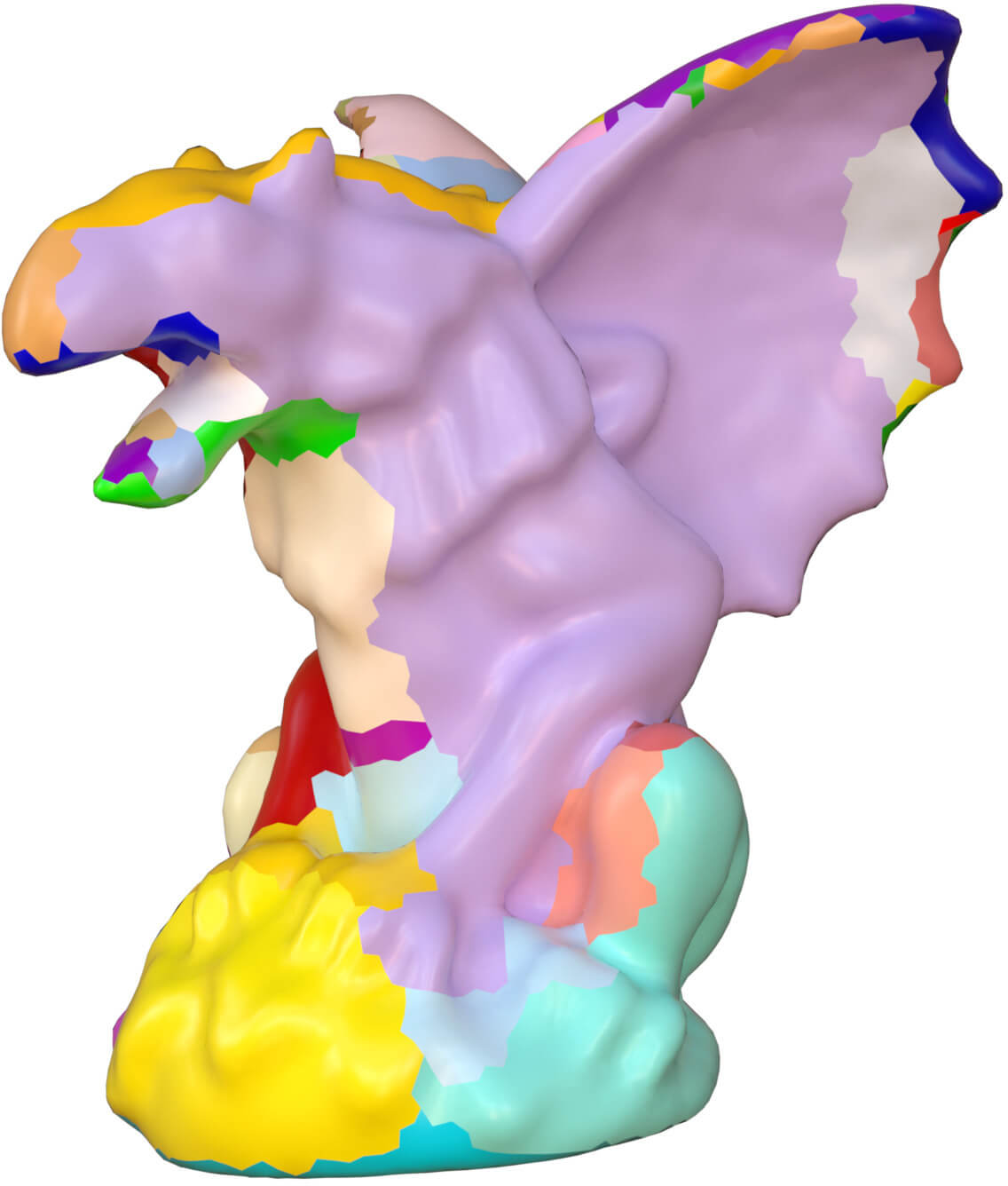}
		\caption*{ }     
	\end{subfigure}         
	\begin{subfigure}[b]{0.18\textwidth}
		\includegraphics[width=\textwidth]{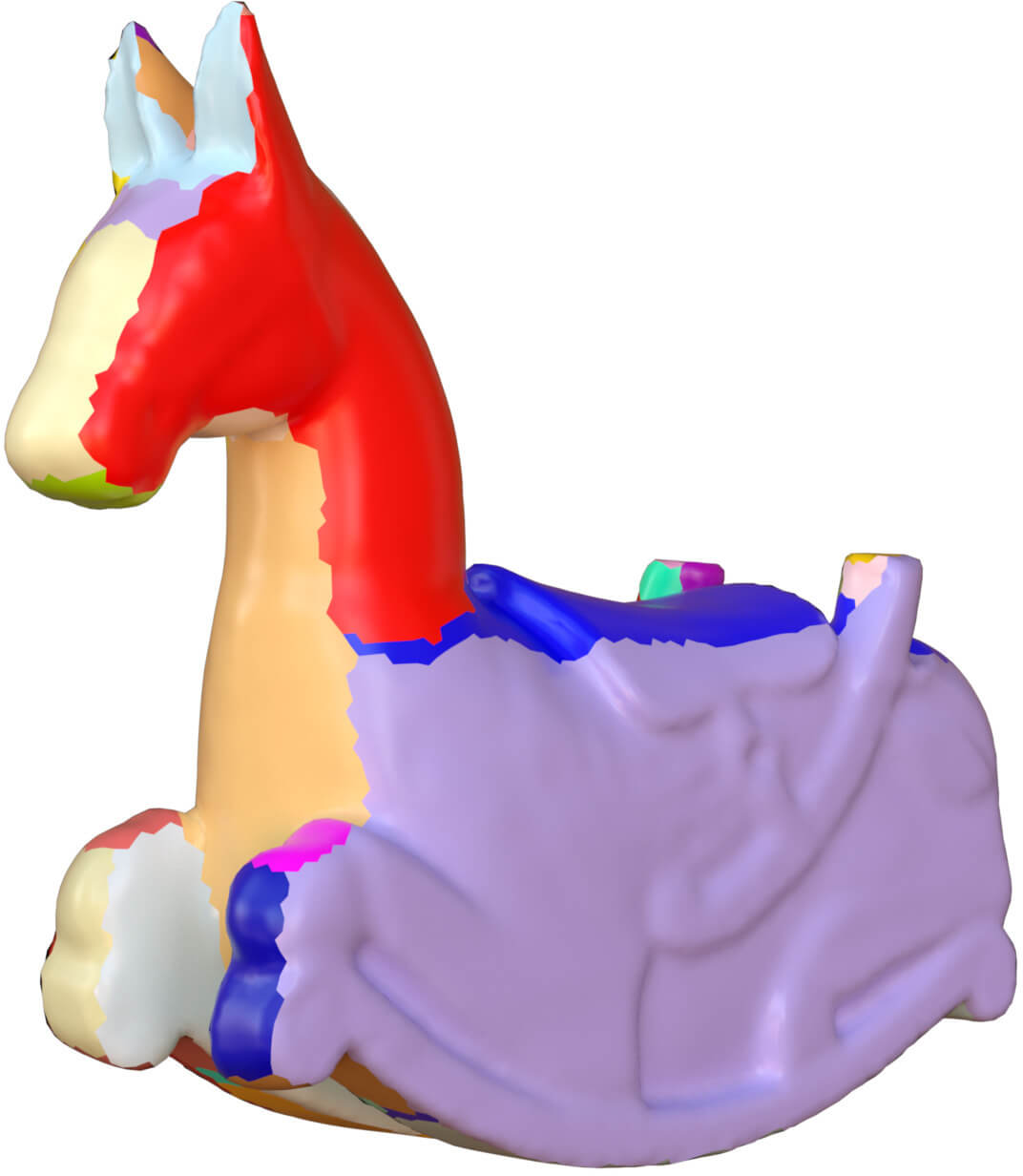}
		\caption*{ }       
	\end{subfigure}
	\begin{subfigure}[b]{0.18\textwidth}
		\includegraphics[width=\textwidth]{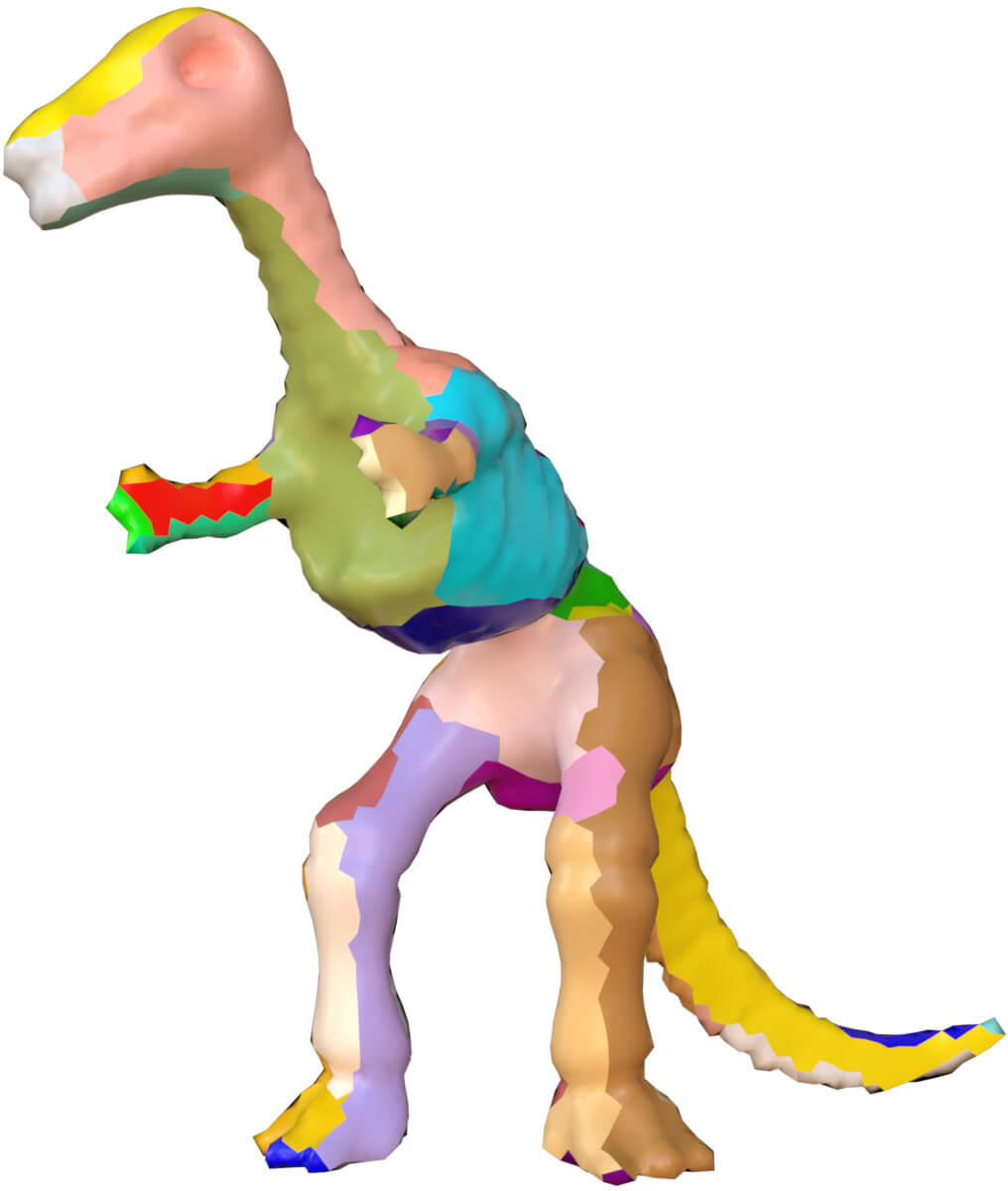}
		\caption*{ }         
	\end{subfigure}
	\begin{subfigure}[b]{0.14\textwidth}
		\includegraphics[width=\textwidth]{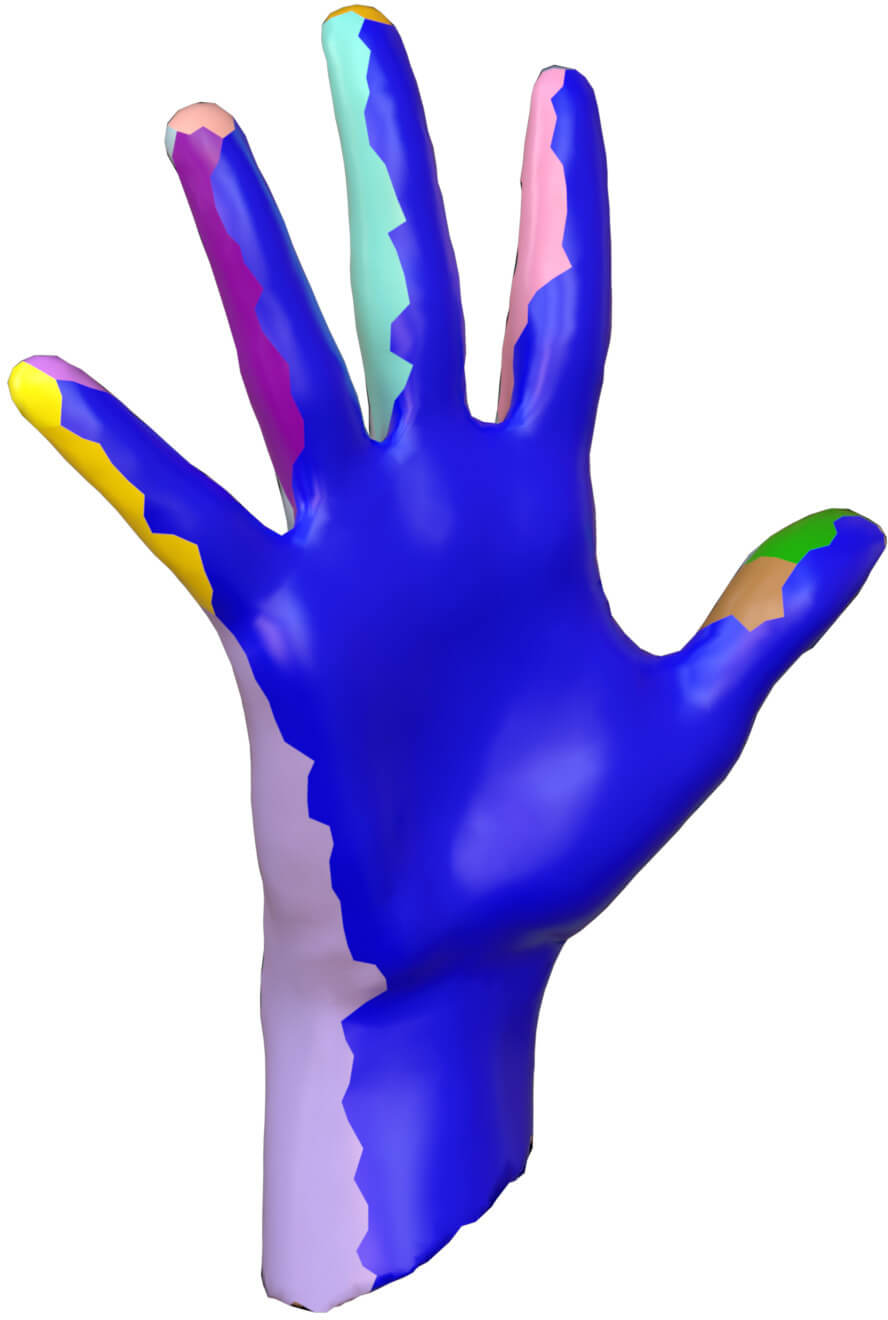}
		\caption*{ }        
	\end{subfigure}

	\begin{subfigure}[b]{0.065\textwidth}
		\includegraphics[width=\textwidth]{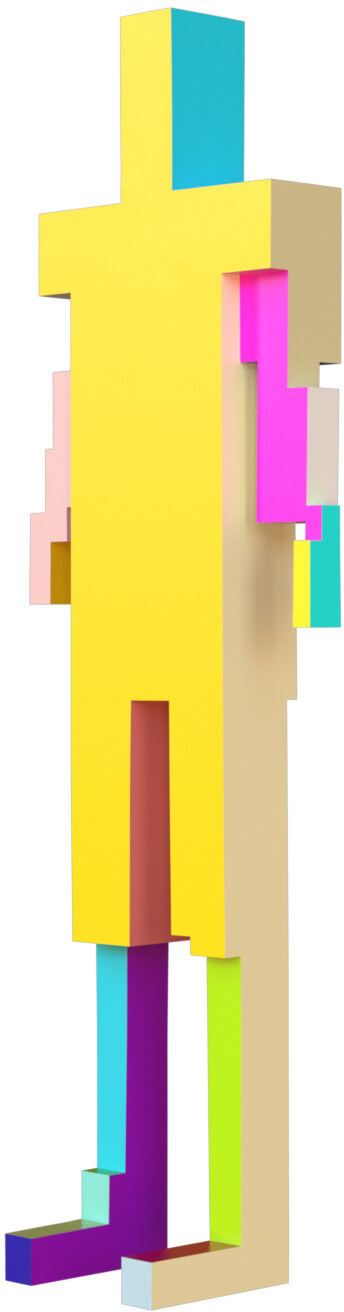}
		\caption{man}  
	\end{subfigure}
	\begin{subfigure}[b]{0.23\textwidth}
		\includegraphics[width=\textwidth]{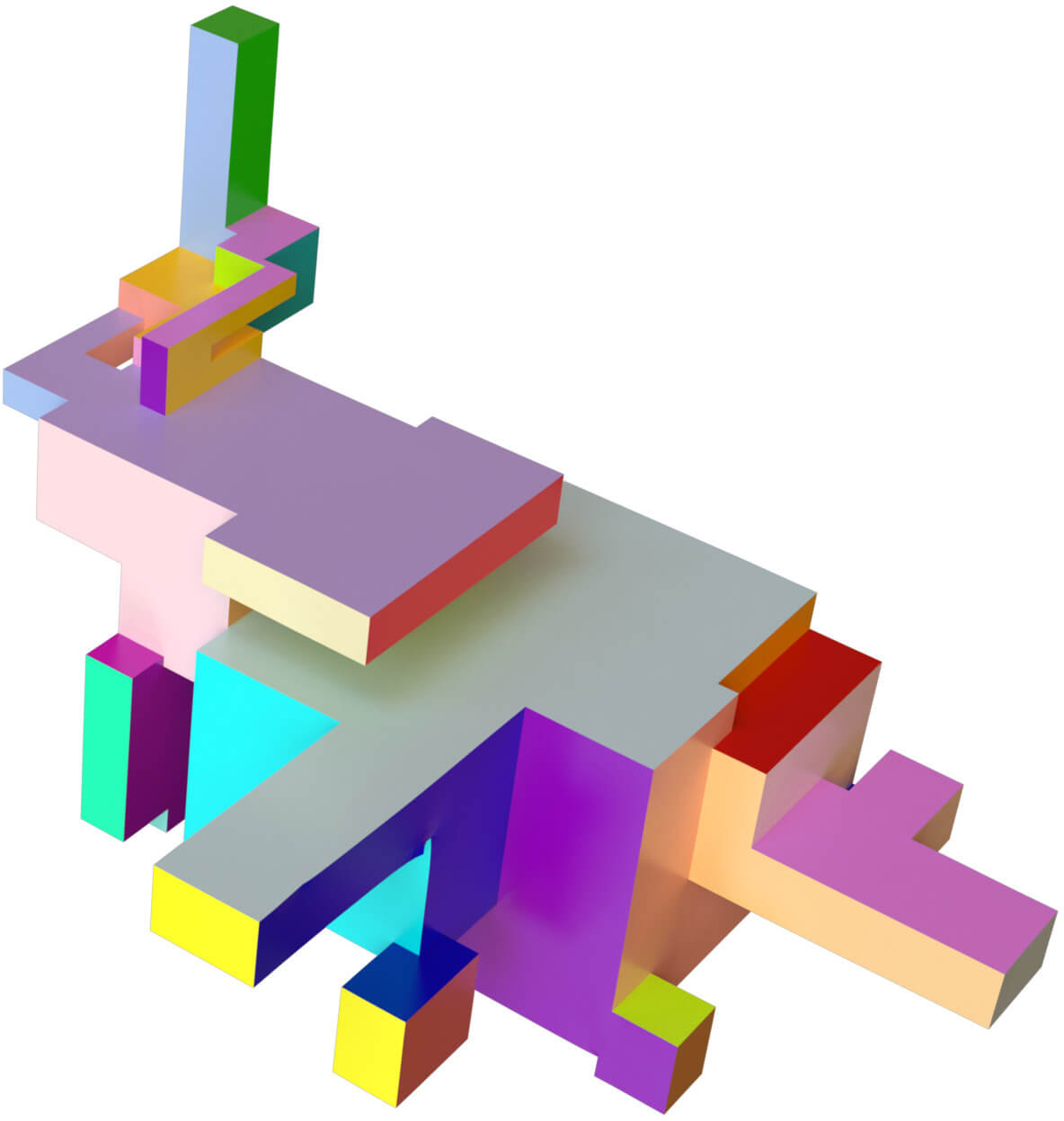}
		\caption{elephant}  
	\end{subfigure}           
	\begin{subfigure}[b]{0.21\textwidth}
		\includegraphics[width=\textwidth]{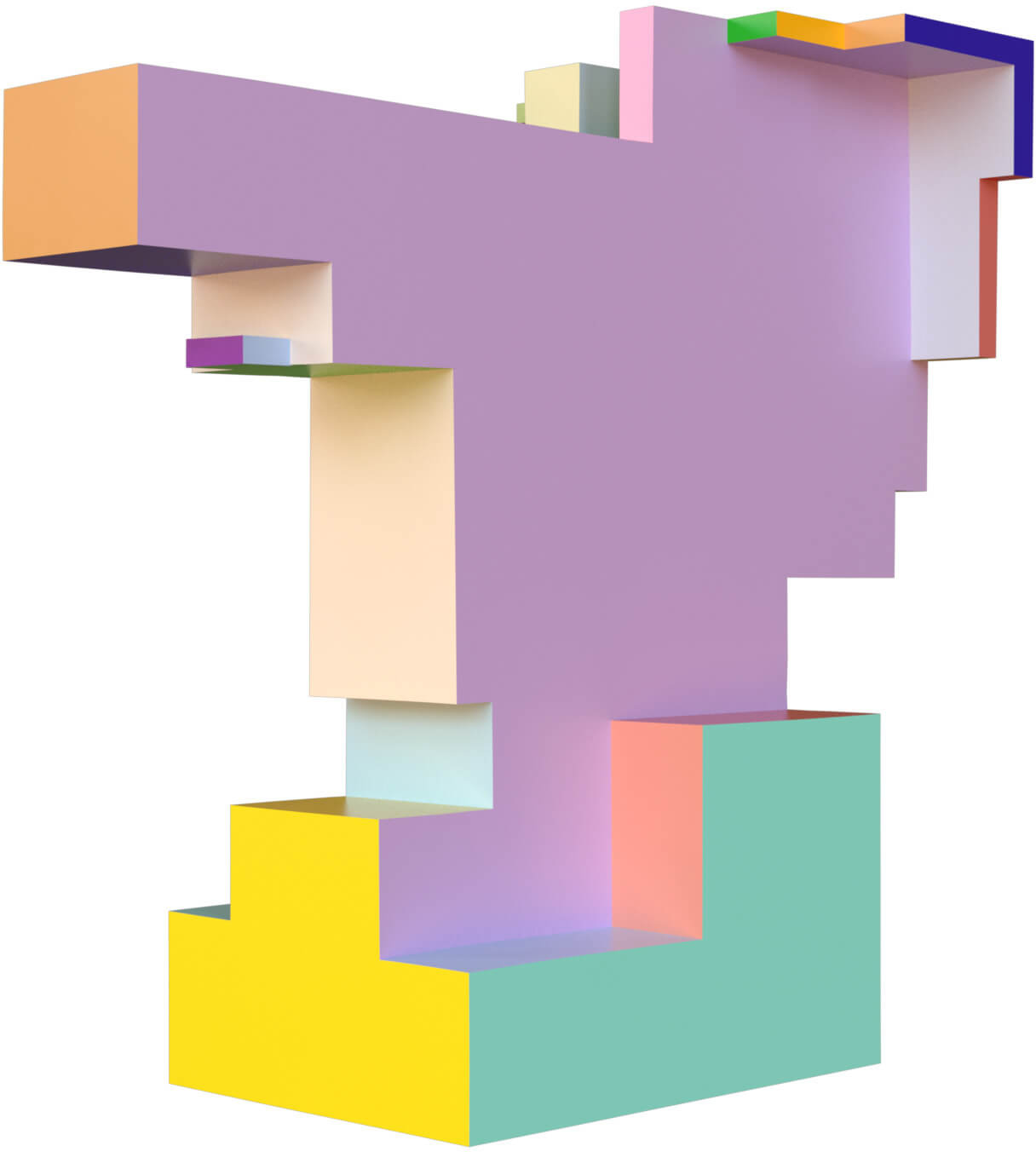}
		\caption{gargoyle}      
	\end{subfigure}         
	\begin{subfigure}[b]{0.205\textwidth}
		\includegraphics[width=\textwidth]{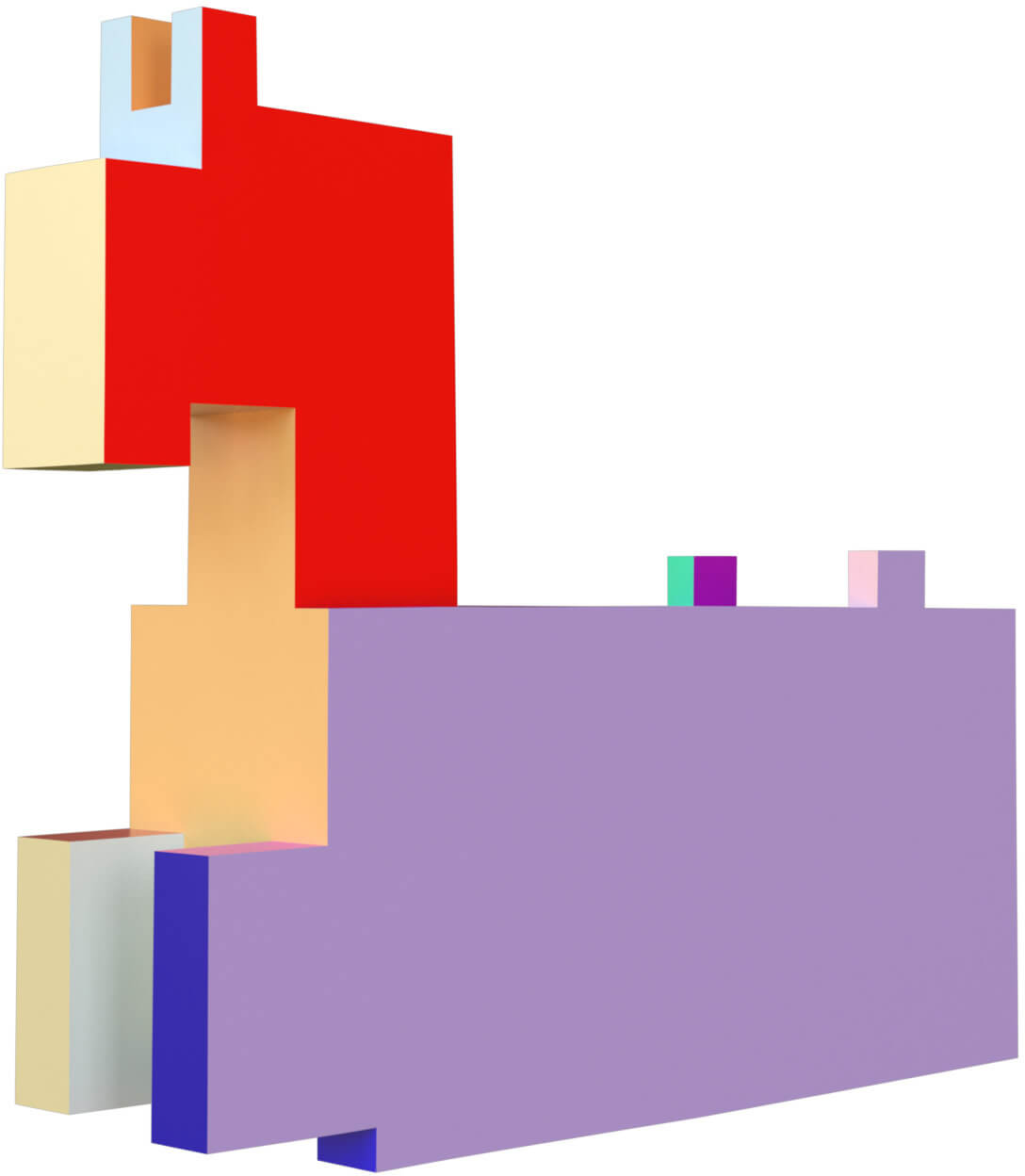}
		\caption{isidorehorse}        
	\end{subfigure}
	\begin{subfigure}[b]{0.12\textwidth}
		\includegraphics[width=\textwidth]{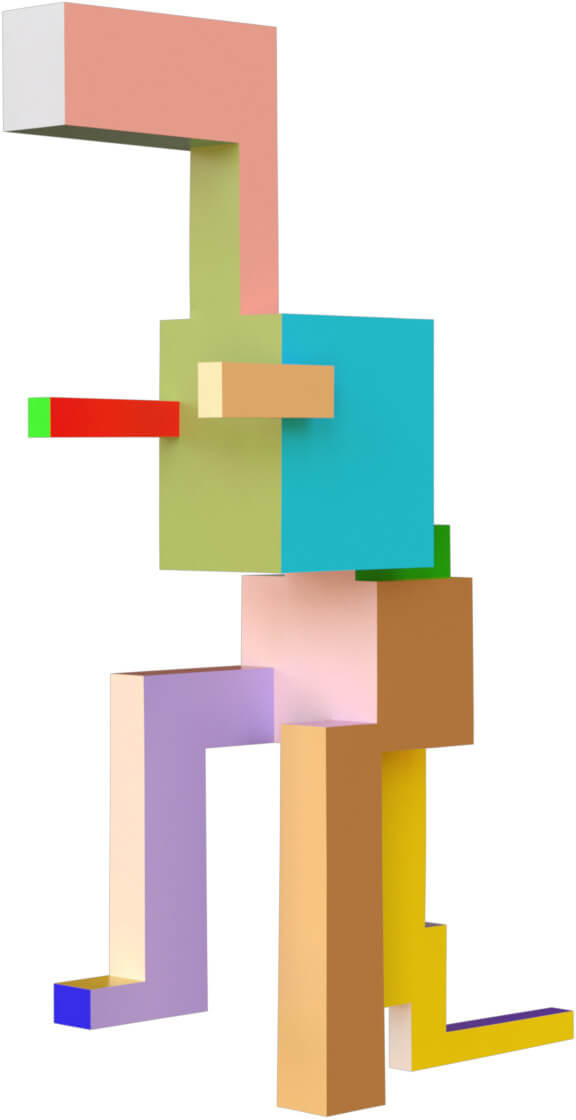}
		\caption{dinosaur}       
	\end{subfigure}
	\begin{subfigure}[b]{0.105\textwidth}
		\includegraphics[width=\textwidth]{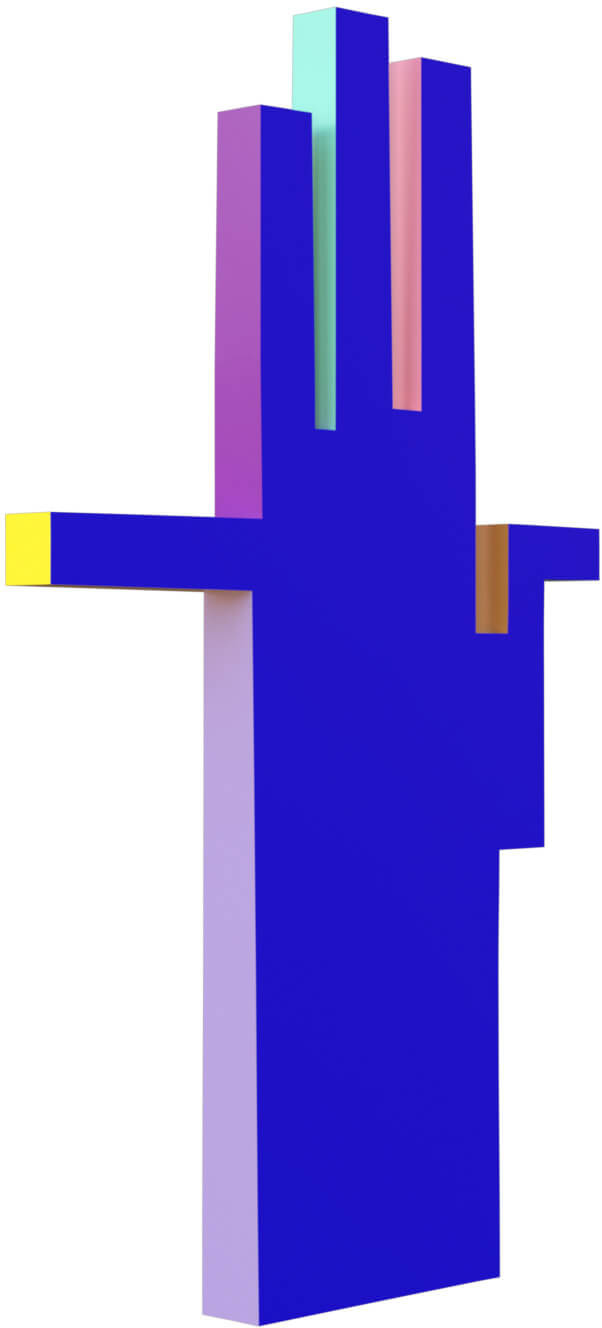}
		\caption{hand}        
	\end{subfigure}  
	\caption{The six models and their polycube shape.}\label{fig:polycubeDemo}
\end{figure*}

\subsection{Polycube Geometry}
 The second step aligns and reorients the triangles in each chart with one axis direction. And every chart should be mapped into a plane rectangle and all of chart boundaries are straight lines.
 
 When creating a polycube from a mesh, we wish that  the  parameterization distortion is low. At the same time, the number of singularity, i.e. the chart corners and the number of charts is kept low.  The algorithm of polycube construction should provide an optimal trade-off between parametrization distortion and chart or singularity numbers.
 
 Our polycube geometry method is based on a Poisson system which reconstructs the deformed polycube mesh to satisfies the current assigned face normals of triangles. As the Poisson system can only approximate the  input normal requirements, therefore we use an iterative Poisson systems. After several iteration, our system converges and outputs the corresponding polycube shape whose boundary between two patches  is  exactly straight automatically,  and the triangles in every chart fall on a plane automatically without the need of any extra planarity constraints.
 
 Changing a model into its polycube shape fundamentally is a surface deformation process. Previously all kinds of deformation algorithms  \cite{Hui:2016:Deform,Chao:2010:SGM,DeformationSurvey:2008,ARAP_modeling:2007} focus on preserving the local features of the original model as much as possible. In polycube case, we propose that the target of the deformation is preserving the metric instead of local features. 
 In \cite{Hui:2016:Deform}, the deformation energy is separated into two kinds of energy explicitly: stretching and bending energy. The stretching energy tries to preserve the metric, and the bending energy preserves the mean curvatures.  Motived by their explanation and derivation on stretching energy \cite{Hui:2016:Deform}.  Our algorithm  adopts and modifies their stretching energy   under the constraints of the target normal direction of every  triangle of the polycube topology. 
 In their original method \cite{Hui:2016:Deform}, the rotations of faces are   unknown variables and changed in every iterative step, however in our configuration, the rotations are known in advance and kept the same in our iteration.
 In fact, our deformation does not preserve the metric in theory, but in practice we observe the change of edge length is small.

Let $S$ be  an original surface and $S'$ be its deformed surface embedded in 3-dimension. And denote a 3-vector $x_v$ be the position associated with vertex $v$ of
$S$, and a 3-vector $x'_v$  with vertex $v$ of of $S'$. On every triangle of the mesh, we define one rotation matrix variable referred to  as $R(t)$. The stretching energy \cite{Hui:2016:Deform} is defined as 
\begin{align} 
	E(x',R)  =\sum_{he_{vw}} \cot (a_{vw})
	\| (x^{\prime}_{v}-x^{\prime}_{w} )-R(t_{vw})(x_{v}-x_{w})\|^{2} .
\label{func:StretchingEnergy}
\end{align}

In above,  $\|\cdot\|^2$ is the standard 3-vector norm, $he_{vw}$ represents the half edge from
the vertex $v$ to $w$. We denote the angle of the corner opposite to the half edge $he_{vw}$ in its triangle with $a_{vw}$.
Finally $R(t_{vw})$ represents the $3 \times 3$  rotation matrix associated with the
triangle face whose the half edge is $he_{vw}$.

It is proved in \cite{Hui:2016:Deform} that the $E(x')$ measures
the quantity
\begin{equation}
\int_{s} 
(\sigma_1(p) -1)^2 + (\sigma_2(p)-1)^2
\;\; dA_{g}(p) ,
\end{equation}
where the symbol $\sigma_1(p)$ and  $\sigma_2(p)$ represent the
the maximum and  minimal stretching ratios
of a tangent vectorof $S$ at a point $p$ under the differential mapping $dx'$ from $S$ to $R^3$.
Therefore $E(x')$ is one reasonable quantity to measure the stretching of a deforming surface.

This stretching energy is quadratic in $x'$ with a fixed rotation matrix $R$ per triangle. 
Take the gradient of the stretching energy and set it to zero. We can obtain the optimal variables $x'$ by solving a single linear system as the following: 
 \begin{align}  \label{eqn:PseudoPoisson} 
 &\sum_{w \in N(v)} \big[\cot(a_{vw})+\cot(a_{wv})\big]
 (x^{\prime}_{v} -  x^{\prime}_{w})   \nonumber \\
 =     
 &\sum_{w \in N(v)} 
 \big[
 \cot(a_{vw})R(t_{vw})
 -  \cot(a_{wv})R(t_{wv})
 \big]
 (x_{v} -  x_{w})    .     
 \end{align} 
 
By defining the 3-vector at vertex $v$ as:
 \begin{equation}
 b_v := 
 \sum_{w \in N(v)} 
 \big[
 \cot(a_{vw})R(t_{vw})
 +\cot(a_{wv})R(t_{wv})
 \big]
 (x_{v} - x_{w})   ,      
 \end{equation}
we can change the above system into matrix format as:
 \begin{equation} 
 \label{eqn:one}
 L x' =  b ,
  \end{equation}
 where $L$ is the 
 $n$-by-$n$  Laplacian matrix, $x'$ and
 $b$ are n-vectors of 3-vectors.

\textbf{Rotation per triangle.}
In the above system, we need know the rotation matrix of every triangle face of the mesh. Although we do not know the exact vertex positions of the polycube in advance, 
the face normals of the polycube are determined and fixed  by its polycube topology. Therefore we can calculate the rotation matrix for every triangle from its  unit normal on the original mesh and the target normal from its polycube topology without knowing the target polycube shape.  Given two unit normals, the rotation between them can be computed by the algorithm of Rodrigues' rotation formula.

\textbf{Iteration.}
Poisson system basically is an approximation method.  The system \ref{eqn:one} cannot result in an exact polycube, such as the models with a nearly polycube shape shown in Figure \ref{fig:iterfirst}. We fix the problem with an iteration method. In every step $i$, we recompute the rotation matrix $R_i(t)$ for the triangle $t$ according the face normal of the current model and the target normal from its polycube topology. With the new  $R_i(t)$, we update the  $L_i$ and $b_i$. Then the iteration system is as the following:
 \begin{equation} 
\label{eqn:iter}
L_i  x_i' =  b_i \, .
\end{equation}

Figure \ref{fig:polycubeIteration} demonstrates  the iteration process of the two polycube deformation. We observe that the polycube shapes get better and better after each iteration. And the planarity and straightness of the polycube are realized in the converge of the iterations. In these experiments, It is shown that the polycube shape in the fifth step is almost the same as the one from the hundredth step. Therefore  the converge speed of our polycube  deformation is very fast. In practice, the speed varies according to the different models.

The stretching energy defined in equation \ref{func:StretchingEnergy} can measure the stretching ratio if it is a function of both rotation $R$ and unknown position vectors $x'$.  In our framework, we fix the variable $R$,   our system is only a function of  unknown position vectors $x'$. 
Therefore our method does not minimize the stretching energy, it is  a simple Poisson system. 
The explanation of  "stretching energy" in \cite{Hui:2016:Deform} gives us a hint why our simple Poisson system changes edge length not so much in practice.

\section{Results and Demonstrations}
We test our method in a variety of meshes with  complicated topology, with or without boundary. Our experiments demonstrate that  our algorithms can work on all kinds of shapes. 

In Figure \ref{fig:polycubeDemo}, we show   several models and their polycube shapes. It shows that our algorithms can obtain perfect polycubes no matter the complexity of the model shapes.

\begin{figure*} [th!] 
 	\centering   
 	\begin{subfigure}[b]{0.155\textwidth}
 		\includegraphics[width=\textwidth]{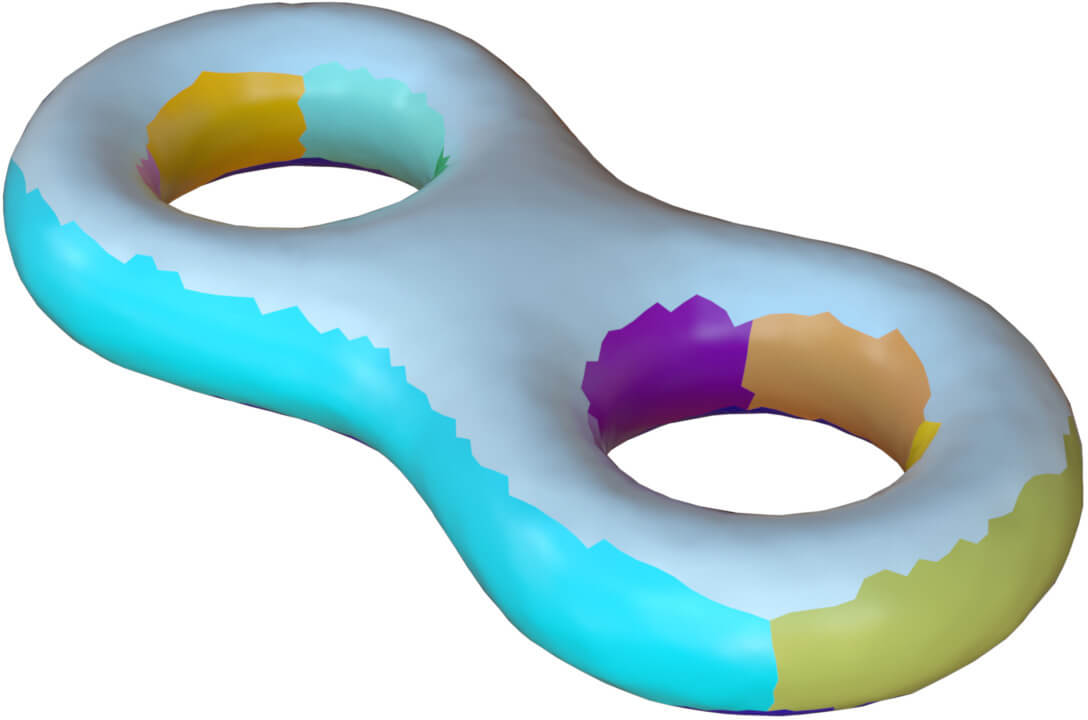}
 		\caption*{ }   
 	\end{subfigure}	
 	\begin{subfigure}[b]{0.125\textwidth}
 		\includegraphics[width=\textwidth]{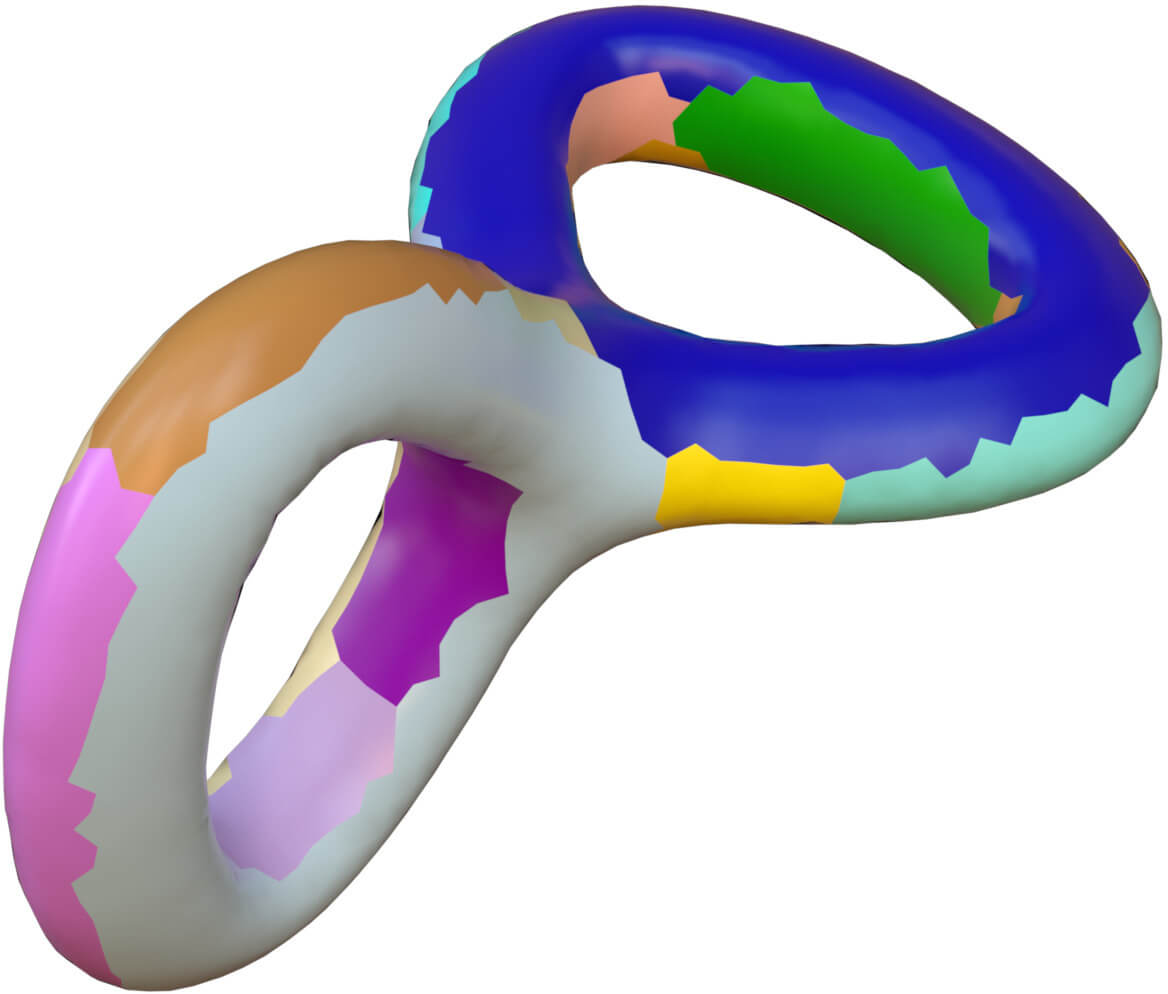}
 		\caption*{ }   
 	\end{subfigure}
 	\begin{subfigure}[b]{0.165\textwidth}
 		\includegraphics[width=\textwidth]{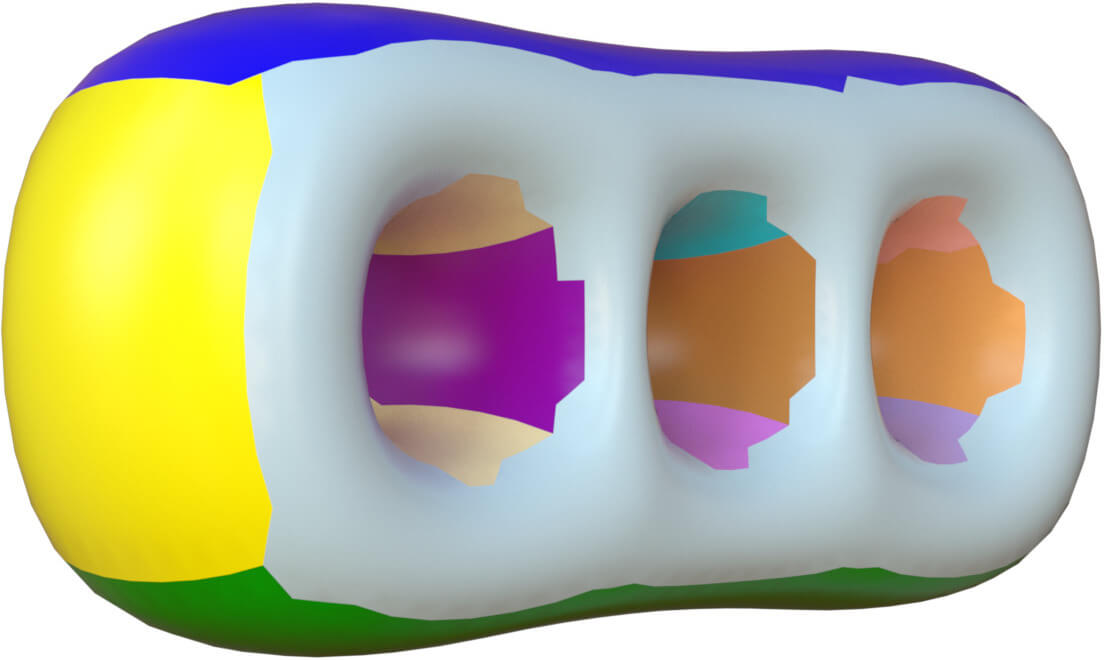}
 		\caption*{ }   
 	\end{subfigure}           
 	\begin{subfigure}[b]{0.165\textwidth}
 		\includegraphics[width=\textwidth]{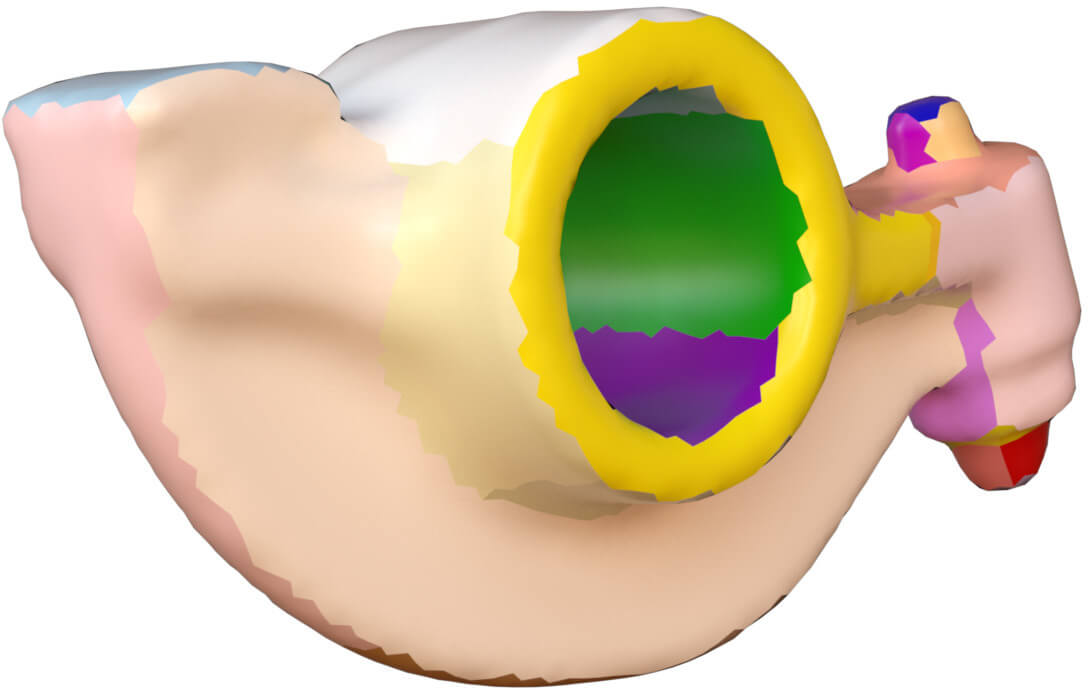}
 		\caption*{ }     
 	\end{subfigure}         
 	\begin{subfigure}[b]{0.085\textwidth}
 		\includegraphics[width=\textwidth]{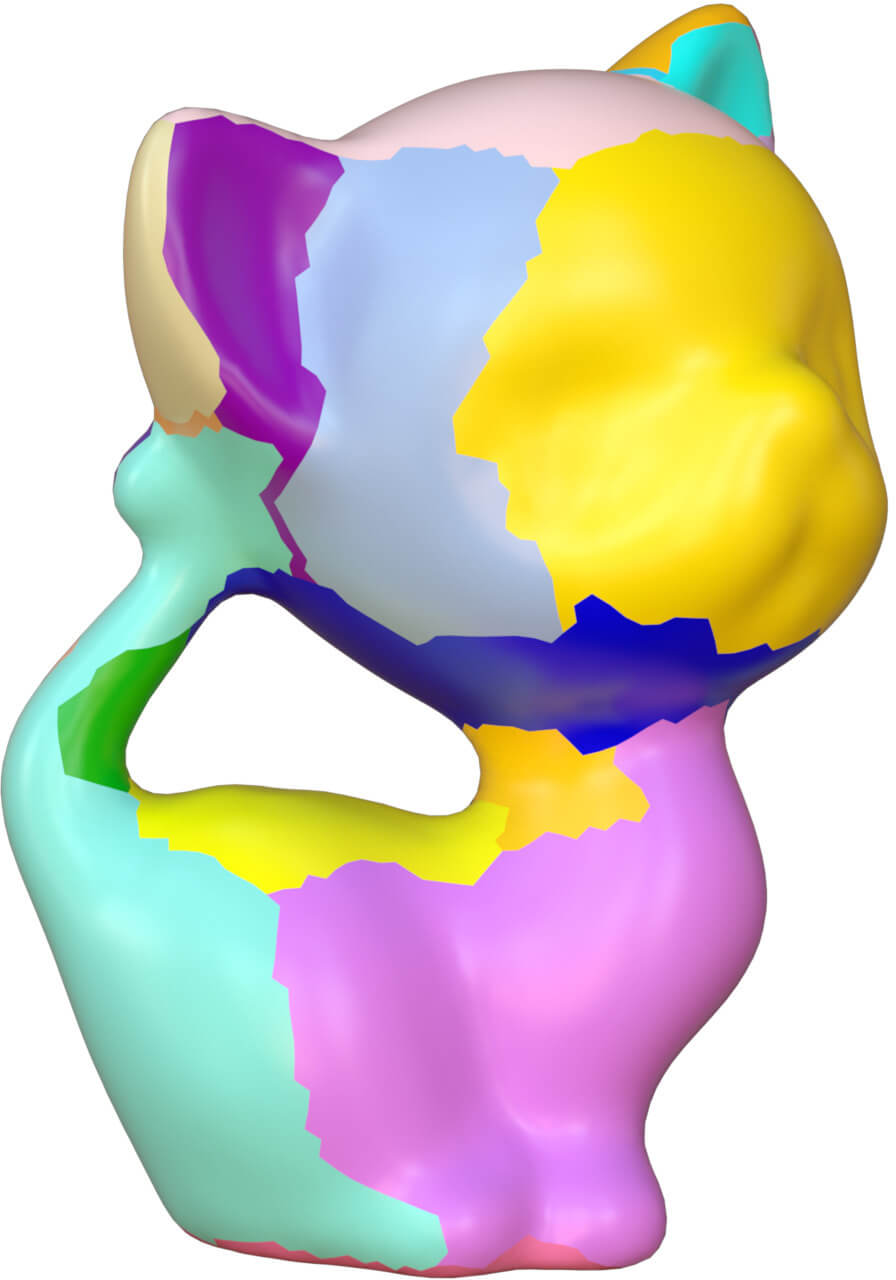}
 		\caption*{ }       
 	\end{subfigure}
 	\begin{subfigure}[b]{0.115\textwidth}
 		\includegraphics[width=\textwidth]{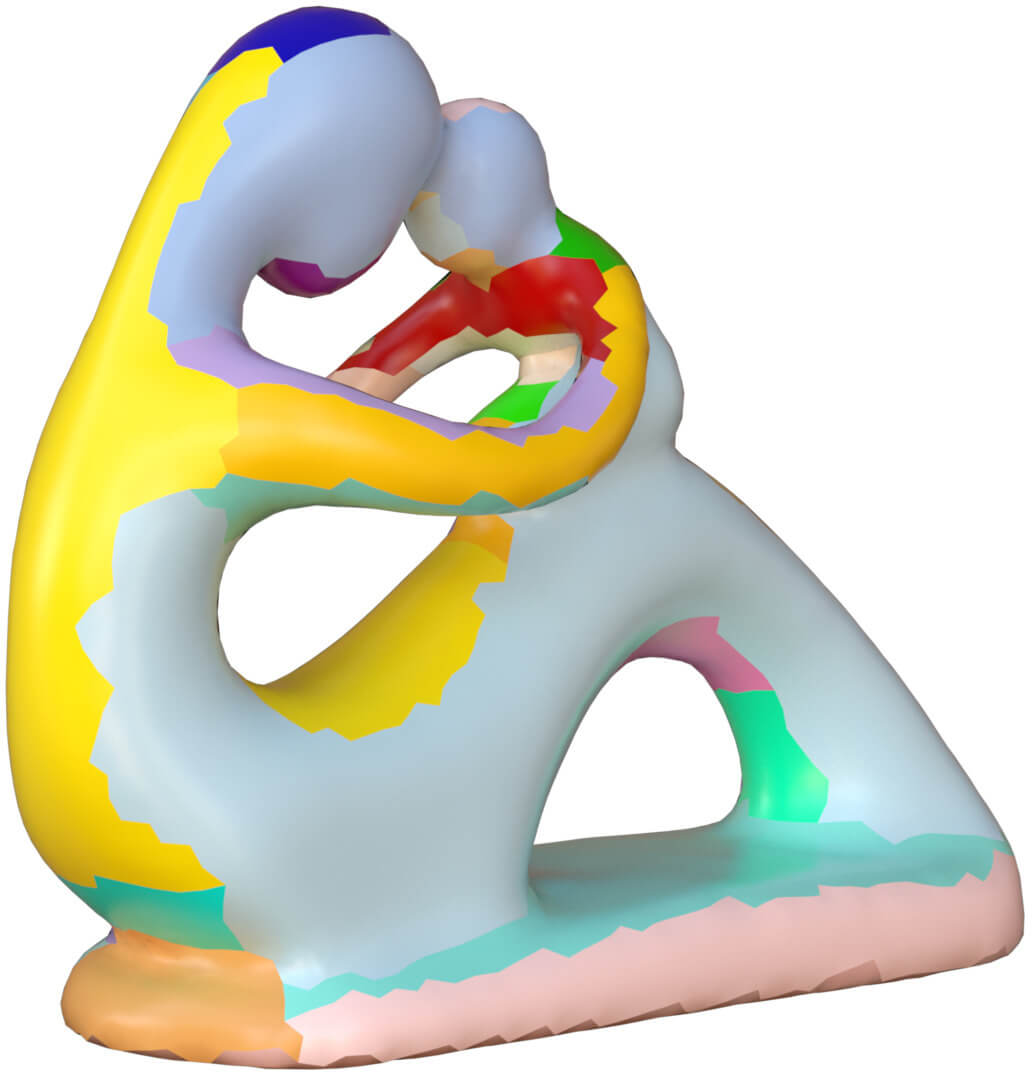}
 		\caption*{ }         
 	\end{subfigure}
 	\begin{subfigure}[b]{0.165\textwidth}
 		\includegraphics[width=\textwidth]{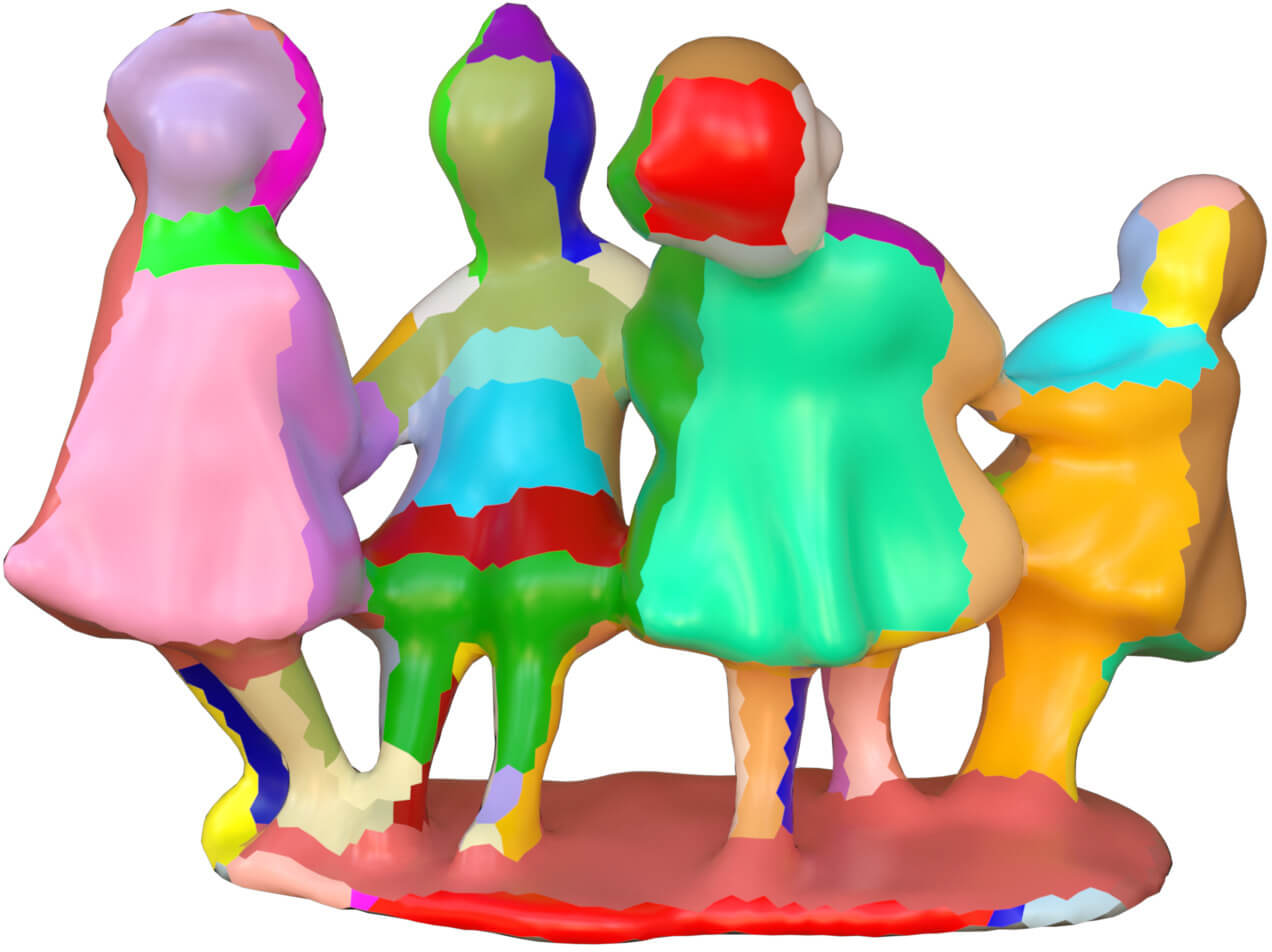}
 		\caption*{ }        
 	\end{subfigure}    
 	
 	\begin{subfigure}[b]{0.155\textwidth}
 		\includegraphics[width=\textwidth]{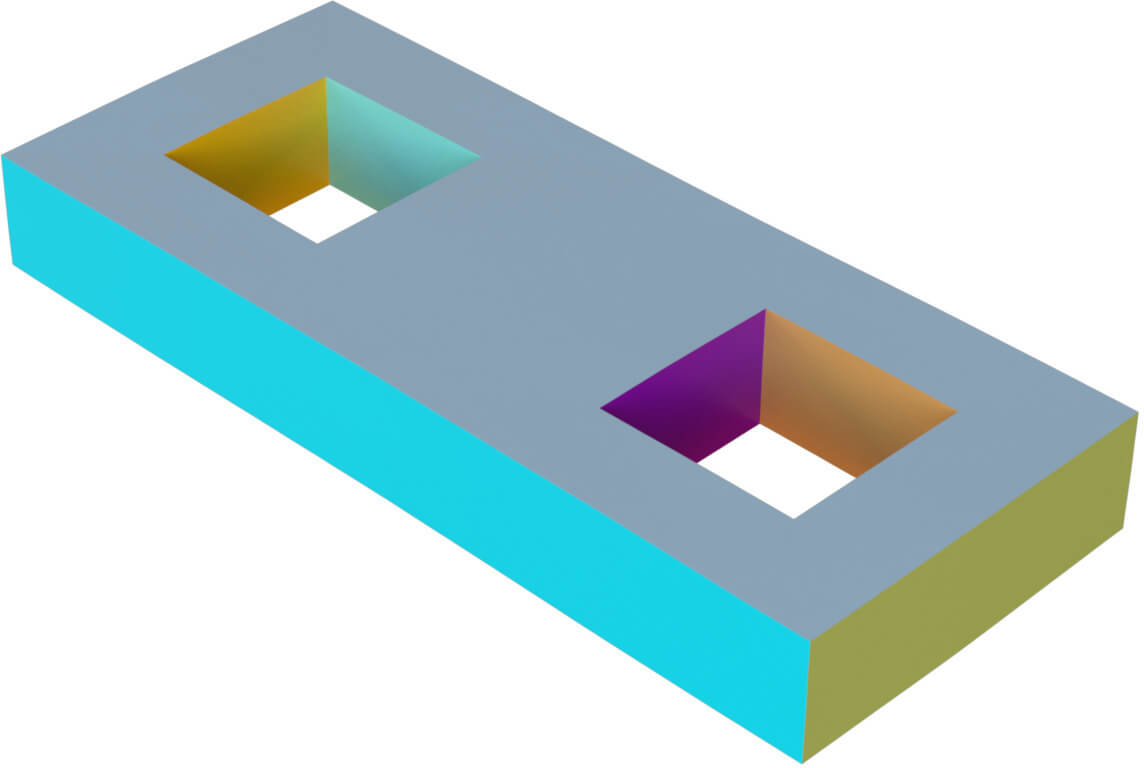}
 		\caption{eight}  
 	\end{subfigure} 	
 	\begin{subfigure}[b]{0.145\textwidth}
 		\includegraphics[width=\textwidth]{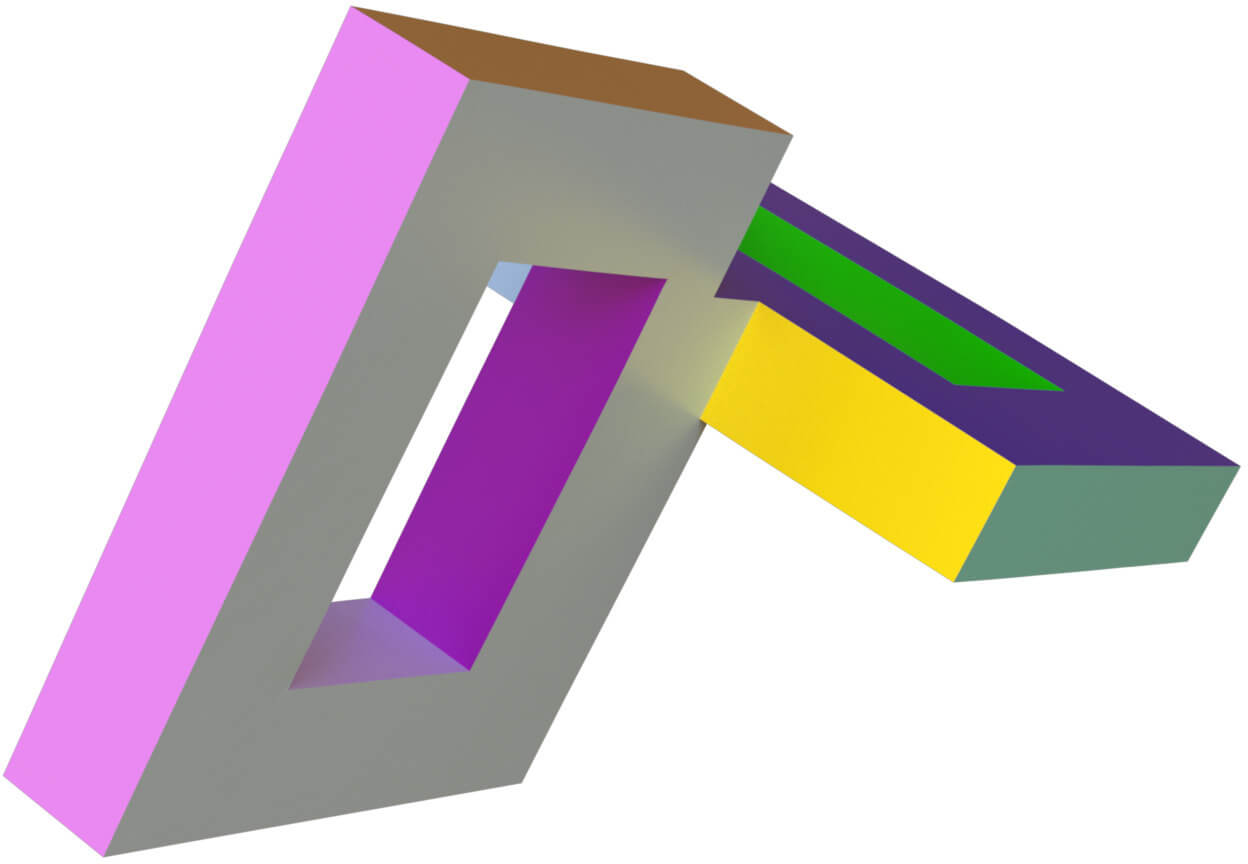}
 		\caption{ }  
 	\end{subfigure}
 	\begin{subfigure}[b]{0.185\textwidth}
 		\includegraphics[width=\textwidth]{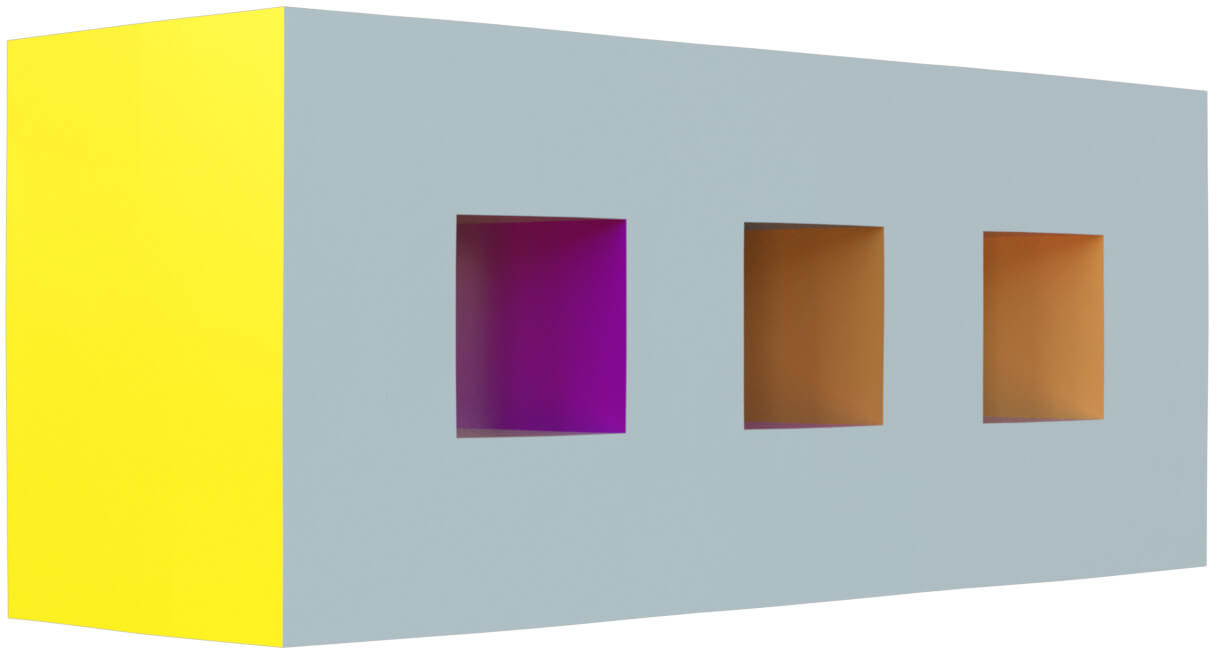}
 		\caption{ }  
 	\end{subfigure}           
 	\begin{subfigure}[b]{0.155\textwidth}
 		\includegraphics[width=\textwidth]{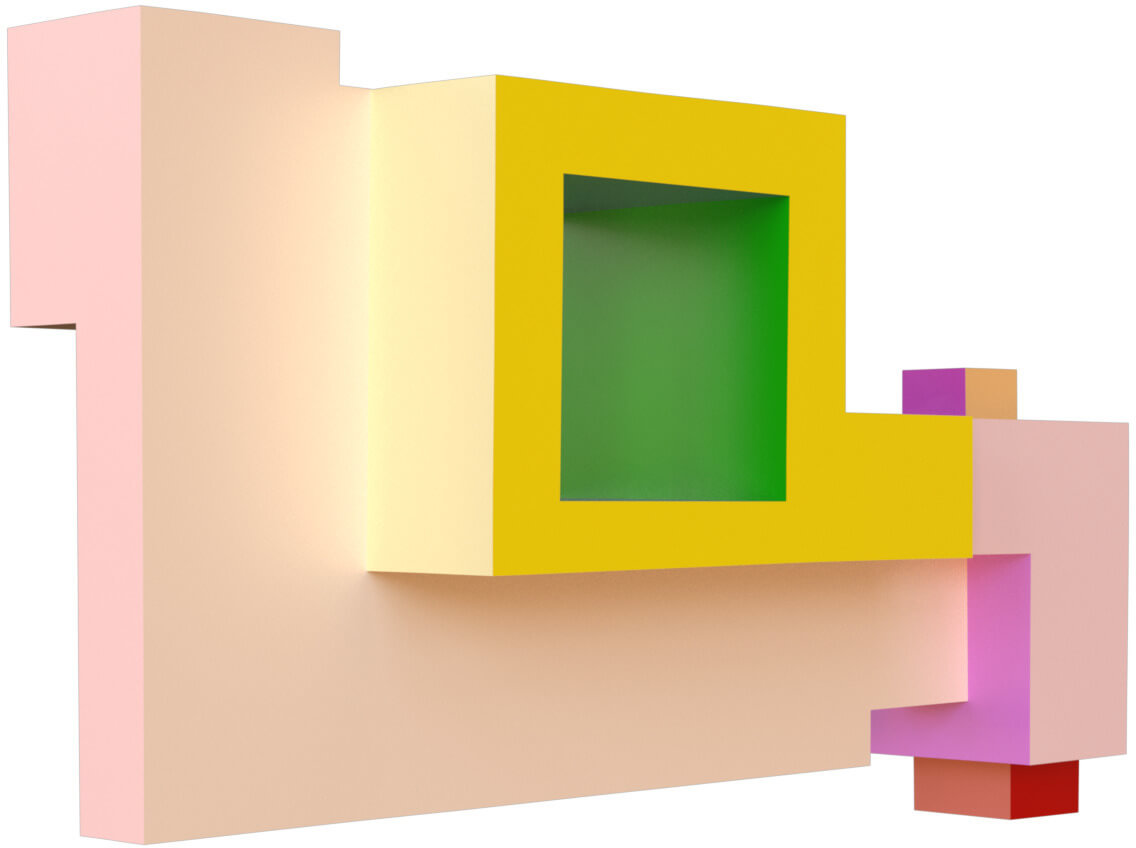}
 		\caption{ }      
 	\end{subfigure}         
 	\begin{subfigure}[b]{0.070\textwidth}
 		\includegraphics[width=\textwidth]{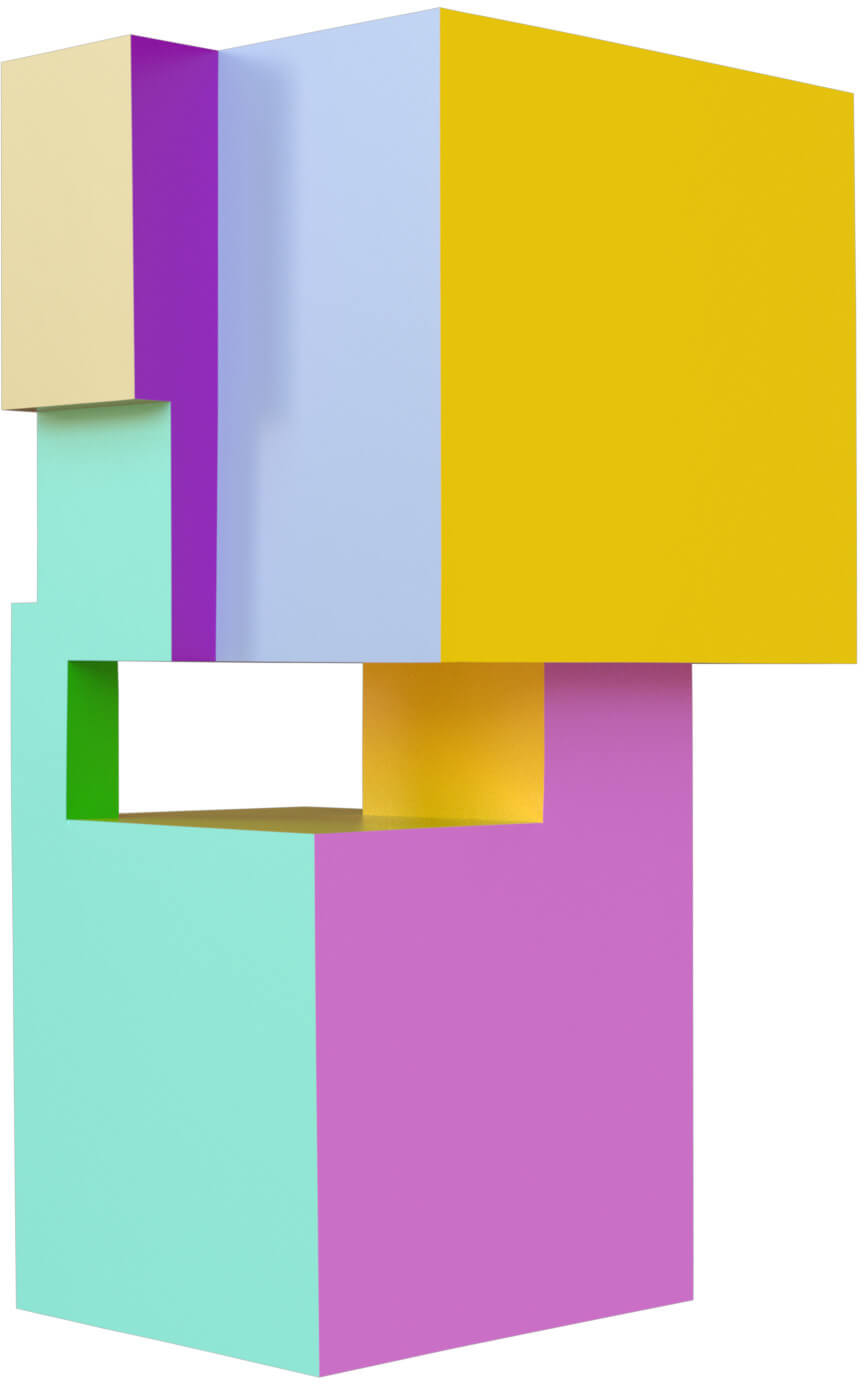}
 		\caption{ }         
 	\end{subfigure}
 	\begin{subfigure}[b]{0.120\textwidth}
 		\includegraphics[width=\textwidth]{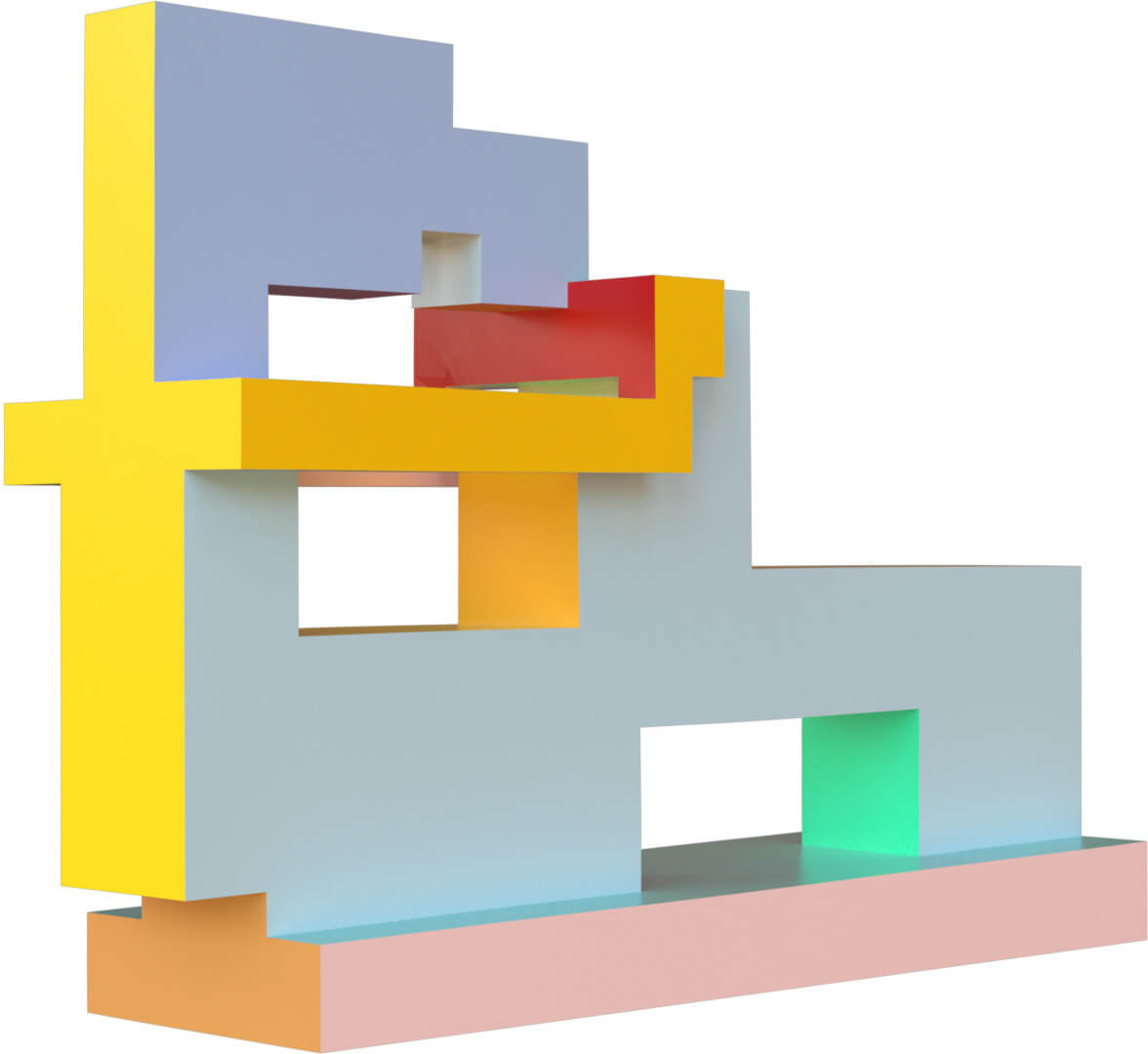}
 		\caption{ }          
 	\end{subfigure}
 	\begin{subfigure}[b]{0.145\textwidth}
 		\includegraphics[width=\textwidth]{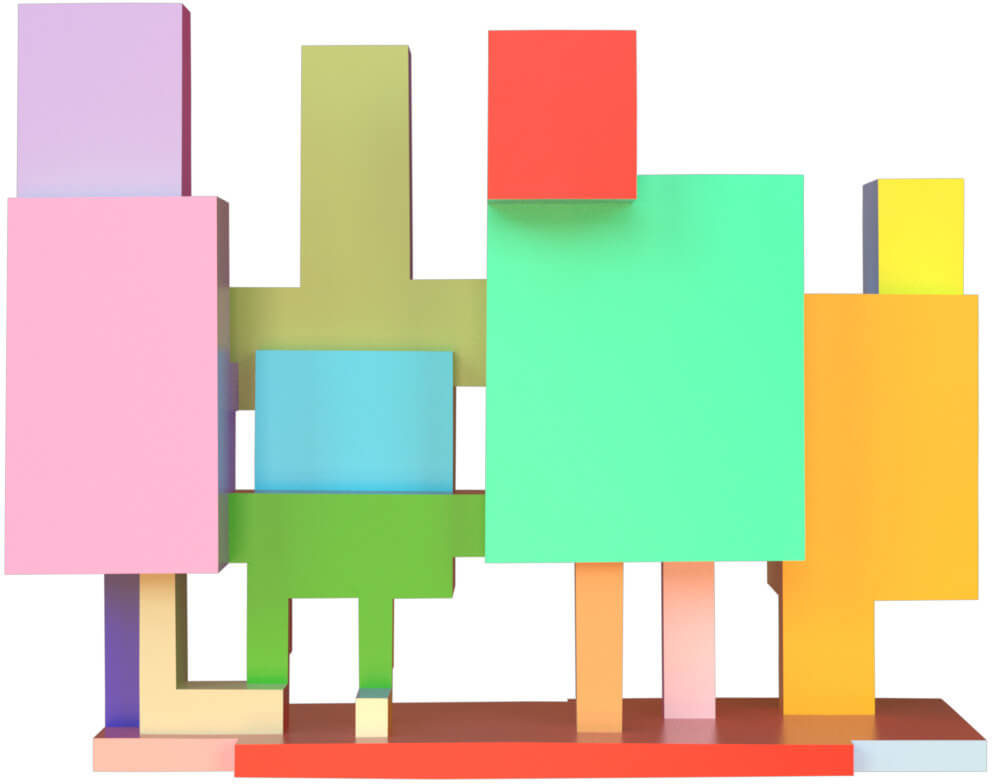}
 		\caption{ }         
 	\end{subfigure}  
 	\caption{The models of high genus and their polycube shapes.}\label{fig:polycubeGenus}
 \end{figure*}

The polycubes are also affected by the genus of the models. The technique we propose can manipulate models with high genus directly, such as shown in Figure \ref{fig:polycubeGenus}.
All of these models are deformed into their polycubes. Therefore our algorithm  is robust to the topologies of meshes.

\begin{figure} [th!] 
	\centering   
	\begin{subfigure}[b]{0.115\textwidth}
		\includegraphics[width=\textwidth]{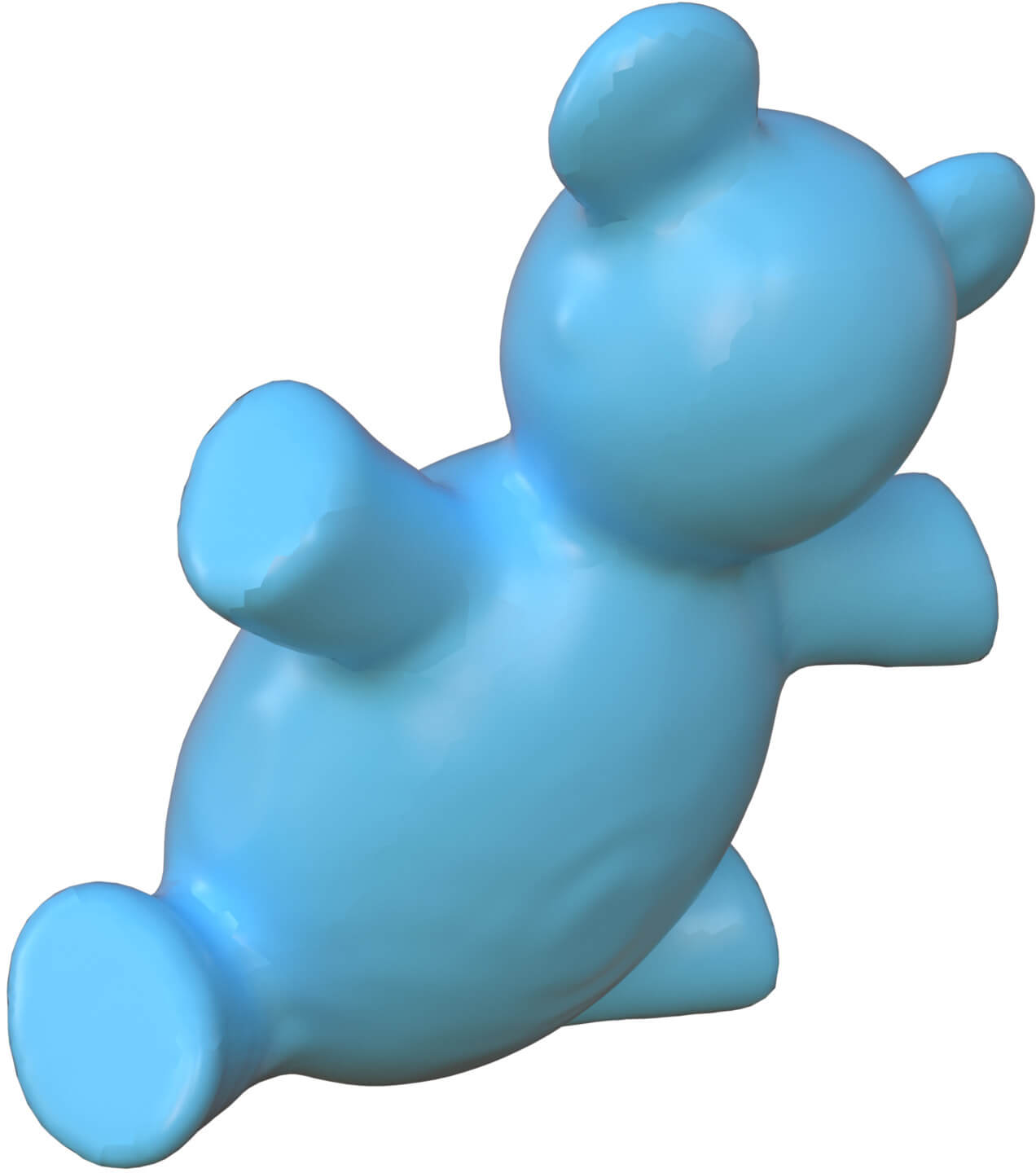}
		\caption*{ }   \label{fig:toy1}
	\end{subfigure}
	\begin{subfigure}[b]{0.115\textwidth}
		\includegraphics[width=\textwidth]{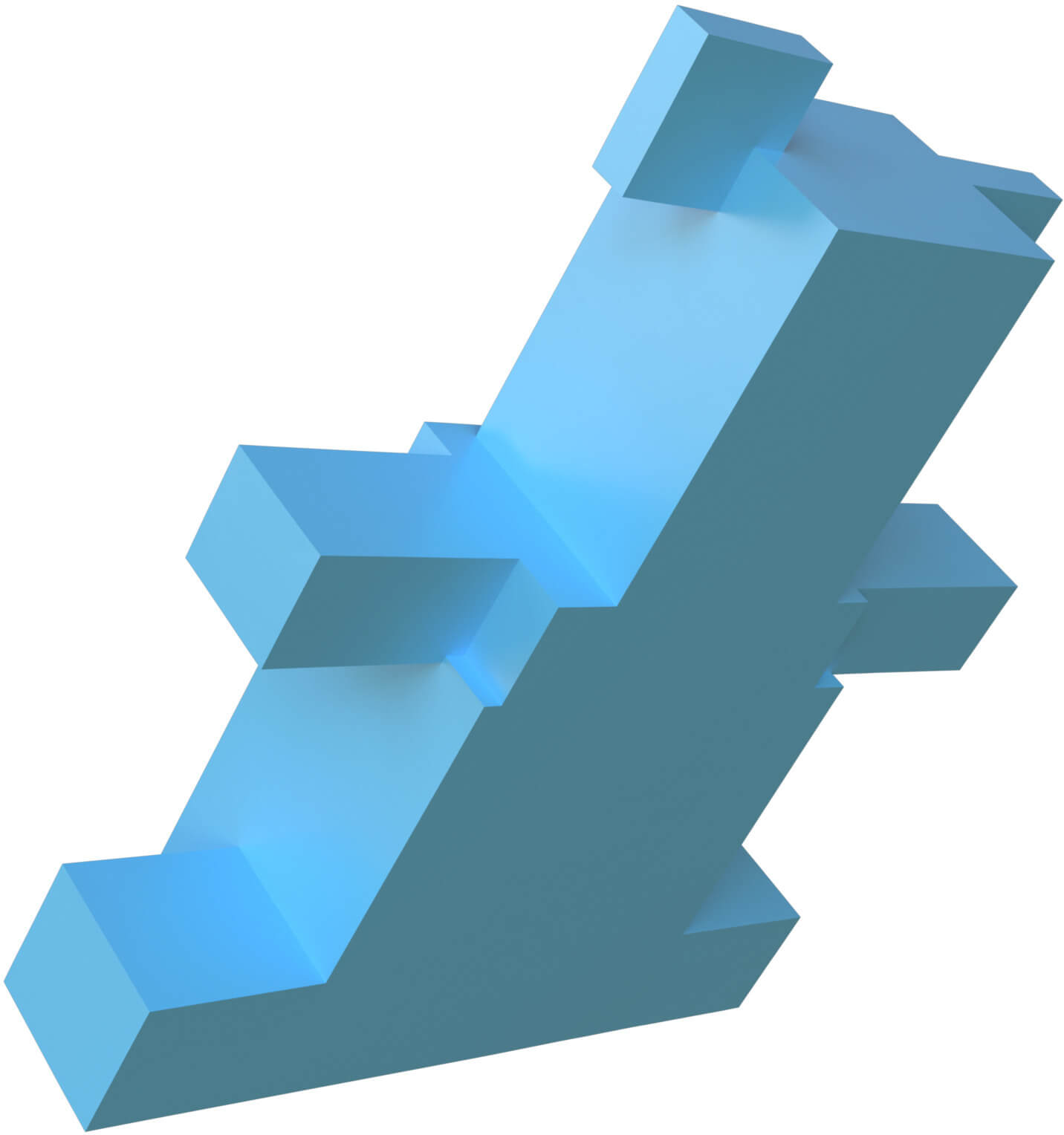}
		\caption*{ }   
	\end{subfigure}   
	\begin{subfigure}[b]{0.115\textwidth}
		\includegraphics[width=\textwidth]{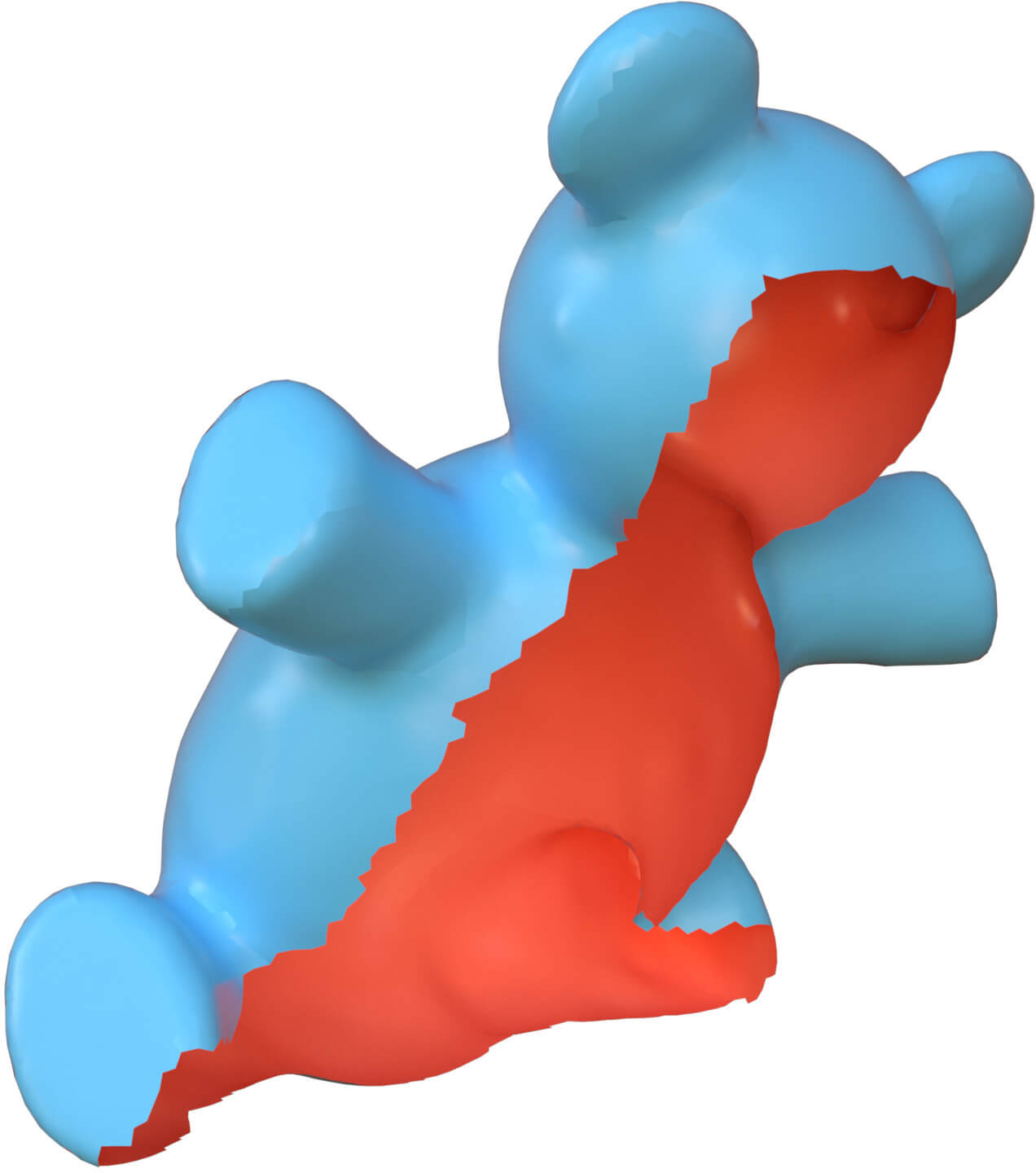}
		\caption*{ }     
	\end{subfigure}
	\begin{subfigure}[b]{0.115\textwidth}
		\includegraphics[width=\textwidth]{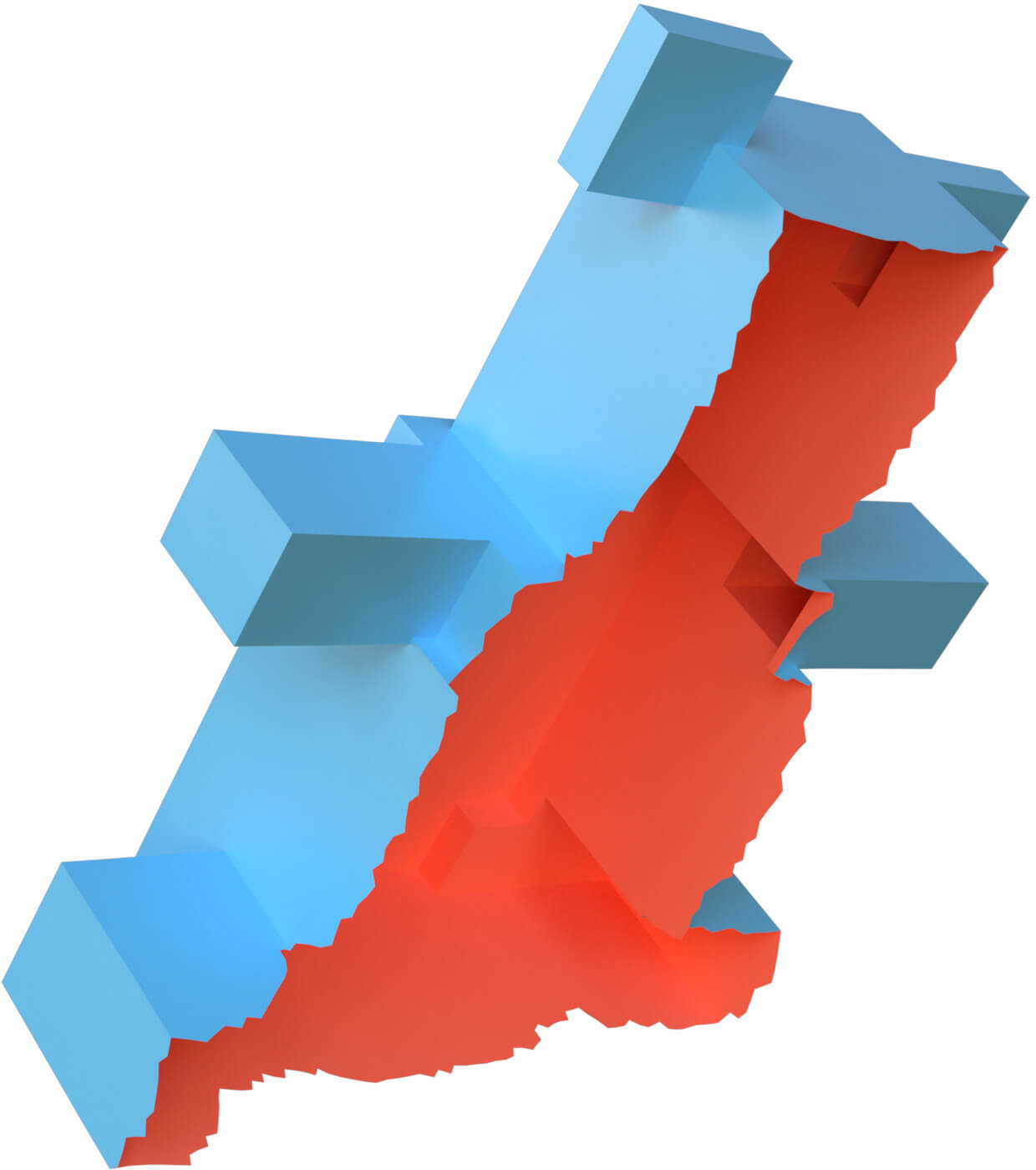}
		\caption*{ }    
	\end{subfigure}  

   	\begin{subfigure}[b]{0.120\textwidth}
   	\includegraphics[width=\textwidth]{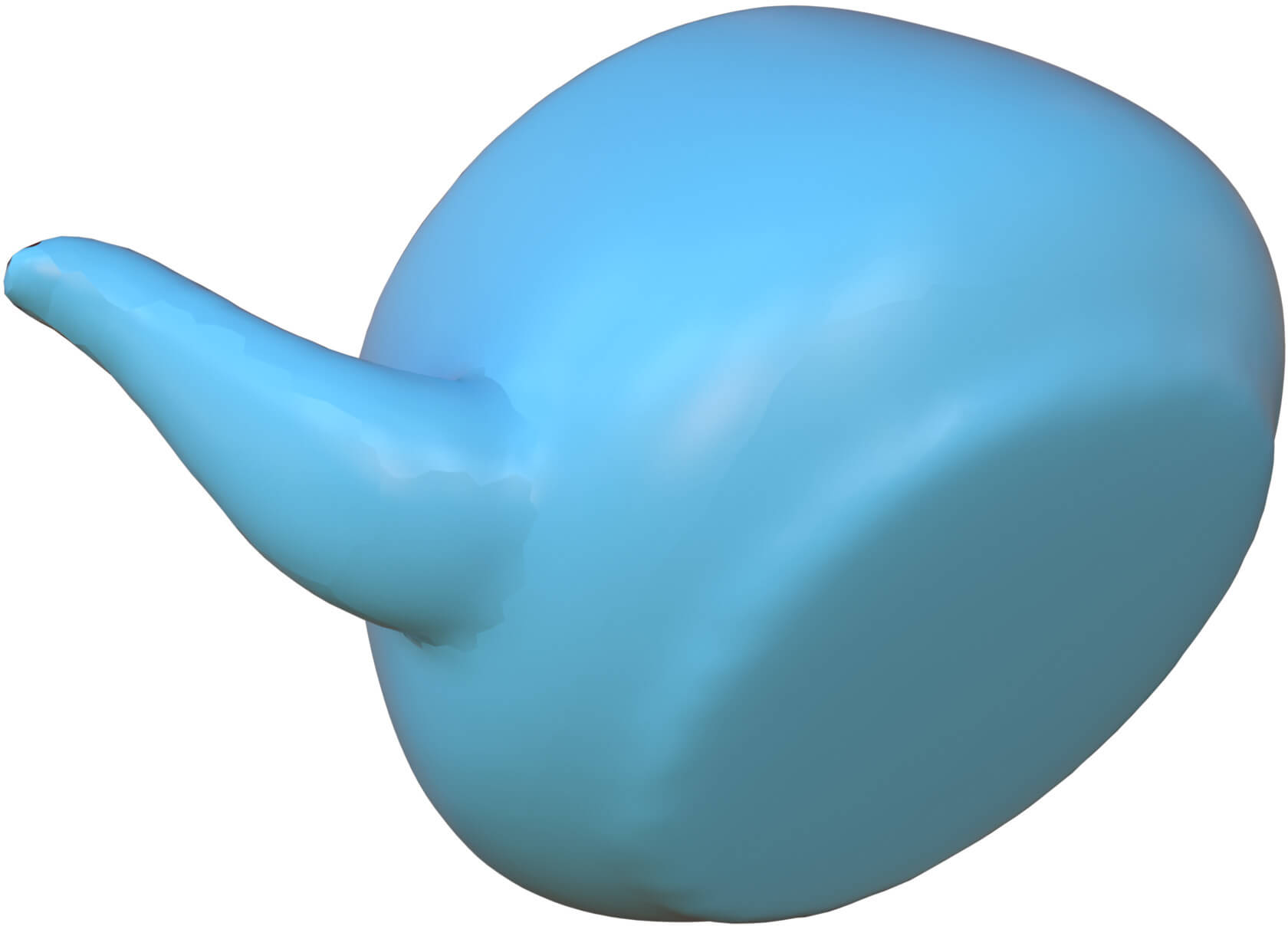}
   	\caption{ }   
   \end{subfigure}
   \begin{subfigure}[b]{0.110\textwidth}
   	\includegraphics[width=\textwidth]{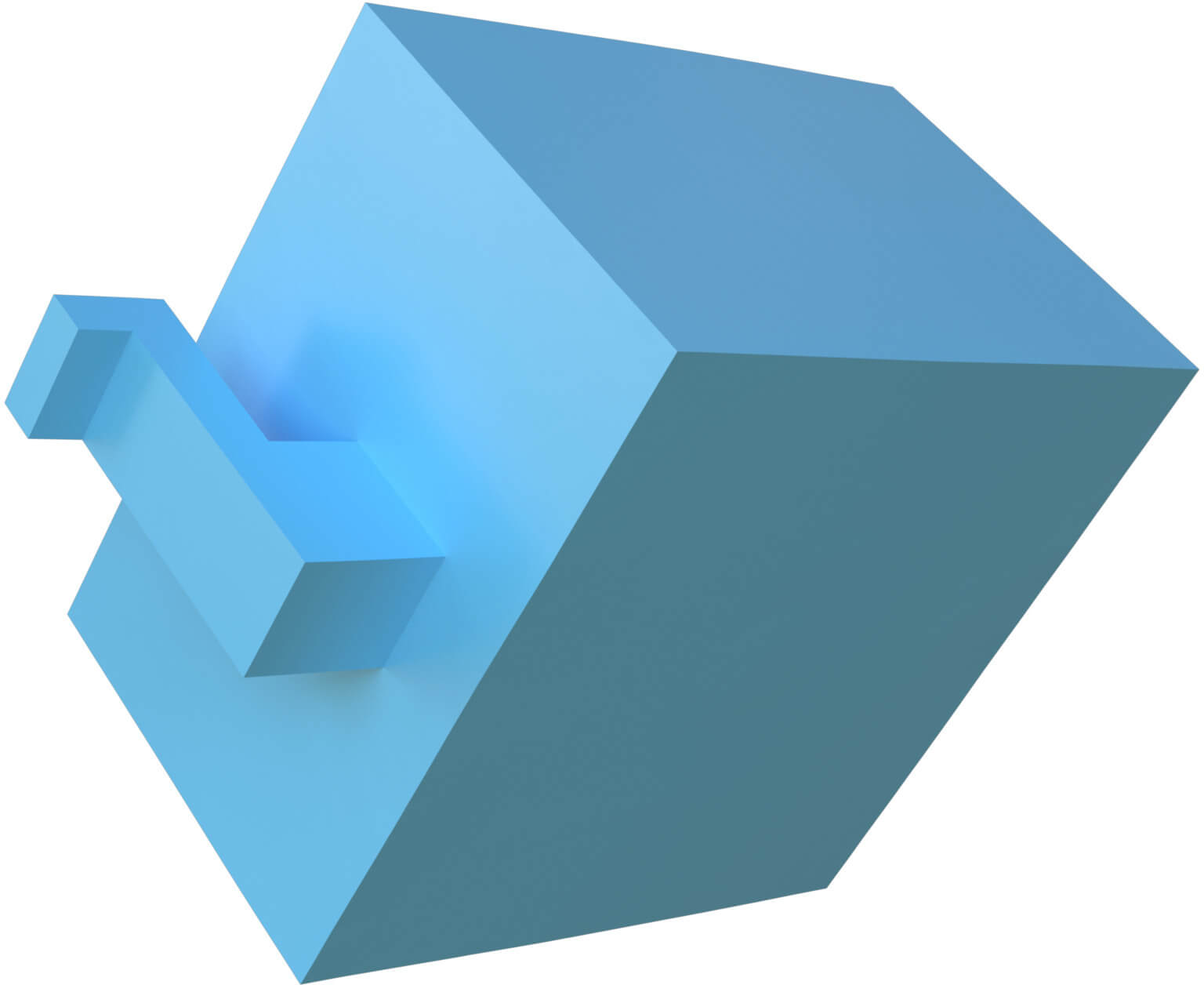}
   	\caption{ }   
   \end{subfigure}   
   \begin{subfigure}[b]{0.120\textwidth}
   	\includegraphics[width=\textwidth]{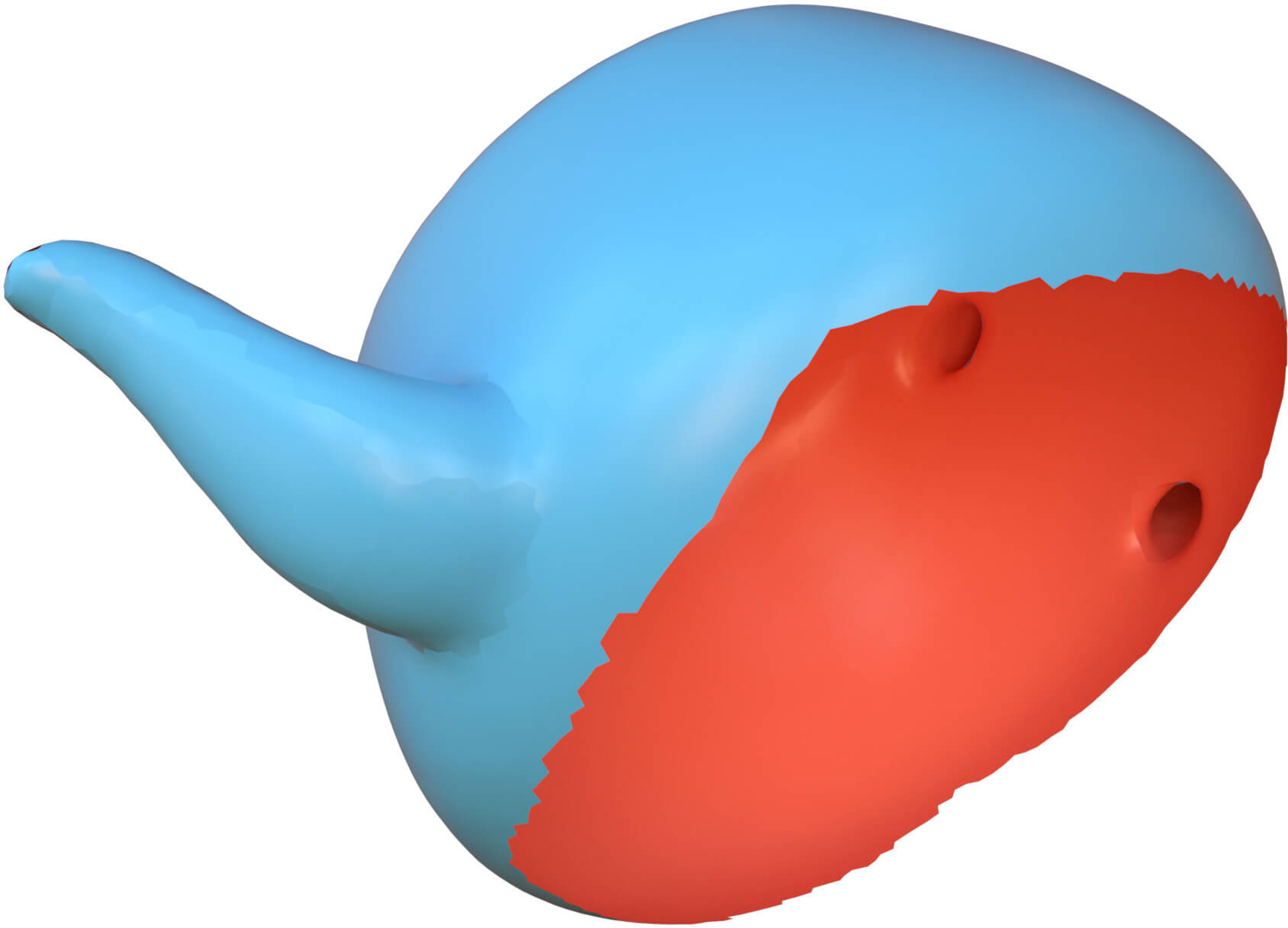}
   	\caption{ }     
   \end{subfigure}
   \begin{subfigure}[b]{0.110\textwidth}
   	\includegraphics[width=\textwidth]{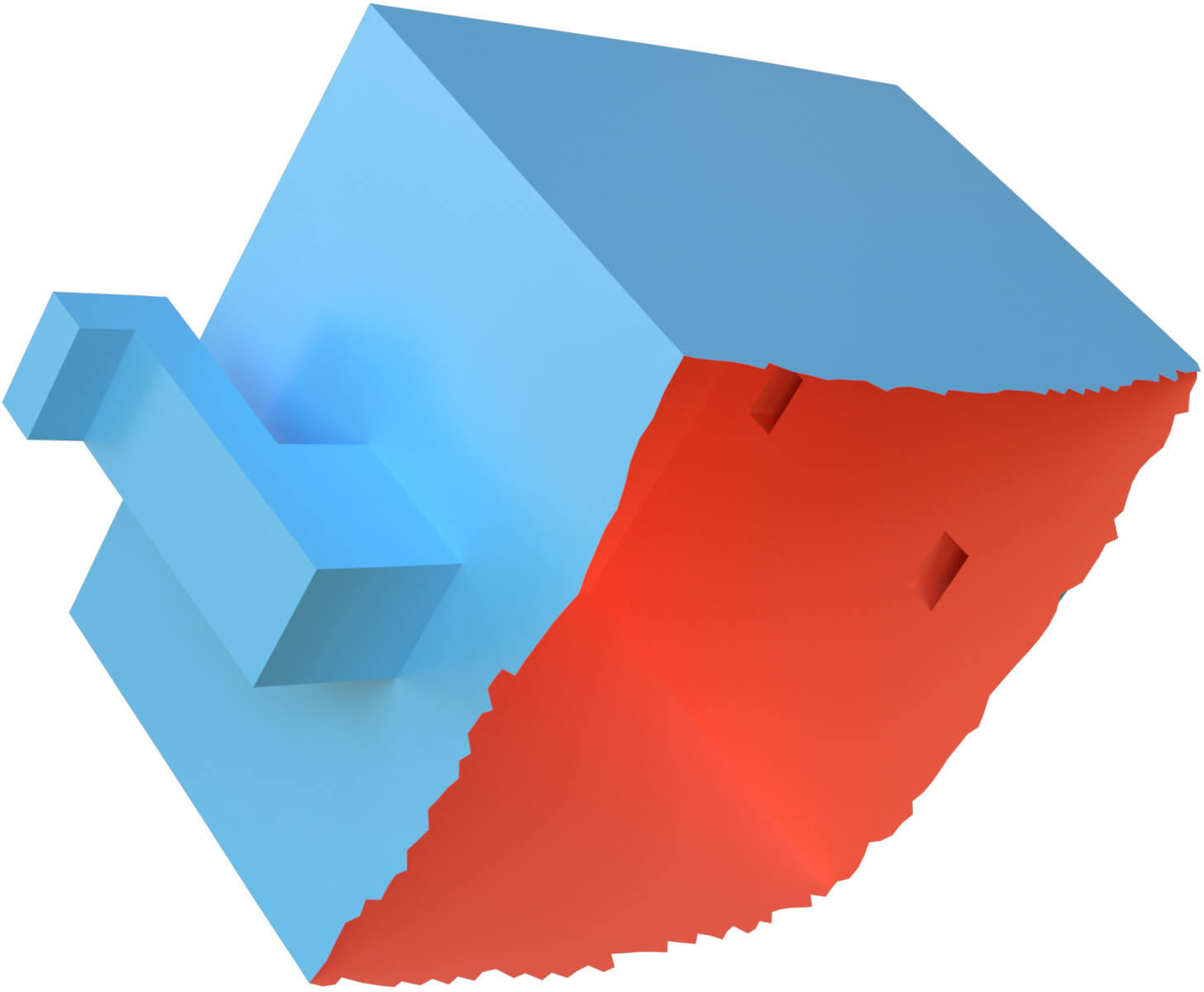}
   	\caption{ }    
   \end{subfigure}     
	\caption{The models with  or without boundary, and their polycube shapes.}\label{fig:polyBoundary}
\end{figure}

Our method is insensitive to the boundaries. 
The figures in \ref{fig:polyBoundary} exhibit some meshes with boundaries and their polycubes. 
As there are implicit  constraints of polycube topology, the edges on the boundary can not be deformed into  the straight lines. 

\begin{figure} [th!] 
	\centering   
	\begin{subfigure}[b]{0.23\textwidth}
		\includegraphics[width=\textwidth]{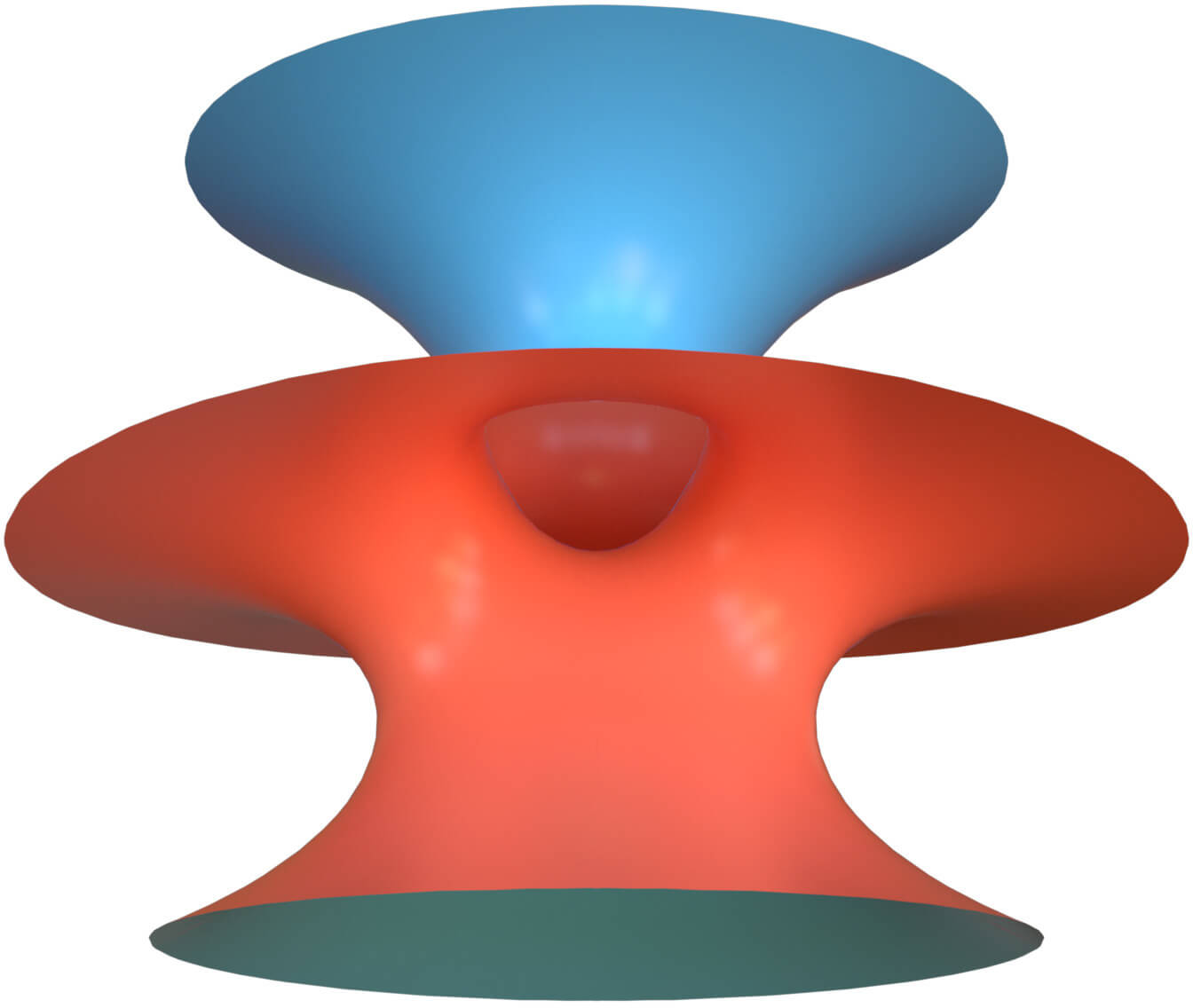}
		\caption{costa}   
	\end{subfigure}
	\begin{subfigure}[b]{0.23\textwidth}
		\includegraphics[width=\textwidth]{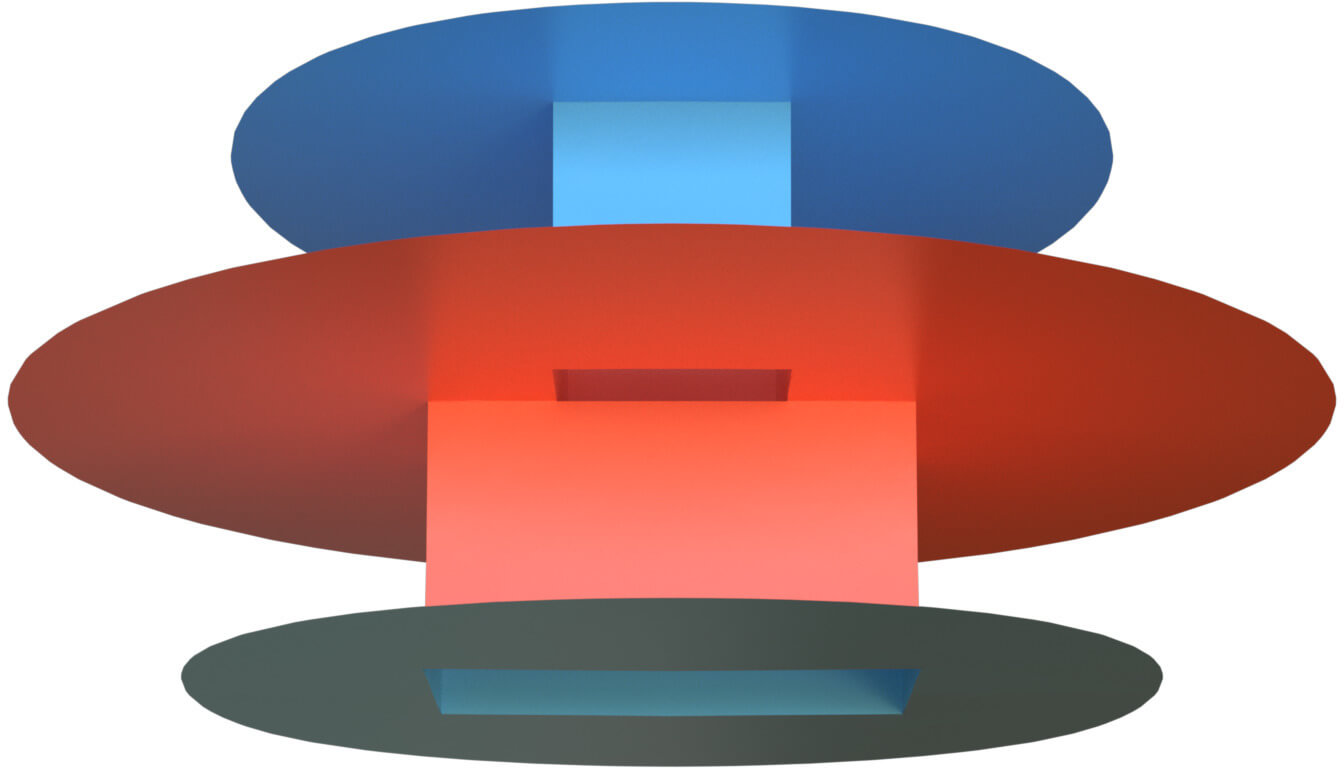}
		\caption{polycube}   
	\end{subfigure} 
	\caption{The non-orientable surface "costa" and its polycube.}\label{fig:polycubeCosta}
\end{figure}

Even on non-orientable meshes, our algorithm can also deform them successfully. In Figure \ref{fig:polycubeCosta}, the well-known "costa" surface mesh is deformed into a polycube.

After obtaining the segmentation and the polycube topology, the PolyCut \cite{Livesu:2013:PolyCut} method also presents a deformation algorithm to obtain polycube geometry. Their deformation is based on the vertex normal rotations, our rotations are applied on face normals. Therefore their method and the algorithm in \cite{Fu2016PC} can not process the  non-orientable surfaces, and the meshes with boundaries.

\begin{figure} [th!] 
	\centering   
	\begin{subfigure}[b]{0.115\textwidth}
		\includegraphics[width=\textwidth]{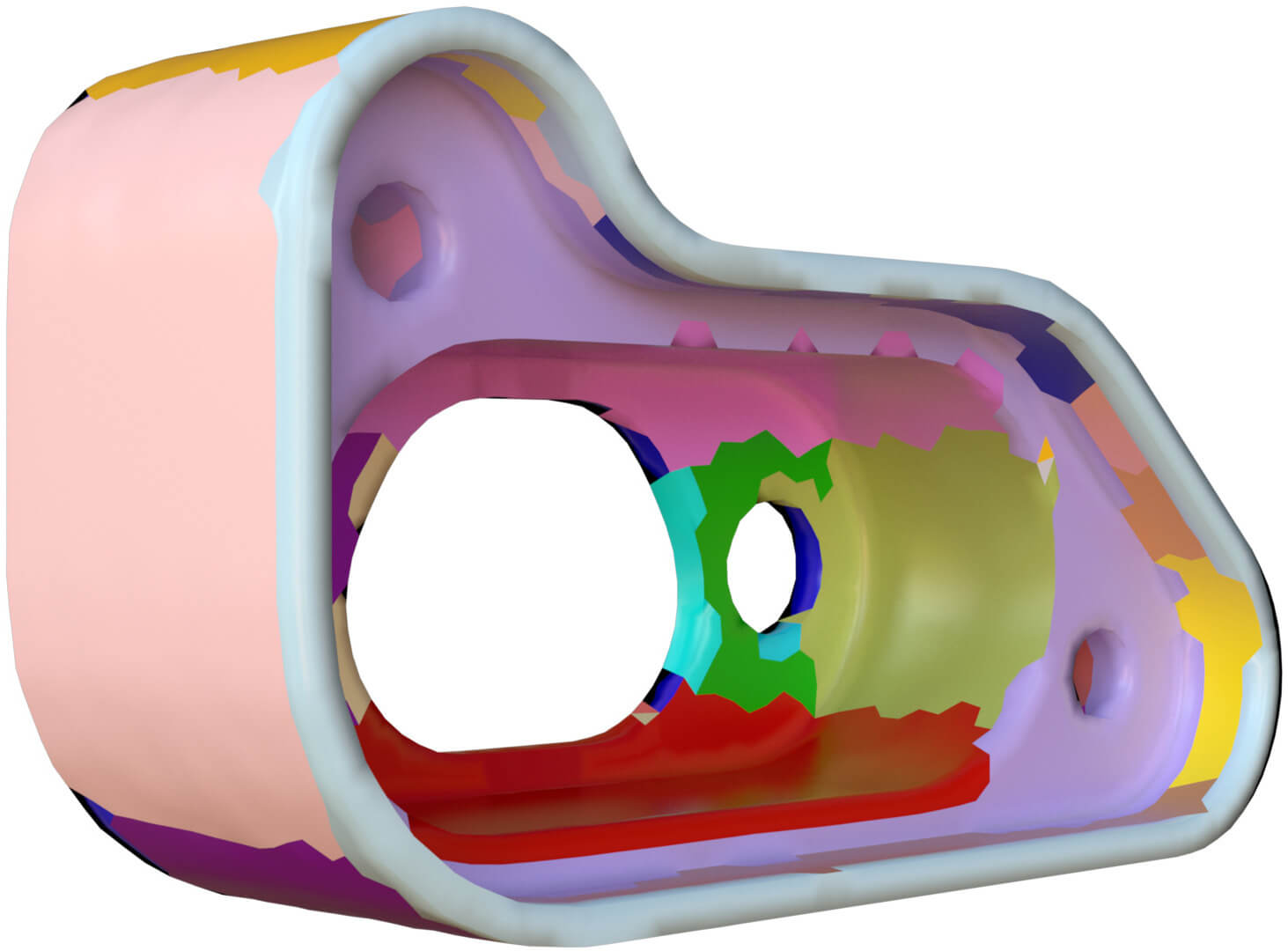}
		\caption*{ }     
	\end{subfigure}
	\begin{subfigure}[b]{0.115\textwidth}
		\includegraphics[width=\textwidth]{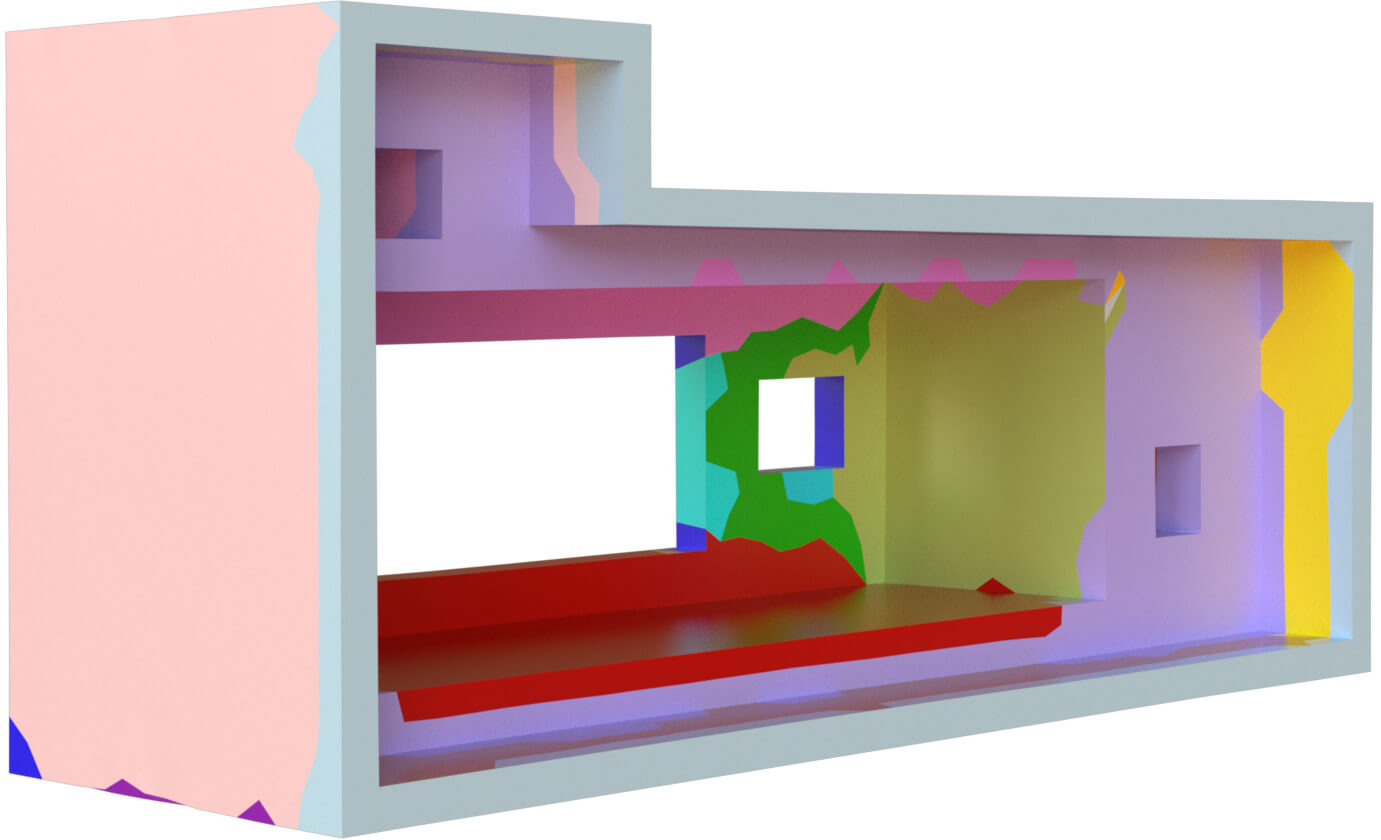}
		\caption*{ }
	\end{subfigure}
	\begin{subfigure}[b]{0.115\textwidth}
		\includegraphics[width=\textwidth]{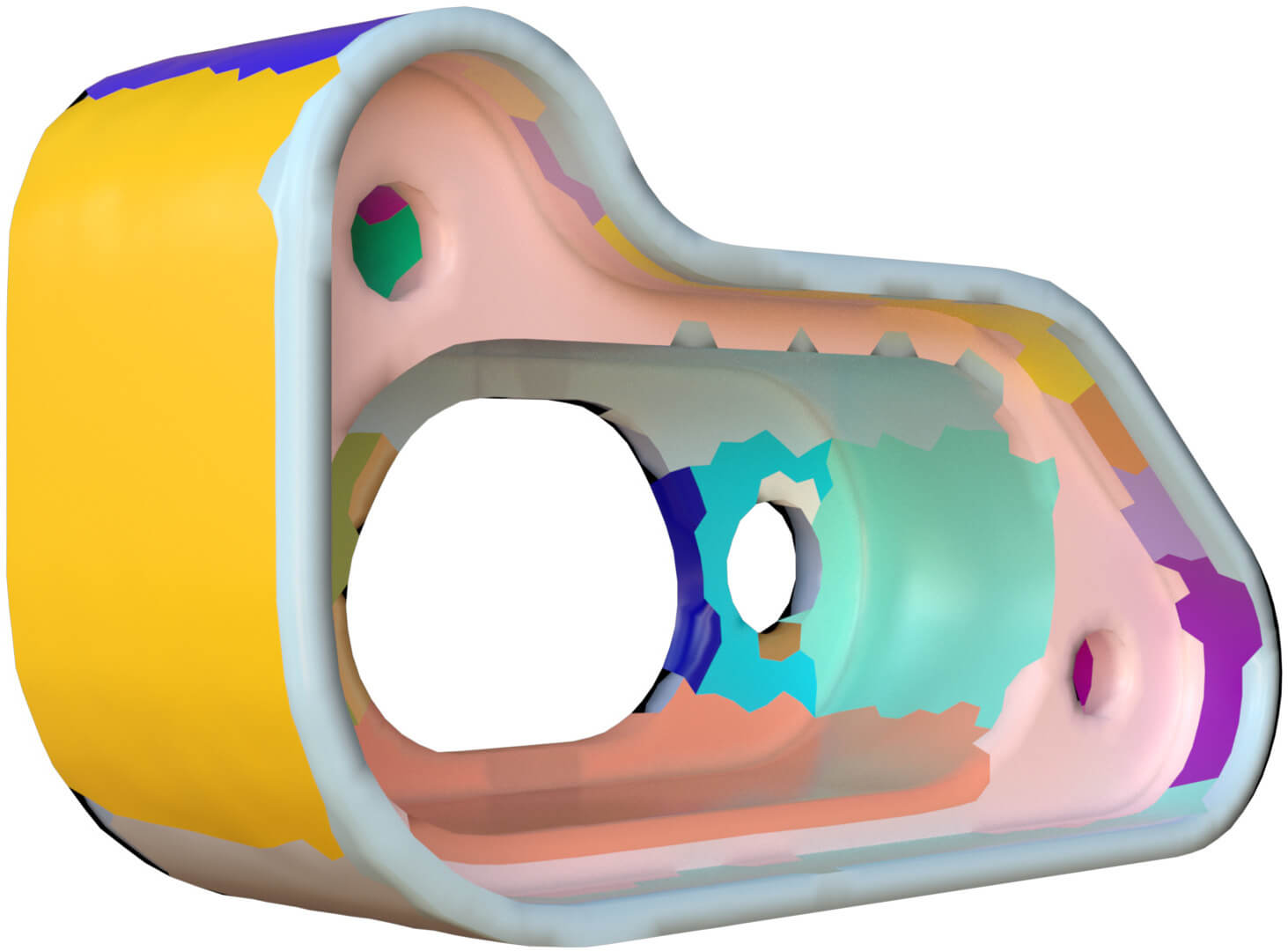}
		\caption*{ }
	\end{subfigure}   
	\begin{subfigure}[b]{0.115\textwidth}
		\includegraphics[width=\textwidth]{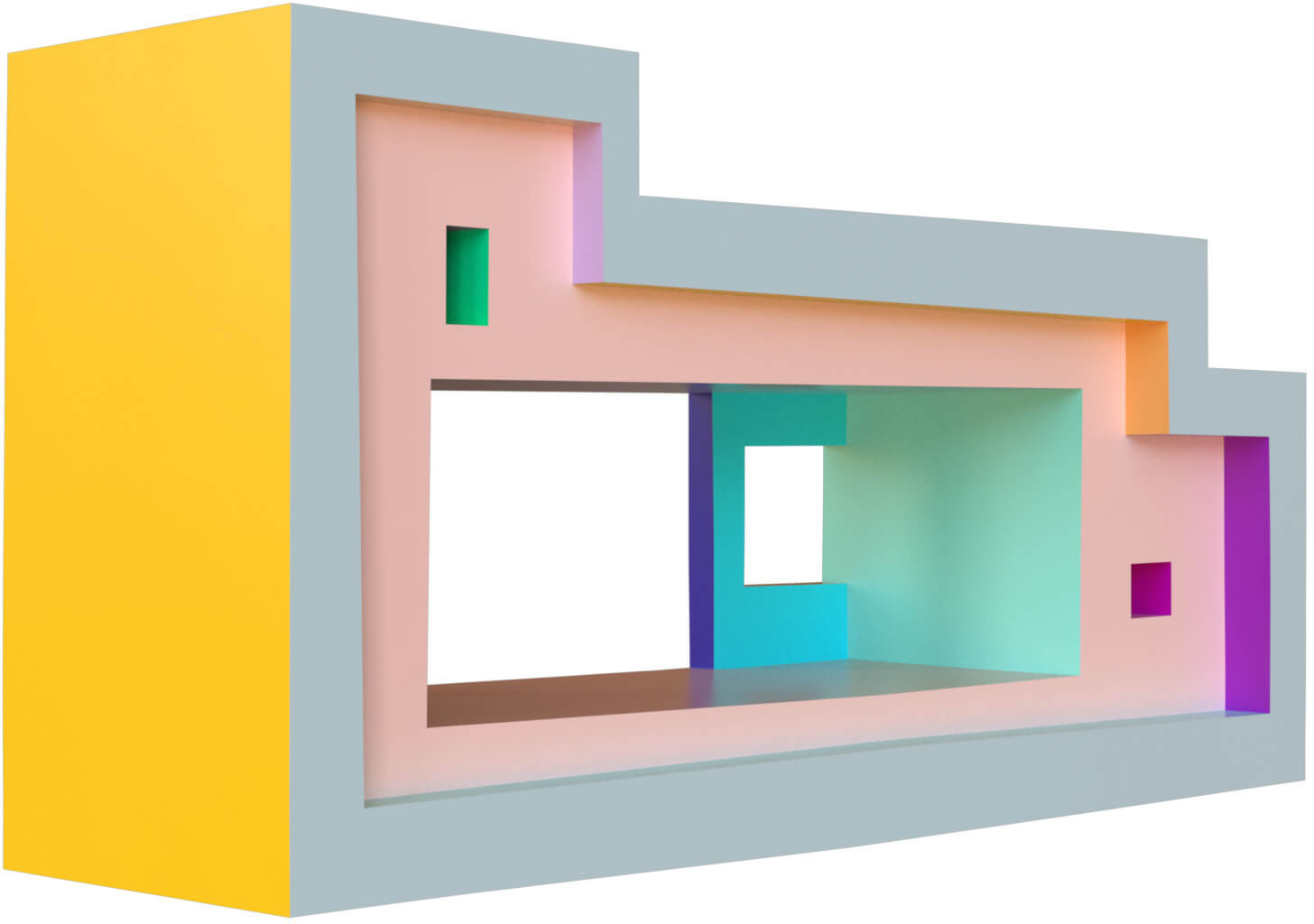}
		\caption*{ }
	\end{subfigure}  

    \begin{subfigure}[b]{0.115\textwidth}
    	\includegraphics[width=\textwidth]{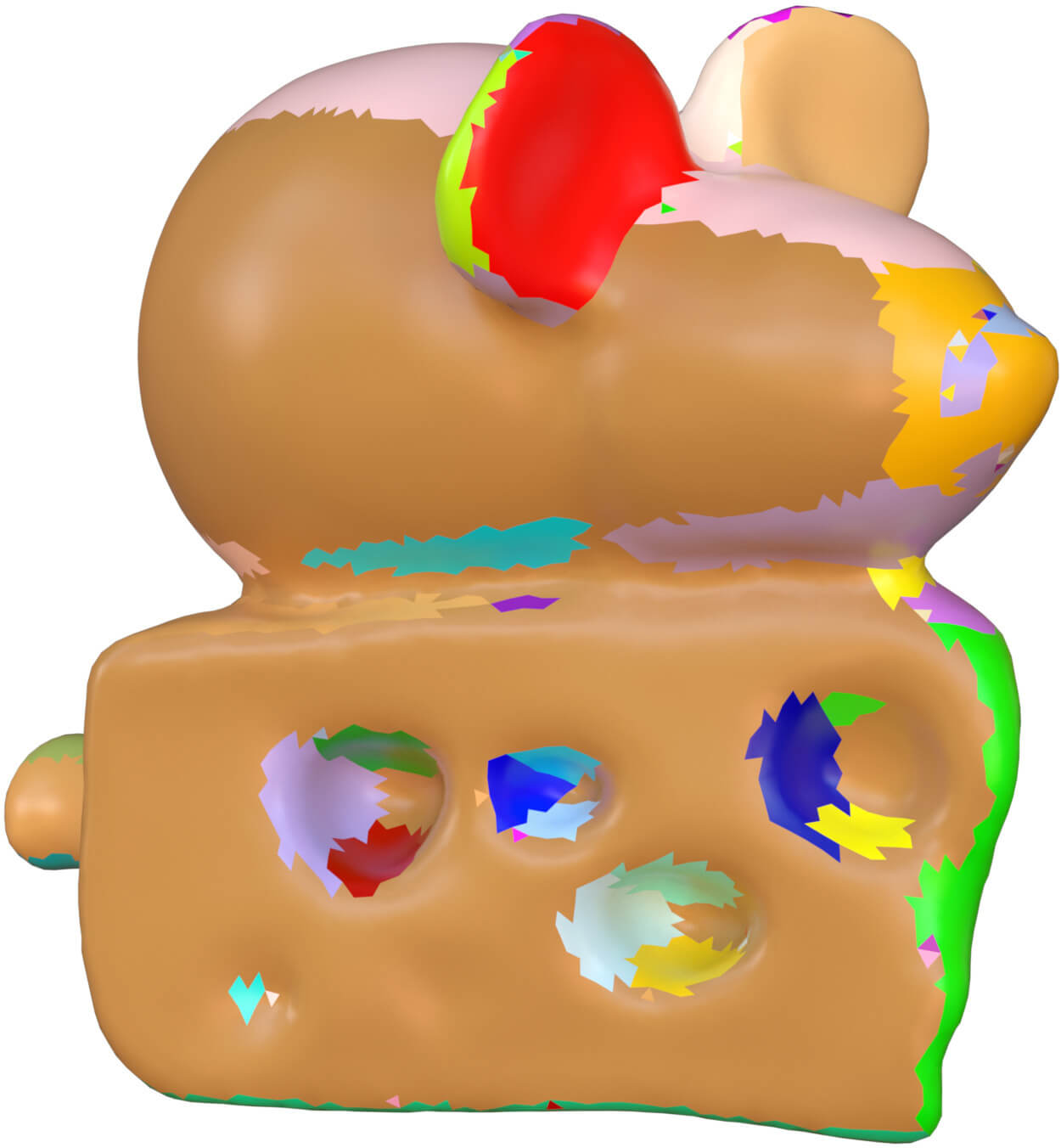}
    	\caption{ }    \label{fig:WithDefects}
    \end{subfigure}
    \begin{subfigure}[b]{0.115\textwidth}
    	\includegraphics[width=\textwidth]{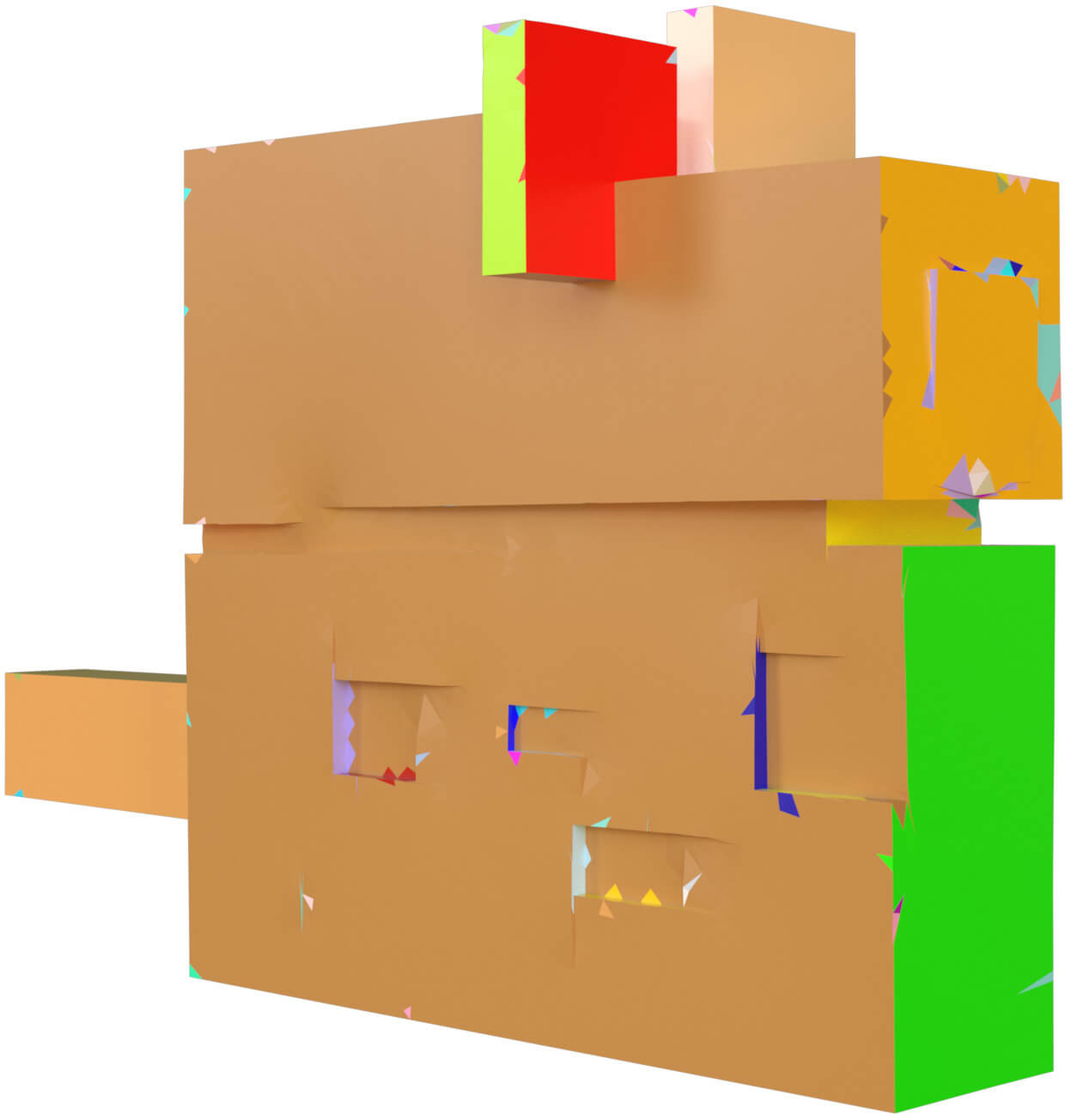}
    	\caption{ }     \label{fig:WithDefectspolycube}
    \end{subfigure}
    \begin{subfigure}[b]{0.115\textwidth}
    	\includegraphics[width=\textwidth]{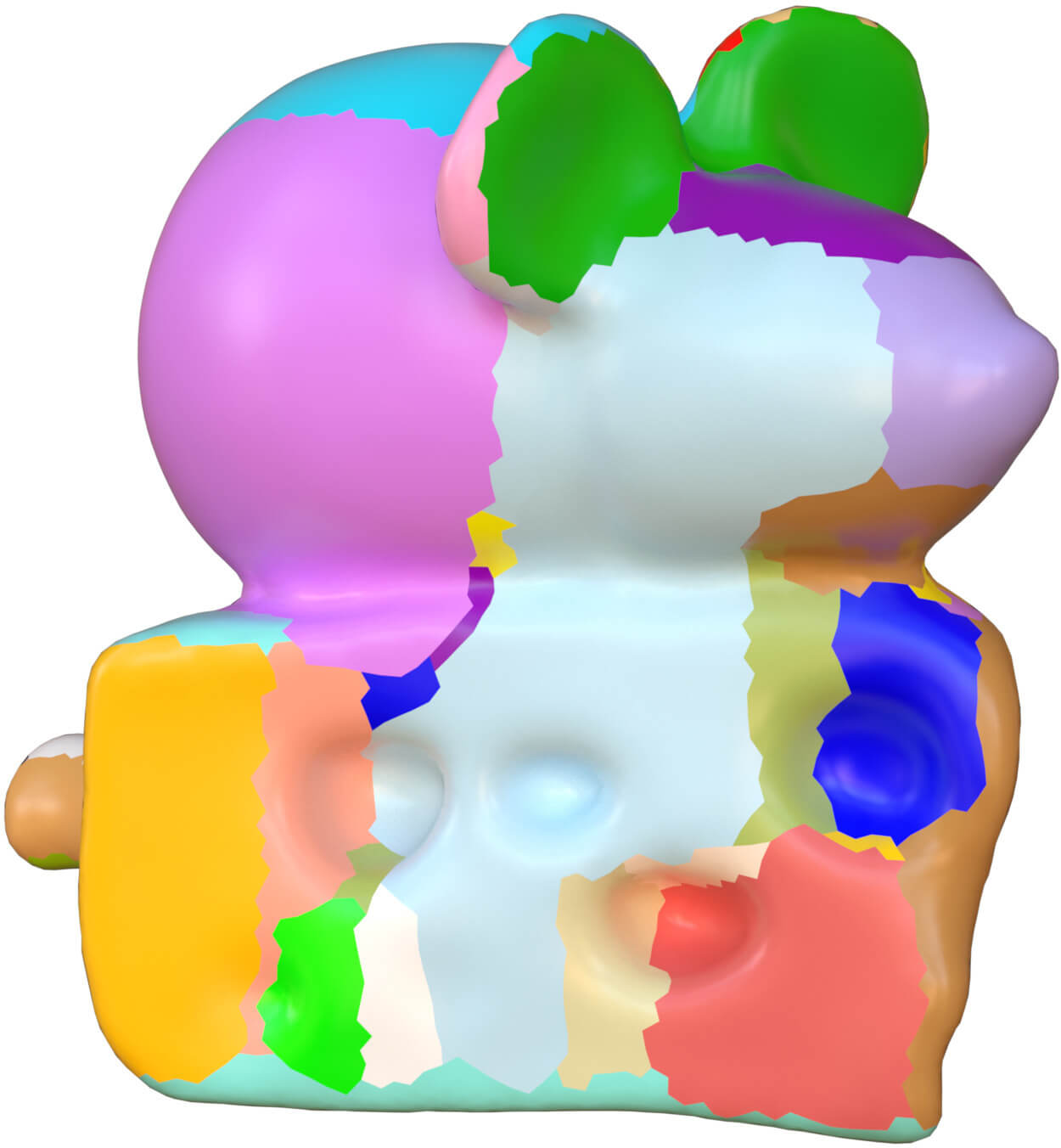}
    	\caption{ }   \label{fig:WithoutDefects}
    \end{subfigure}   
    \begin{subfigure}[b]{0.115\textwidth}
    	\includegraphics[width=\textwidth]{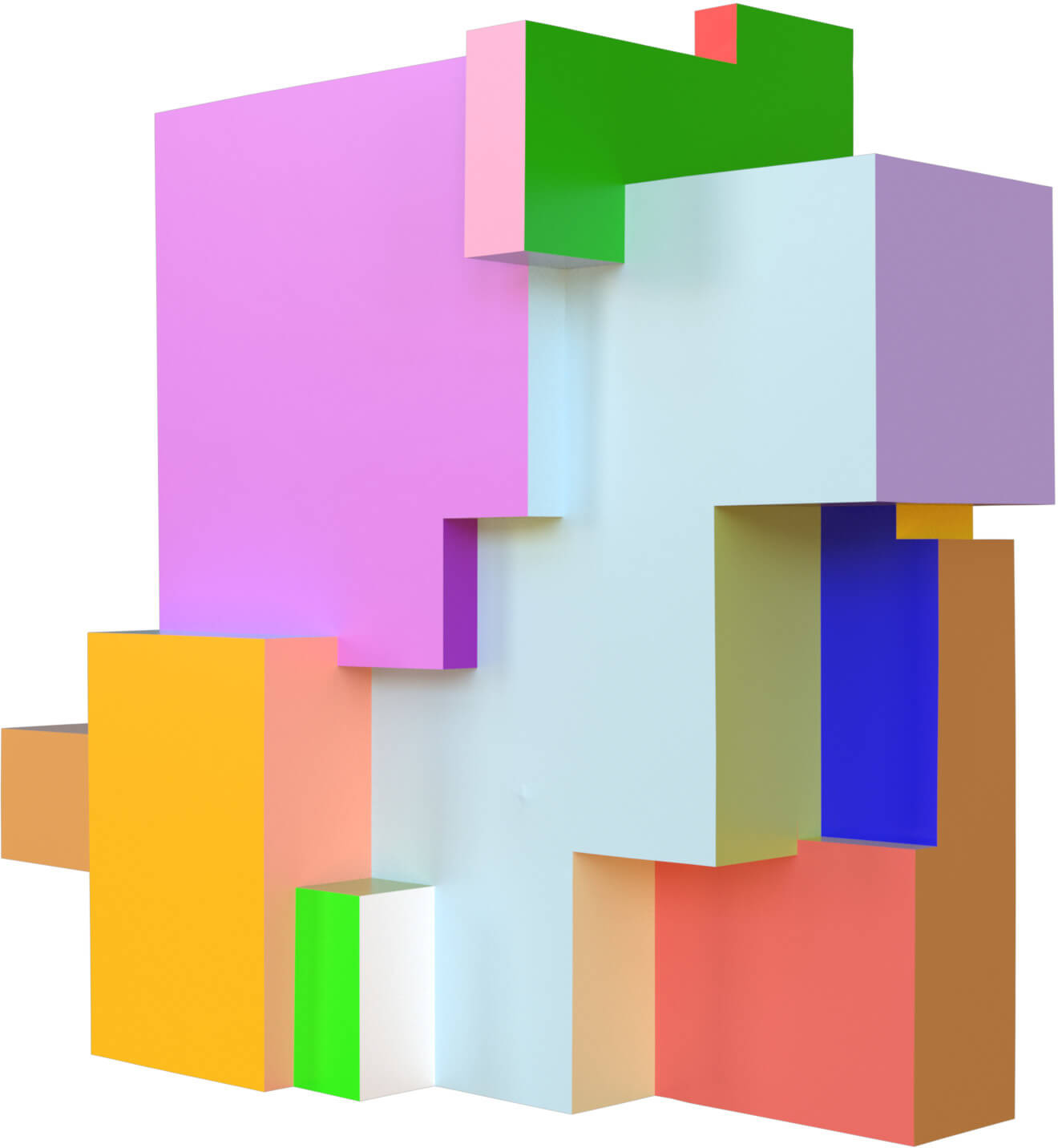}
    	\caption{ }     \label{fig:WithoutDefectspolycube}
    \end{subfigure}   
	\caption{The polycube deformations with and without polycube topological defects.}\label{fig:Defects}
\end{figure}

The algorithm we present is also robust to the polycube topological defects. When the polycube topology is not valid. 
Our method can still process the model and output a polycube-like shape with the definitely same polycube topological defects.
In Figure \ref{fig:Defects}, the  two models with and without topological defects are shown in Figure \ref{fig:WithDefects} and \ref{fig:WithoutDefects}.
Their corresponding polycube deformation results are displayed in Figure \ref{fig:WithDefectspolycube} and \ref{fig:WithoutDefectspolycube}.

\begin{figure} [th!] 
    	\centering   
    	\begin{subfigure}[b]{0.145\textwidth}
    		\includegraphics[width=\textwidth]{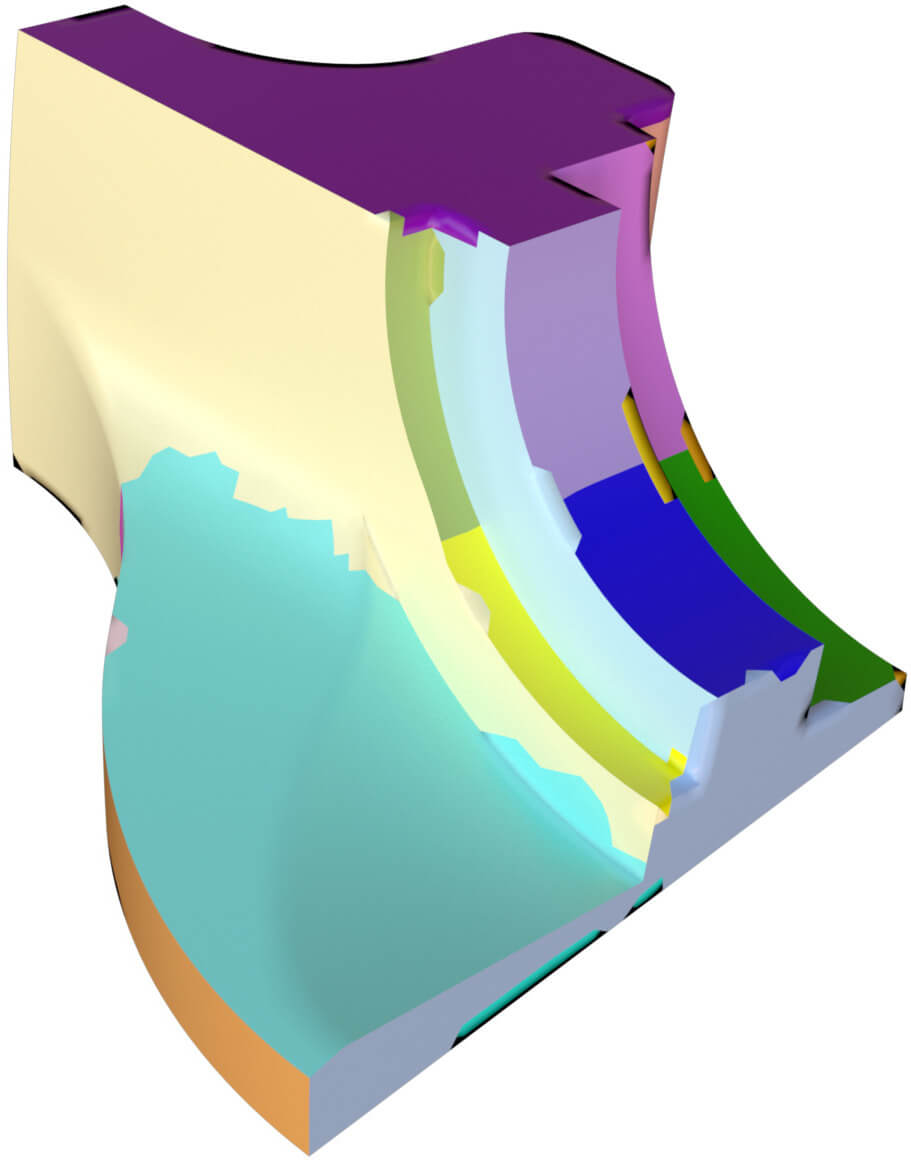}
    		\caption{fandisk}   
    	\end{subfigure}
    	\begin{subfigure}[b]{0.16\textwidth}
    		\includegraphics[width=\textwidth]{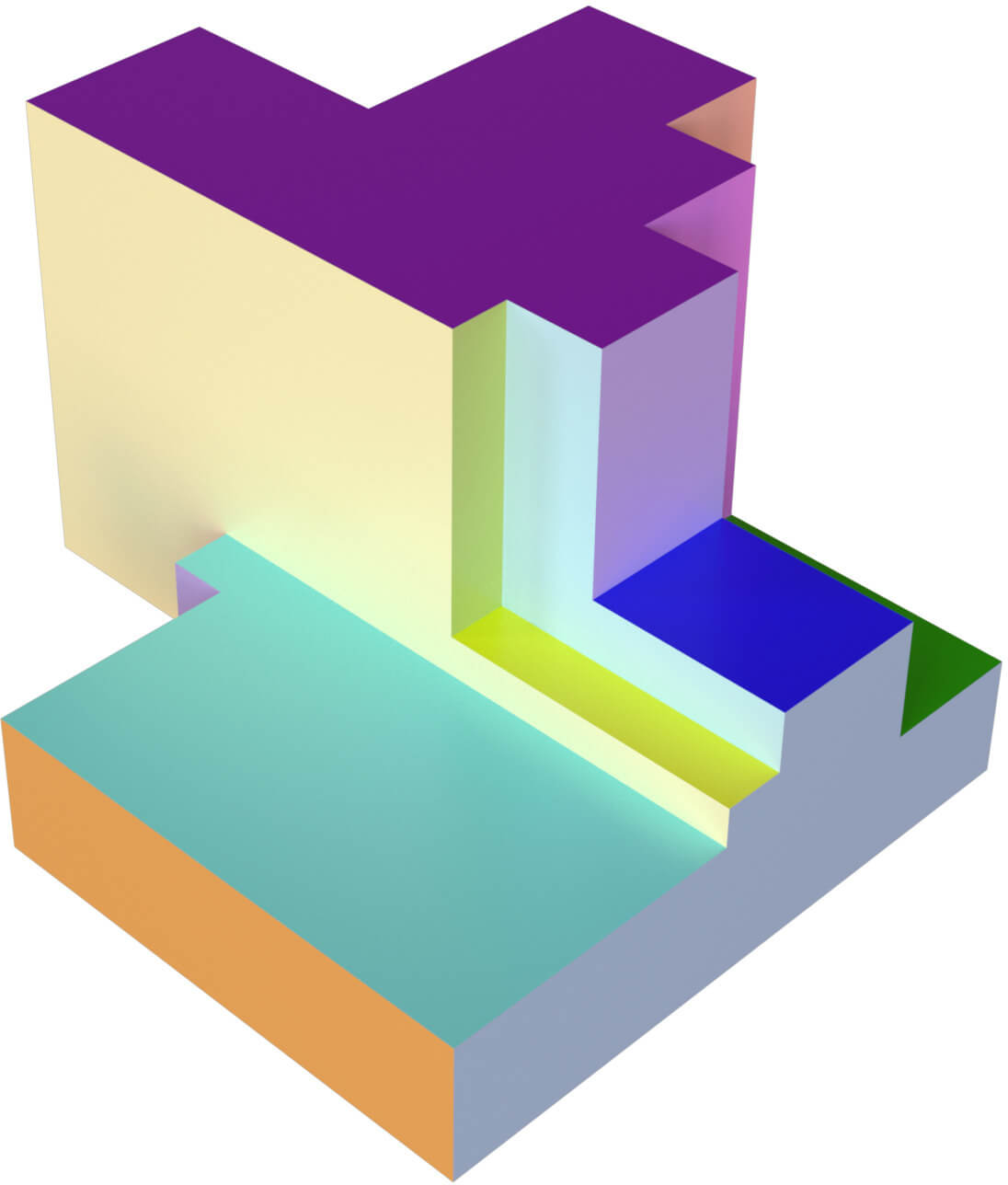}
    		\caption{\cite{Fu2016PC} }   
    	\end{subfigure}   
    	\begin{subfigure}[b]{0.16\textwidth}
    		\includegraphics[width=\textwidth]{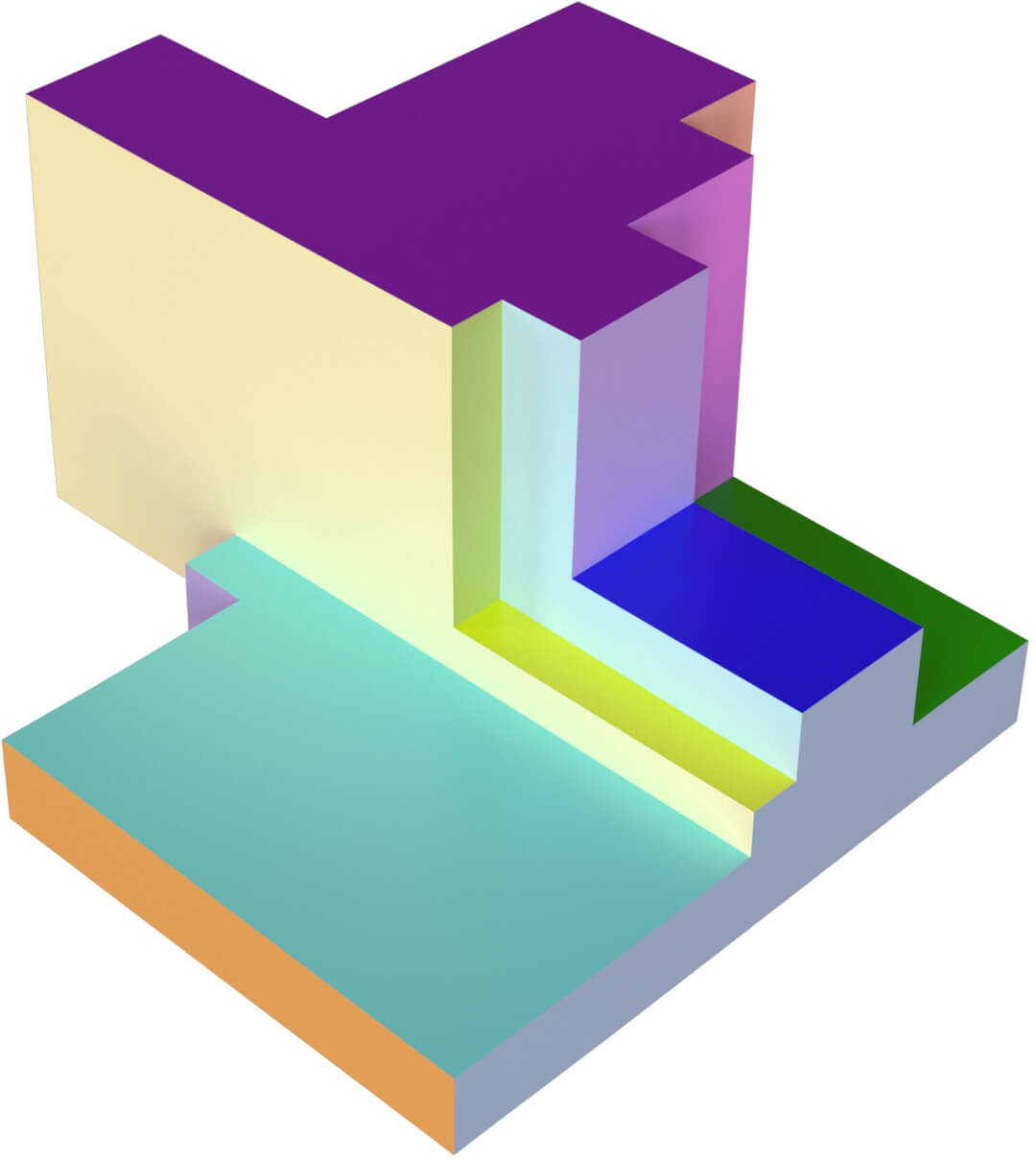}
    		\caption{ours}     
    	\end{subfigure}    
    	\caption{The polycube comparison of \cite{Fu2016PC}  and ours.}\label{fig:comparison}
    \end{figure}

\begin{figure} [th!] 
	\centering  
	\begin{subfigure}[b]{0.205\textwidth}
		\includegraphics[width=\textwidth]{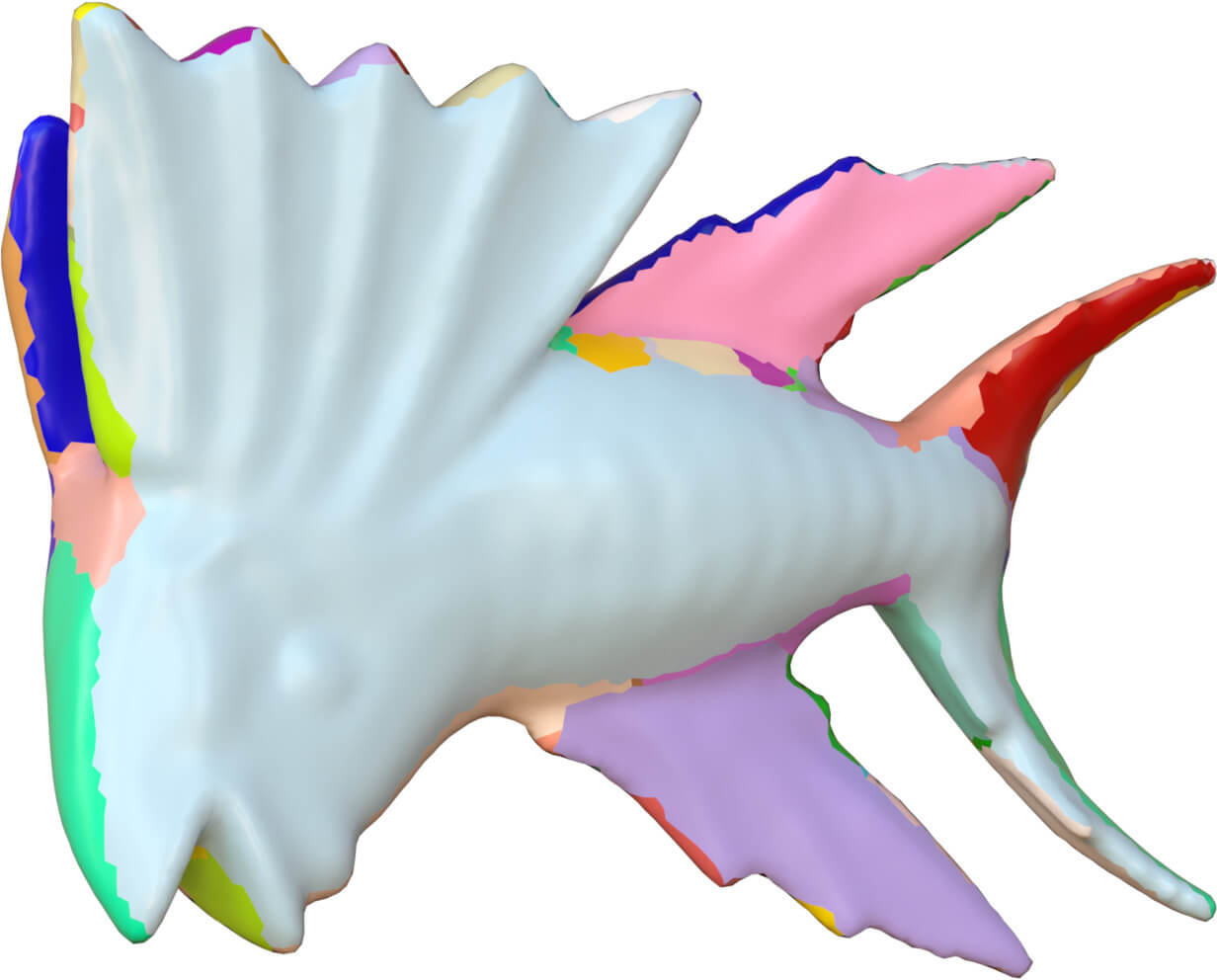}
		\caption{woodenfish}   
	\end{subfigure}
	\begin{subfigure}[b]{0.130\textwidth}
		\includegraphics[width=\textwidth]{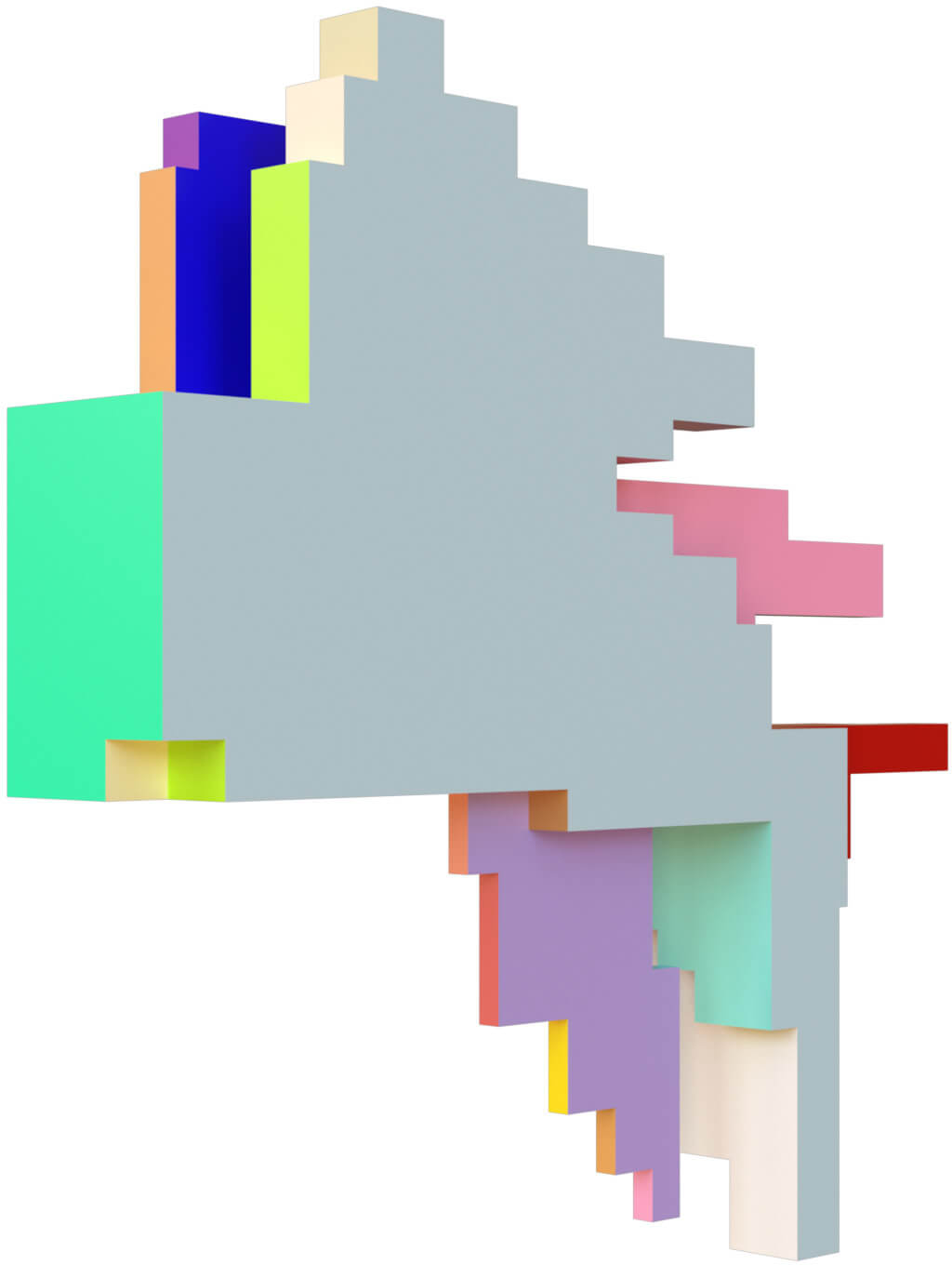}
		\caption{\cite{Fu2016PC} }   
	\end{subfigure}   
	\begin{subfigure}[b]{0.130\textwidth}
		\includegraphics[width=\textwidth]{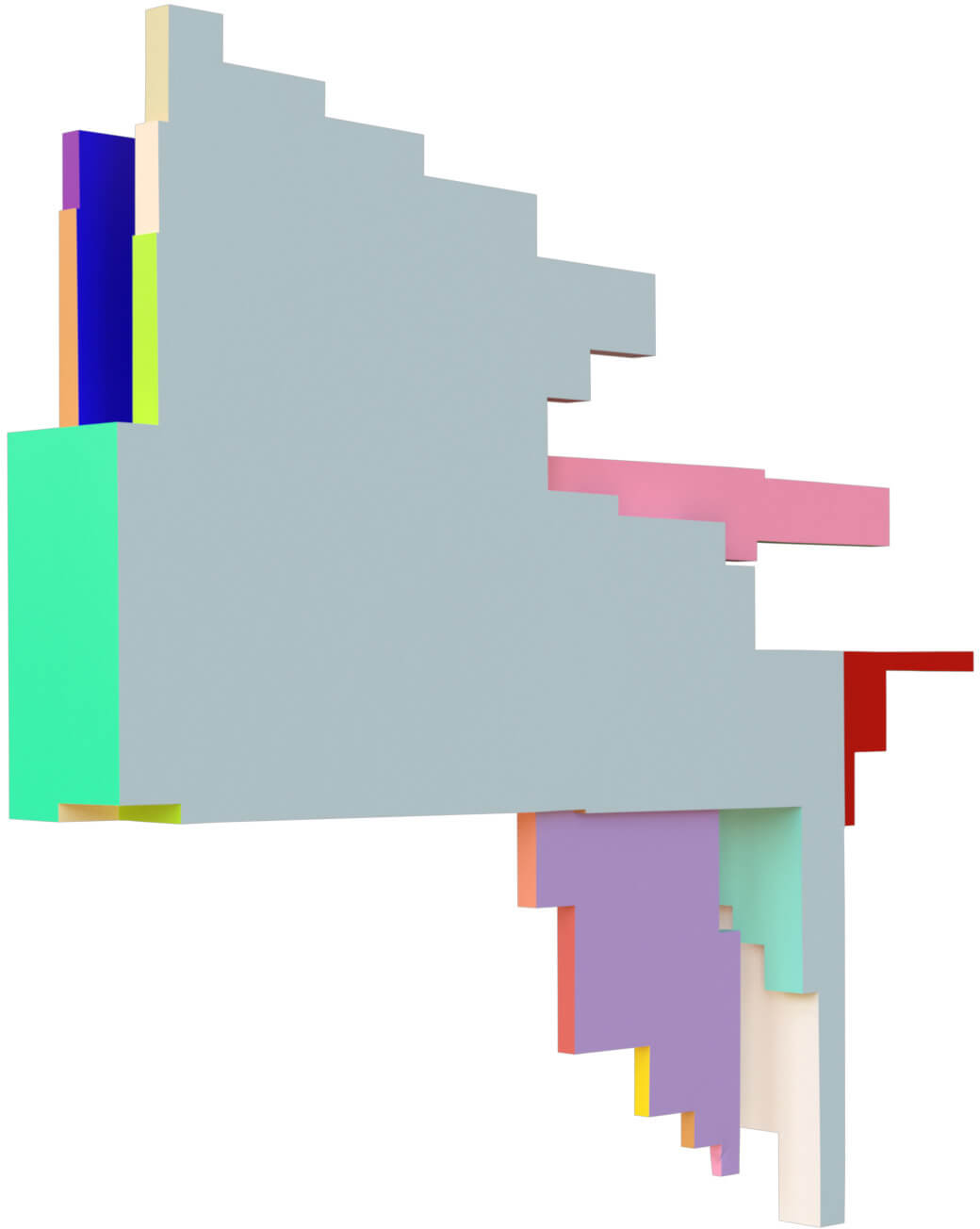}
		\caption{ours}     
	\end{subfigure}    
	\caption{The polycube comparison of \cite{Fu2016PC}  and ours.}\label{fig:comparison2}
\end{figure}

\textbf{Comparison}.
Many recent algorithms \cite{huang2014ell} can not guarantee to obtain a perfect polycube shape without the topological defects except the only one method proposed in \cite{Fu2016PC}.
In this part, we compare the methods between ours and theirs \cite{Fu2016PC}. We use the same models, the same segmentation charts and the same polycube topologies of the data in \cite{Fu2016PC}. 
We run our algorithms in a hundred of models, and exhibit several results in Figure \ref{fig:comparison} and Figure \ref{fig:comparison2}. 

In speed, our algorithm just solves several linear systems, which is faster than theirs. The polycube shapes from our and their algorithms are almost the same. However the area and the size of every polycube face is slightly different.  

\begin{figure} [th!] 
    	\centering   
    	\begin{subfigure}[b]{0.5\textwidth}
    		\includegraphics[width=\textwidth]{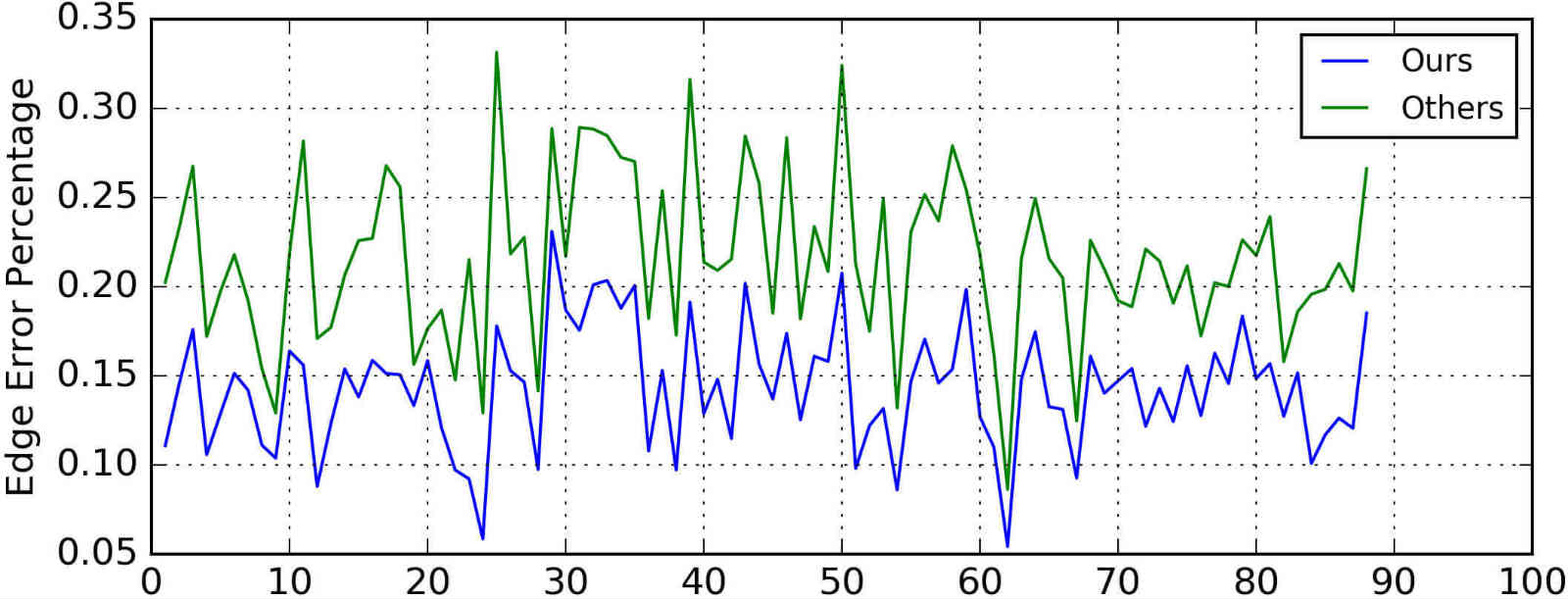}
    		\caption*{ }    
    	\end{subfigure}
    	\caption{The edge   error diagram of our algorithm and \cite{Fu2016PC}.}\label{fig:errorEdge}
    \end{figure}

\begin{figure} [th!] 
	\centering   
	\begin{subfigure}[b]{0.5\textwidth}
		\includegraphics[width=\textwidth]{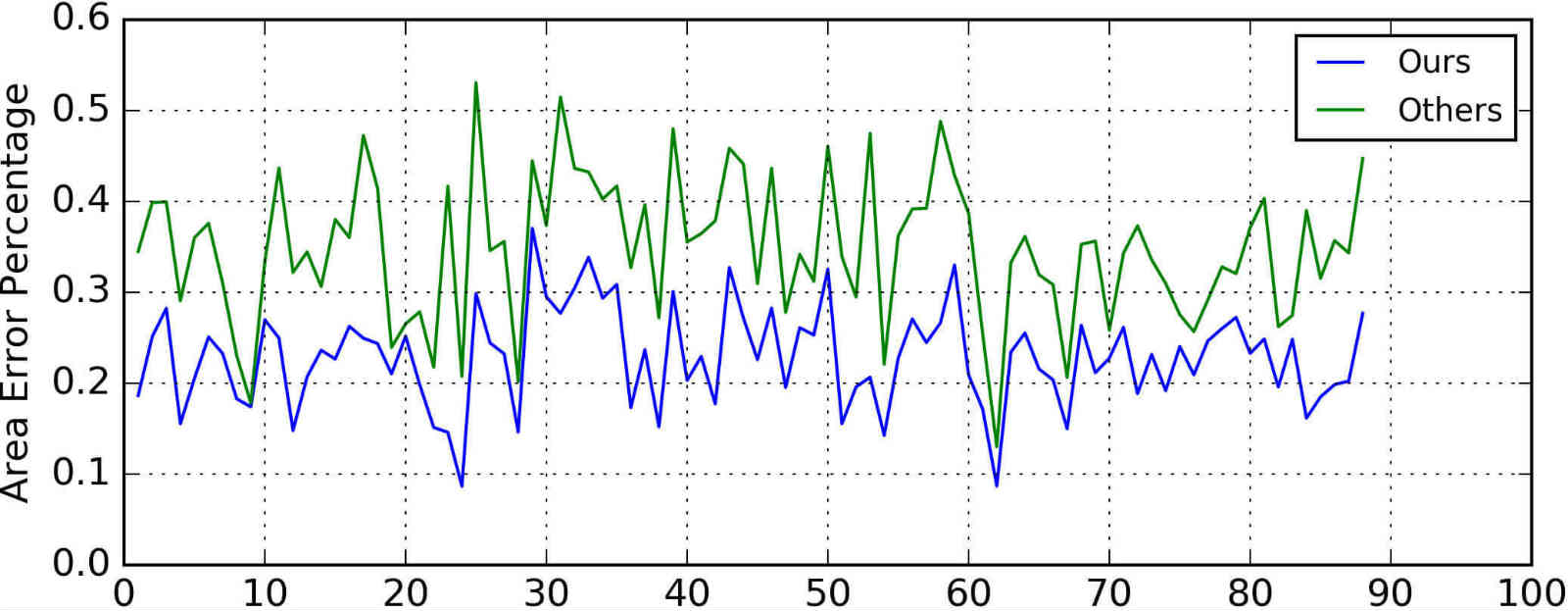}
		\caption*{ }   
	\end{subfigure}
	\caption{The  area error diagram our algorithm and \cite{Fu2016PC}.}\label{fig:errorArea}
   \end{figure}

We compute the edge and area errors for each model with our and their polycube results. 
The error ratios of one hundred models from two methods are displayed in Figure \ref{fig:errorEdge} and \ref{fig:errorArea}. 
We can conclude that our algorithm can preserve the edge and area much better than the method in \cite{Fu2016PC}.

  \begin{figure*} [th!] 
 	\centering
 	 \begin{subfigure}[b]{0.8\textwidth}  
 		\includegraphics[width=\textwidth]{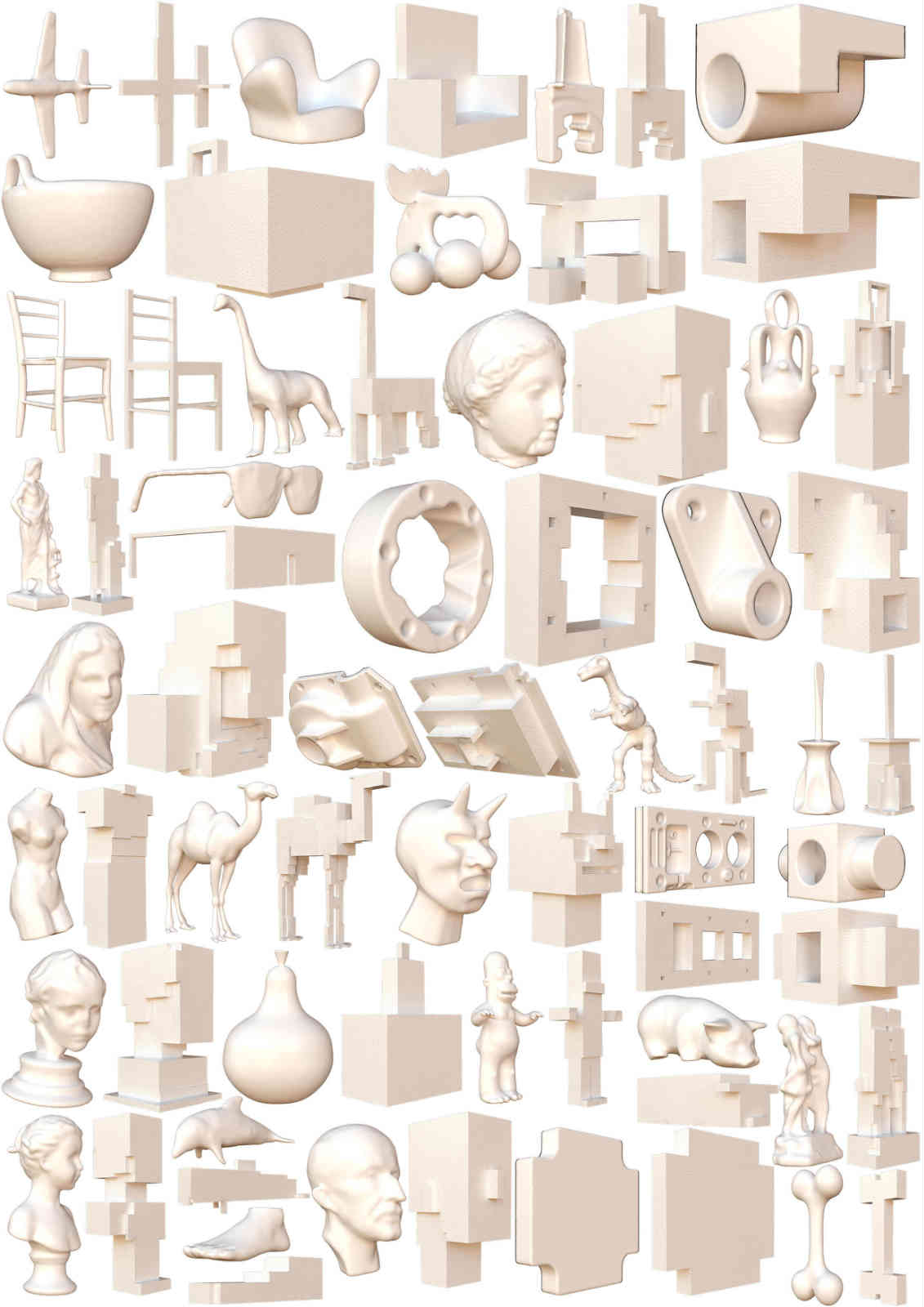}
 		\caption*{ }  
 	\end{subfigure} 
 	\caption{Gallery of our polycube deformations.}\label{fig:gallary}
 \end{figure*}

\section{Conclusion and Future work}

In this paper, we present a robust, efficient polycube deformation algorithm.
Our method is based on the explicitly separating the whole process into three steps. Evey step can be processed with different methods.
 Our method outperforms previous ones in speed, robustness, simplicity, diversity and quality. Although this deformation technique leads to a direct cross-map between original mesh and its polycube,
 we can not guarantee that the map is bijection, and one-to-one. In the future, we plan to contain more constraints in the polycube geometry step to obtain a bijection map.

The quadrangulation and hexahedral meshing from a surface mesh is a crucial problem in graphics community, it is a promising direction to exploit our method in these kinds of applications.

 \section*{Acknowledgments}

We wish to thank anonymous reviewers for encouragements and thoughtful
suggestions.
We are grateful for Professor Steven J. Gortler for the motivation and the insightful guide  which make this paper possible.
  We also thanks Yue Li for the help in our experiments.
Mesh models are courtesy of the Aim@Shape Repository, the Stanford
3D Scanning Repository and the dataset of \cite{Fu2016PC}.  We used
Mitsuba\cite{Mitsuba} for rendering images. Our algorithms are implemented on MeshDGP \cite{Hui:MeshDGP:2016} framework. We also thank Libigl \cite{libigl} for reference.

	
	\bibliographystyle{eg-alpha-doi}
	
	\bibliography{PolyCube}

\newcommand{\etalchar}[1]{$^{#1}$}
\begin{thebibliography}{\uppercase{HWFQ09}}

\bibitem[ACOL00]{alexa2000rigid}
\textsc{Alexa M., Cohen-Or D., Levin D.}:
\newblock As-rigid-as-possible shape interpolation.
\newblock In \emph{Proceedings of the 27th annual conference on Computer
  graphics and interactive techniques} (2000), ACM Press/Addison-Wesley
  Publishing Co., pp.~157--164.

\bibitem[BS08]{DeformationSurvey:2008}
\textsc{Botsch M., Sorkine O.}:
\newblock On linear variational surface deformation methods.
\newblock \emph{IEEE Transactions on Visualization and Computer Graphics 14}, 1
  (2008), 213--230.

\bibitem[CL{\etalchar{*}}10]{chang2010texture}
\textsc{Chang C.-C., Lin C.-Y., et~al.}:
\newblock Texture tiling on 3d models using automatic polycube-maps and wang
  tiles.
\newblock \emph{Journal of Information Science and Engineering 26}, 1 (2010),
  291--305.

\bibitem[CLS16]{cherchi2016polycube}
\textsc{Cherchi G., Livesu M., Scateni R.}:
\newblock Polycube simplification for coarse layouts of surfaces and volumes.
\newblock In \emph{Computer Graphics Forum} (2016), vol.~35, Wiley Online
  Library, pp.~11--20.

\bibitem[CPSS10]{Chao:2010:SGM}
\textsc{Chao I., Pinkall U., Sanan P., Schr{\"o}der P.}:
\newblock A simple geometric model for elastic deformations.
\newblock \emph{ACM Transactions on Graphics 29}, 4 (July 2010), 38:1--38:6.

\bibitem[EM10]{eppstein2010steinitz}
\textsc{Eppstein D., Mumford E.}:
\newblock Steinitz theorems for orthogonal polyhedra.
\newblock In \emph{Proceedings of the twenty-sixth annual symposium on
  Computational geometry} (2010), ACM, pp.~429--438.

\bibitem[FBL16]{Fu2016PC}
\textsc{Fu X., Bai C., Liu Y.}:
\newblock Efficient volumetric polycube-map construction.
\newblock \emph{Computer Graphics Forum (Pacific Graphics) 35}, 7 (2016).

\bibitem[FJFS05]{fan2005mesh}
\textsc{Fan Z., Jin X., Feng J., Sun H.}:
\newblock Mesh morphing using polycube-based cross-parameterization.
\newblock \emph{Computer Animation and Virtual Worlds 16}, 3-4 (2005),
  499--508.

\bibitem[GGH02]{gu2002geometry}
\textsc{Gu X., Gortler S.~J., Hoppe H.}:
\newblock Geometry images.
\newblock In \emph{ACM Transactions on Graphics (TOG)} (2002), vol.~21, ACM,
  pp.~355--361.

\bibitem[GSZ11]{gregson2011all}
\textsc{Gregson J., Sheffer A., Zhang E.}:
\newblock All-hex mesh generation via volumetric polycube deformation.
\newblock In \emph{Computer graphics forum} (2011), vol.~30, Wiley Online
  Library, pp.~1407--1416.

\bibitem[GXH{\etalchar{*}}13]{garcia2013interactive}
\textsc{Garcia I., Xia J., He Y., Xin S.-Q., Patow G.}:
\newblock Interactive applications for sketch-based editable polycube map.
\newblock \emph{Visualization and Computer Graphics, IEEE Transactions on 19},
  7 (2013), 1158--1171.

\bibitem[HJS{\etalchar{*}}14]{huang2014ell}
\textsc{Huang J., Jiang T., Shi Z., Tong Y., Bao H., Desbrun M.}:
\newblock L1-based construction of polycube maps from complex shapes.
\newblock \emph{ACM Transactions on Graphics (TOG) 33}, 3 (2014), 25.

\bibitem[HWFQ09]{he2009divide}
\textsc{He Y., Wang H., Fu C.-W., Qin H.}:
\newblock A divide-and-conquer approach for automatic polycube map
  construction.
\newblock \emph{Computers \& Graphics 33}, 3 (2009), 369--380.

\bibitem[HXH10]{han2010hexahedral}
\textsc{Han S., Xia J., He Y.}:
\newblock Hexahedral shell mesh construction via volumetric polycube map.
\newblock In \emph{Proceedings of the 14th ACM Symposium on Solid and Physical
  Modeling} (2010), ACM, pp.~127--136.

\bibitem[HYLG08]{he2008harmonic}
\textsc{He Y., Yin X., Luo F., Gu X.}:
\newblock Harmonic volumetric parameterization using green’s functions on
  star shapes.
\newblock In \emph{Symposium on geometry processing, Copenhagen, Denmark}
  (2008).

\bibitem[Jak10]{Mitsuba}
\textsc{Jakob W.}:
\newblock Mitsuba renderer, 2010.
\newblock http://www.mitsuba-renderer.org.

\bibitem[JP{\etalchar{*}}16]{libigl}
\textsc{Jacobson A., Panozzo D., et~al.}:
\newblock {libigl}: A simple {C++} geometry processing library, 2016.
\newblock http://libigl.github.io/libigl/.

\bibitem[LGW{\etalchar{*}}07]{li2007harmonic}
\textsc{Li X., Guo X., Wang H., He Y., Gu X., Qin H.}:
\newblock Harmonic volumetric mapping for solid modeling applications.
\newblock In \emph{Proceedings of the 2007 ACM symposium on Solid and physical
  modeling} (2007), ACM, pp.~109--120.

\bibitem[LJFW08]{lin2008automatic}
\textsc{Lin J., Jin X., Fan Z., Wang C.~C.}:
\newblock Automatic polycube-maps.
\newblock In \emph{Advances in Geometric Modeling and Processing}. Springer,
  2008, pp.~3--16.

\bibitem[LVS{\etalchar{*}}13]{Livesu:2013:PolyCut}
\textsc{Livesu M., Vining N., Sheffer A., Gregson J., Scateni R.}:
\newblock Polycut: Monotone graph-cuts for polycube base-complex construction.
\newblock \emph{Transactions on Graphics (Proc. SIGGRAPH ASIA 2013) 32}, 6
  (2013).
\newblock \href {http://dx.doi.org/10.1145/2508363.2508388}
  {\path{doi:10.1145/2508363.2508388}}.

\bibitem[LZLW15]{liu2015feature}
\textsc{Liu L., Zhang Y., Liu Y., Wang W.}:
\newblock Feature-preserving t-mesh construction using skeleton-based
  polycubes.
\newblock \emph{Computer-Aided Design 58} (2015), 162--172.

\bibitem[SA07]{ARAP_modeling:2007}
\textsc{Sorkine O., Alexa M.}:
\newblock As-rigid-as-possible surface modeling.
\newblock In \emph{Proceedings of EUROGRAPHICS/ACM SIGGRAPH Symposium on
  Geometry Processing} (2007), pp.~109--116.

\bibitem[THCM04]{tarini2004polycube}
\textsc{Tarini M., Hormann K., Cignoni P., Montani C.}:
\newblock Polycube-maps.
\newblock \emph{ACM Transactions on Graphics (TOG) 23}, 3 (2004), 853--860.

\bibitem[WHL{\etalchar{*}}08]{wang2008polycube}
\textsc{Wang H., He Y., Li X., Gu X., Qin H.}:
\newblock Polycube splines.
\newblock \emph{Computer-Aided Design 40}, 6 (2008), 721--733.

\bibitem[WJH{\etalchar{*}}08]{wang2008user}
\textsc{Wang H., Jin M., He Y., Gu X., Qin H.}:
\newblock User-controllable polycube map for manifold spline construction.
\newblock In \emph{Proceedings of the 2008 ACM symposium on Solid and physical
  modeling} (2008), ACM, pp.~397--404.

\bibitem[WYZ{\etalchar{*}}11]{wan2011topology}
\textsc{Wan S., Yin Z., Zhang K., Zhang H., Li X.}:
\newblock A topology-preserving optimization algorithm for polycube mapping.
\newblock \emph{Computers \& Graphics 35}, 3 (2011), 639--649.

\bibitem[XHY{\etalchar{*}}10]{xia2010direct}
\textsc{Xia J., He Y., Yin X., Han S., Gu X.}:
\newblock Direct-product volumetric parameterization of handlebodies via
  harmonic fields.
\newblock In \emph{Shape Modeling International Conference (SMI), 2010} (2010),
  IEEE, pp.~3--12.

\bibitem[XZWB06]{xu2006poisson}
\textsc{Xu D., Zhang H., Wang Q., Bao H.}:
\newblock Poisson shape interpolation.
\newblock \emph{Graphical Models 68}, 3 (2006), 268--281.

\bibitem[YL08]{yao2008adaptive}
\textsc{Yao C.-Y., Lee T.-Y.}:
\newblock Adaptive geometry image.
\newblock \emph{Visualization and Computer Graphics, IEEE Transactions on 14},
  4 (2008), 948--960.

\bibitem[YZX{\etalchar{*}}04]{yu2004mesh}
\textsc{Yu Y., Zhou K., Xu D., Shi X., Bao H., Guo B., Shum H.-Y.}:
\newblock Mesh editing with poisson-based gradient field manipulation.
\newblock \emph{ACM Transactions on Graphics (TOG) 23}, 3 (2004), 644--651.

\bibitem[ZG16]{Hui:2016:Deform}
\textsc{Zhao H., Gortler S.~J.}:
\newblock A report on shape deformation with a stretching and bending energy.
\newblock \emph{CoRR abs/1603.06821} (2016).
\newblock URL: \url{http://arxiv.org/abs/1603.06821}.

\bibitem[Zha16]{Hui:MeshDGP:2016}
\textsc{Zhao H.}:
\newblock {MeshDGP}: A {C Sharp} mesh processing framework, 2016.
\newblock http://meshdgp.github.io/.

\bibitem[ZRKS05]{zayer2005harmonic}
\textsc{Zayer R., R{\"o}ssl C., Karni Z., Seidel H.-P.}:
\newblock Harmonic guidance for surface deformation.
\newblock In \emph{Computer Graphics Forum} (2005), vol.~24, Wiley Online
  Library, pp.~601--609.

\end{thebibliography}
	
	\newpage

\end{document}